\documentclass[aps,prc,twocolumn,amsmath,amssymb,floatfix]{revtex4-1}

\usepackage{graphicx}
\usepackage{dcolumn}
\usepackage{bm}
\usepackage{changepage}
\usepackage{threeparttable}

\newcommand{\be}{\begin{eqnarray}}
\newcommand{\ee}{\end{eqnarray}}
\newcommand{\bc}{\begin{center}}
\newcommand{\ec}{\end{center}}
\newcommand{\nn}{\nonumber \\}

\begin{document}


\title{Two-Pion production in the second resonance region in $\pi^- p$ collisions with HADES}

\author{J.~Adamczewski-Musch$^{4}$, O.~Arnold$^{10,9}$, E.T.~Atomssa$^{16}$, C.~Behnke$^{8}$, 
A.~Belounnas$^{16}$, A.~Belyaev$^{7}$, J.C.~Berger-Chen$^{10,9}$, J.~Biernat$^{3}$, A.~Blanco$^{1}$, 
C.~~Blume$^{8}$
% ...
}

\affiliation{(HADES Collaboration)$^{f}$}
\author{A.~V. Sarantsev$^{20,21}$, V.~A. Nikonov$^{20,21}$}

\affiliation{
 \\\mbox{$^{1}$LIP-Laborat\'{o}rio de Instrumenta\c{c}\~{a}o e F\'{\i}sica Experimental de Part\'{\i}culas , 3004-516~Coimbra, Portugal}\\
\mbox{$^{2}$Institute of Nuclear Physics, Polish Academy of Sciences, 31342~Krak\'{o}w, Poland}\\
\mbox{$^{3}$Smoluchowski Institute of Physics, Jagiellonian University of Cracow, 30-059~Krak\'{o}w, Poland}\\
\mbox{$^{4}$GSI Helmholtzzentrum f\"{u}r Schwerionenforschung GmbH, 64291~Darmstadt, Germany}\\
\mbox{$^{5}$Technische Universit\"{a}t Darmstadt, 64289~Darmstadt, Germany}\\
\mbox{$^{6}$Institut f\"{u}r Strahlenphysik, Helmholtz-Zentrum Dresden-Rossendorf, 01314~Dresden, Germany}\\
\mbox{$^{7}$Joint Institute of Nuclear Research, 141980~Dubna, Russia}\\
\mbox{$^{8}$Institut f\"{u}r Kernphysik, Goethe-Universit\"{a}t, 60438 ~Frankfurt, Germany}\\
\mbox{$^{9}$Excellence Cluster 'Origin and Structure of the Universe' , 85748~Garching, Germany}\\
\mbox{$^{10}$Physik Department E62, Technische Universit\"{a}t M\"{u}nchen, 85748~Garching, Germany}\\
\mbox{$^{11}$II.Physikalisches Institut, Justus Liebig Universit\"{a}t Giessen, 35392~Giessen, Germany}\\
\mbox{$^{12}$Institute for Nuclear Research, Russian Academy of Science, 117312~Moscow, Russia}\\
\mbox{$^{13}$Institute of Theoretical and Experimental Physics, 117218~Moscow, Russia}\\
\mbox{$^{14}$National Research Nuclear University MEPhI (Moscow Engineering Physics Institute), 115409~Moscow, Russia}\\
\mbox{$^{15}$Department of Physics, University of Cyprus, 1678~Nicosia, Cyprus}\\
\mbox{$^{16}$Institut de Physique Nucl\'{e}aire, CNRS-IN2P3, Univ. Paris-Sud, Universit\'{e} Paris-Saclay, F-91406~Orsay Cedex, France}\\
\mbox{$^{17}$Nuclear Physics Institute, The Czech Academy of Sciences, 25068~Rez, Czech Republic}\\
\mbox{$^{18}$LabCAF. F. F\'{\i}sica, Univ. de Santiago de Compostela, 15706~Santiago de Compostela, Spain}\\
\mbox{$^{19}$Uniwersytet Warszawski - Instytut Fizyki Do\'{s}wiadczalnej, 02-093~Warszawa, Poland}\\ 
\mbox{$^{20}$NRC "Kurchatov Institute", PNPI, 188300, Gatchina, Russia}\\ 
\mbox{$^{21}$Helmholtz-Institut f\"{u}r Strahlen- und Kernphysik, Universit\"{a}t Bonn, Germany}\\ 
\\
\mbox{$^{a}$ also at Coimbra Polytechnic - ISEC, ~Coimbra, Portugal}\\
\mbox{$^{b}$ also at ExtreMe Matter Institute EMMI, 64291~Darmstadt, Germany}\\
\mbox{$^{c}$ also at Technische Universit\"{a}t Dresden, 01062~Dresden, Germany}\\
\mbox{$^{d}$ also at Frederick University, 1036~Nicosia, Cyprus}\\
\mbox{$^{e}$ also at Dipartimento di Fisica and INFN, Universit\`{a} di Torino, 10125 Torino, Italy}\\
\mbox{$^{f}$ e-mail: hades-info@gsi.de}\\
\mbox{$\dagger$ deceased}
} 

\date{\today}

\begin{abstract}
\vspace{12 mm}
Pion induced reactions provide unique opportunities for an 
unambiguous description of baryonic resonances and their coupling channels
 by means of a partial wave analysis. Using the secondary pion beam at SIS18,
 the two pion production in the second resonance region has been investigated 
 to unravel the role of the $N(1520) \frac{3}{2}^-$ resonance in the
 intermediate $\rho$ production. Results on exclusive channels with one
 pion ($\pi^{-}p$) and two pions ($\pi^{+}\pi^{-}n$, $\pi^{0}\pi^{-}p$) in
 the final state measured in the $\pi^{-}-p$ reaction at four different 
pion beam momenta (0.650, 0.685, 0.733, and 0.786 GeV/c) are presented.
 The excitation function of the different partial waves 
and $\Delta\pi$, $N\sigma$ and $N\rho$ isobar configurations is 
obtained, using the Bonn-Gatchina partial wave analysis. 
The $N(1520) \frac{3}{2}^-$ resonance is found to dominate
 the $N\rho$ final state with the branching ratio  $BR=12.2 \pm1.9\%$.  


\end{abstract}

\maketitle


\section{Introduction}
\label{sec1}

Pion ($\pi$) scattering on a nucleon ($N$) is an ideal tool to study baryon resonance production and their decays. In such experiments, pion-nucleon resonances are excited at a fixed mass, defined by the energy ($\sqrt{s}$ ) of the $\pi N$ collision system, and can be analysed via their decay products. The analysis of pion elastic scattering data taken in the 1980s was the primary source of our understanding of the nucleon excitation spectrum till the end of the 1990s \cite{Hoh:83,Cut:79,Arndt:1990bp,Arn:06}. The results revealed too few states as compared to the predictions of various quark models, leading to the famed "missing resonances" problem (for a review see Ref. \cite{Crede:2013sze}). It motivated the next round of high precision experiments conducted with electron and photon beams to search for new states, particularly those which could be missed due to a small pion-nucleon coupling. Indeed, several new resonances have been established in photoproduction of various meson final states using the advantage of polarization observables. The  analysis of the new data led to the observation of six new $N^*$ and $\Delta^*$ states \cite{Anisovich:2011fc} which were included in the Review of Particle Physics \cite{pdg}. In contrast, the database for pion-induced reactions has not been updated and is of much lesser precision though it provides important information which is not directly accessible from photoproduction reactions. For example, the pion photoproduction data only give access to the product of the photon helicity couplings and couplings to the $\pi N$ channel. Hence, the analysis of the elastic $\pi N$ scattering allows to obtain directly information about the $\pi N$ branching ratios of the resonances and therefore about their $\pi N$ couplings. The combined analysis of the pion induced and  photoproduction data defines both the resonance helicity and pion-nucleon couplings. Moreover, the analysis of the pion-nucleon collision data is notably simpler than the analysis of the meson photoproduction data. For the full reconstruction of the meson photoproduction amplitudes it is indeed necessary to measure at least eight observables with a good precision and angular coverage, while in the pion-induced experiments three observables provide the complete database. Furthermore, modern partial wave data analysis techniques enable a combined multi-channel analysis which fully exploits unitarity constraints and allows to study subtle particle correlations on an event-by-event basis.

Studies of two-pion final states are particularly important because they contribute more than $50\%$ to the total inelasticity. The most extensive analysis of the two-pion production in pion-induced reactions was achieved by Manley $et$ $al.$  (\cite{Man:84}, with an update \cite{Man:92}) within the isobar approximation. The analysis relied on 241214 bubble chamber events collected before 1984 (without the $\pi^{-}p\rightarrow\pi^{0}\pi^{0}n$ channel), in the energy region $\sqrt{s}=1.32-1.93$ GeV. The single energy solutions were extracted for 22 energy bins, providing branching ratios to the $\rho$N, $\Delta\pi$ and $\sigma$N final states for various $N^*/\Delta$ resonances in this mass range. Since then, only a few experimental data have become available for the reaction $\pi^{-}p\rightarrow\pi^{0}\pi^{0}n$ at $\sqrt{s} =$1.213-1.527~GeV \cite{Pra:04} and $\pi^{-}p\rightarrow\pi^{+}\pi^{-}n$ at $\sqrt{s} = 1.257-1.302$~GeV \cite{Ker:98} and $\sqrt{s} = 2.060$~GeV \cite{Ale:98}. The $\gamma p\to 2\pi^0 p$ data were measured by the A2 and CBELSA/TAPS collaborations and analysed together with single meson photoproduction data, $\pi N$ elastic data and  $\pi^{-}p\rightarrow \pi^{0}\pi^{0}n$ data in \cite{Anisovich:2011fc,Shklyar:2014kra,Sokhoyan:2015fra}. These data provided results concerning cascade decay transitions for the resonances with masses above 1700 MeV. However, they could not provide any information about the decay of the baryon resonances into  $\rho N$, which requires final states with charged pions. Some results were obtained for the $N(1520) \frac{3}{2}^-$ and for the $N(1440)\frac{1}{2}^+$ from the analysis of two-pion production in electron scattering experiments \cite{Mokeev:2012vsa} pointing $e.g.$ to smaller branching ratios to the $\rho N$ channels than that in \cite{Man:92}. It should also be noted that the Particle Data Group recently removed all the information concerning $N^*$ and $\Delta$'s branching ratios to the $\rho N$ channel from the Review of Particle Physics \cite{pdg}. Therefore, a precise determination of the $\rho N$ couplings of excited baryons is clearly lacking  new high precision data for two-pion production channels with charged pions in the exit channels in pion induced reactions.


The resonance decay into two-pion final states with at least one charged pion is particularly well suited for studies of the $\rho$ meson-baryon coupling, because of its almost $100\%$ decay branching into a pion pair. The studies of the vector meson-baryon interaction is motivated by the Vector Meson Dominance (VMD) model  \cite{Sakurai:1960ju,Kroll:1967it}, predicting the low- mass $\rho/\omega/\phi$ vector mesons (V) as mediating fields in the hadron-photon interactions. The aim of this model in the baryon sector is to provide a comprehensive description of both radiative resonance decays $R \rightarrow N \gamma, N \gamma^*\rightarrow N e(\mu)^+e(\mu)^-$ and mesonic $R \rightarrow NV$ transitions, with the $\rho$-meson playing the most important role due to its stronger coupling to baryons. It provides a foundation for the description of low-mass ($M_{l^+l^-}< 1$~GeV/c$^2$) dilepton ($l^+l^-$) production in elementary processes involving hadron decays in vacuum as well as from dense and hot nuclear matter. In particular, a successful description of the dilepton spectra measured in heavy-ion collisions from RHIC to GSI/SIS18 (for a recent paper, see \cite{Adamczewski-Musch:2019byl}) calls for special attention. The results show that the virtual photon radiation from the hot and dense zone of the collisions can be described by intermediate $\rho-$mesons with a strongly modified spectral function. The interpretation relies on the microscopic calculations of the in-medium $\rho$-meson spectral function and application of VMD to hot and dense hadron gas radiation. The computation of the $\rho$ spectral function includes interactions with mesons and baryons in the fireball and reveals the leading role of the latter (for a theory review, see \cite{Rapp:1999ej}).
The main effects are driven by  excitation of resonance-nucleon hole states induced by the  $\rho$-meson, with $N^*,\Delta^*$ resonances playing the main role. In particular strong effects on the $\rho$ spectral function for a small relative momentum of the meson w.r.t the medium were extracted for S-wave $N-\rho$ resonances like $N(1520)\frac{3}{2}^-$, $\Delta(1620)\frac{1}{2}^+$  and $\Delta(1700)\frac{3}{2}^-$ \cite{Peters:1997va,Post:2000qi}. 
Alternatively, more phenomenological approaches, assuming two-step processes $R\rightarrow  N\rho \rightarrow Ne^+e^-$ and the meson collision broadening are used in the transport codes GiBUU  \cite{Bus:12,Wei:12}, UrQMD \cite{Bas:98}, HSD \cite{Bratkovskaya:2013vx}  and  SMASH  \cite{Staudenmaier:2017vtq}, showing also the dominant role of the $\rho$ meson in dilepton emission. The calculations are constrained by results on the resonance photo-excitation and the resonance  decays $R\rightarrow N\rho$. More direct tests of the applicability of  VMD to the baryon resonance transitions into dileptons require measurements of $R \rightarrow N l^+l^-$ and are currently studied with the HADES detector (see \cite{Adamczewski-Musch:2017hmp, Adamczewski-Musch:2017oij, HADES:2011ab, Agakishiev:2014wqa}). Indeed, it is not clear whether the formulation of VMD with a  baryon coupling to one meson (monopole approximation) is valid or whether more refined approaches are required. In  \cite{Faessler:2000md}, it is argued that the monopole formulation of the VMD model with the couplings of the baryon resonances deduced from the mesonic decays given in  \cite{Man:84,Man:92} tends to overestimate the branchings for radiative decays. These inconsistencies can be removed using various extensions of VMD.  For example, an additional direct coupling of the resonances to photons  \cite{OConnell:1995nse} can be introduced, in combination with a vanishing $\rho \gamma$ coupling in the limit of real photon which allows to fit radiative and mesonic decays independently  \cite{Friman:1997tc,Post:2000rf}.  Interferences with higher excited vector meson states also allow to solve this problem \cite{Krivoruchenko:2001jk}.  

From the above-mentionned motivations, it is clear that new experimental data on baryon resonance decays into two-pion final states from pion-induced reactions are urgently needed. The HADES experiment at GSI has started a systematic investigation of baryon resonance excitations with an energy scan in the second resonance region, focusing on two types of reactions: (i) two-pion ($\pi^+\pi^-$, $\pi^0\pi^-$) and (ii) dilepton production. The main objective is understanding the role of vector mesons in the baryon resonance transitions, and, in particular, in the dilepton decay channels. The upcoming studies will complement former results on  $\Delta(1232)\rightarrow pe^+e^-$ and higher mass resonance transitions obtained with proton-proton reactions \cite{Adamczewski-Musch:2017hmp, Adamczewski-Musch:2017oij, HADES:2011ab, Agakishiev:2014wqa}. 

In this work, we present the results on the two-pion production. 
The paper is organized as follows: Section II presents the details of the experimental set-up and the characteristics of the secondary pion beam at GSI; Section (III) describes the separation of events into the final states with two pions $\pi^+\pi^-$, $\pi^0\pi^-$, and elastic scattering used for the normalization, acceptance corrections, and a brief introduction to the Bonn-Gatchina Partial Wave Analysis (PWA) framework, used for the data analysis. Section IV presents the results, with the conclusions in Section V.      

\section{Experimental technique}
\label{sec2}
The High-Acceptance Di-Electron Spectrometer (HADES) \cite{Aga:09} is installed at GSI Darmstadt and operates with primary proton-, ion-, and secondary pion-beams. HADES has been optimized for electron and positron detection but it provides excellent charged hadron ($p$/$K$/$\pi$) identification capabilities, too. The spectrometer consists of six identical sectors, separated by coils of a superconducting toroidal magnet, centred around the beam direction. It covers the full azimuthal angle range, with the exception of the gaps between the sectors, and a broad range of polar angles ($18^{\circ}$-$85^{\circ}$), measured relative to the beam direction. 
\begin{figure}[!h]
\centering

\includegraphics[width=0.5\textwidth]{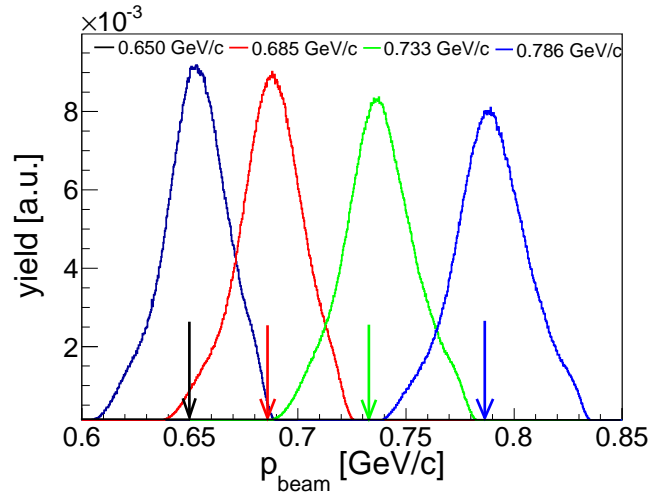}
\caption{Pion beam momentum 
 distributions measured  with the use of the in-beam tracking system CERBEROS. The distributions have been normalized to the area. The arrows indicate the central values of pion beam momenta reconstructed in the analysis, as indicated in the legend. For details see Sec.~\ref{sec3_id}.}
\label{fig1}
\end{figure}

Each sector of the spectrometer is composed of a hadron blind Ring-Imaging Cherenkov Detector (RICH), two Multi-wire Drift Chambers (MDCs) placed in front and two behind the region of the magnetic field, followed by a Multiplicity Electron Trigger Array (META). The latter consists of time-of-flight systems based on: (i) scintillator rods (TOF) (with time resolution $\delta t\simeq 120 $ps) for $\theta > 45^{\circ}$, and (ii) Resistive Plates Counters (RPC) for $\theta < 45^{\circ}$ ($\delta t\simeq 80$ ps), associated with an electromagnetic Pre-Shower detector. Momentum measurement of charged particles is achieved by track reconstruction based on hits detected in the MDCs and the known configuration of the magnetic field with a resolution of $1\%$ to 3$\%$ for pions and protons in the energy range of our experiment.  

 In the experiment, pion projectiles were obtained from a primary $^{14}$N beam provided by the SIS18 synchrotron with an intensity of 0.8-1.0$\times10^{11}$ ions/spill impinging on a beryllium ($^{9}$Be) target \cite{Sim:99,Dia:02}. $\pi^-$ mesons were transported to the HADES target, located 33 m downstream from the production target, within the beam line equipped with 9 quadrupoles and two dipole magnets set for negative particles \cite{Ada:17}.  The pion intensity of $10^{6}$ $\pi^-$/spill reached a maximum for a momentum $p=1.0$ GeV/c and decreased by a factor 2 for the pion momentum range $p=0.65-0.8$ GeV/c covered in this experiment. Four settings of currents in the magnets, corresponding to momenta for the particle on the optical axis of  about $0.650, 0.685, 0.733$, and $0.786 $ GeV/c, were investigated in the measurement. 
 The differential transmission distribution of the pions as a function of the momentum were obtained from dedicated transport calculations and could be approximated by a Gaussian distribution with a typical width of $\sigma \simeq 1.0\%$. Further improvement in the pion beam momentum resolution (down to $\delta p=0.3\%$) was achieved using the dedicated in-beam tracking system CERBEROS (see Refs. \cite{Lal:13,Ada:17}). It was built of silicon strip detectors arranged in two stations, placed upstream of the HADES target: the first one close to the dispersive plane of the beam line ($\simeq 21$m), and the second one $\simeq 3$m before the target, respectively. The detectors provide ($x, y$) coordinates of pion hits, with a precision given by the $780~\mu$m pitch of the strips, which are used to determine the pion beam momentum. Independent measurements using  a proton beam allowed to verify the transport coefficients used in the beamline calculations, as described in detail in Ref. \cite{Ada:17}. The resulting reconstructed momentum distributions of the pion beam are shown in Fig. \ref{fig1} for the four settings with indicated central values (for details see Sec.~\ref{sec3_id}).  Fake track suppression for multi-hit events was achieved by requiring strict correlations on hit positions and timing in both tracking stations. The widths of these distributions correspond to a momentum resolution   $\delta p_{beam}\simeq 1.7\%$ and is larger as compared to the result of simulations. The difference is attributed to the vertical transport coefficients which could not be measured in the dedicated experiment using proton beams (for details see \cite{Ada:17}).
 
 The in-beam detector system included also a segmented START detector (about $14\times 14$  mm$^2$), made of mono-crystalline diamond material, placed in front of the HADES target. It provided a $t_0$ measurement with $\delta t \simeq 100$ ps resolution and was used to monitor the beam flux. $66\%$ of all pions passing the START detector were hitting the target, according to detailed beam transport calculations. A polyethylene target (C$_2$H$_4$)$_n$ of $4.6$ cm length and $12$ mm diameter, containing $4\times 10^{23}$ protons/cm$^2$ and $2\times 10^{23}$ C atoms/cm$^2$, was used for the reported measurements of the $\pi^-p$ reactions. To subtract contributions from the $\pi^-C$ reactions, a segmented carbon target (7 sections with a width of 7.1~mm) of the same length was utilized in separate runs. Details of the pion beamline, the pion beam transport calculations and the in-beam detectors can be found in \cite{Ada:17}.  
 
The $t_0$ signal together with a multiplicity condition $M\ge 2$ measured in the TOF detectors were used for the data acquisition trigger. Since the beam halo at the target position extended up to $\pm 60$ mm in the vertical and up to $\pm 25$ mm in the horizontal direction, $i.e$ well beyond the target diameter ($12$ mm), the trigger reduced the contribution of the off-target reactions. The remaining background  was suppressed in the analysis by a condition demanding that the reconstructed primary vertex must be localized within the target region. Admixtures of electrons and muons from pion decays in the beam were estimated by the beam transport simulations to $9.6\%$ and $0.7\%$, for the $p=0.65-0.8$ GeV/c momentum range, respectively, in fair agreement with previous measurements \cite{Dia:02}.      

\section{Data analysis}
\label{sec3}
\subsection{Extraction of $\pi^- p$ elastic scattering, $\pi^-p \to  n \pi^+ \pi^-$  and $\pi^- p \to p \pi^- \pi^0 p$ signals}
\label{sec3_id}

The particle identification (PID) for pions and protons in HADES is provided by conditions defined by correlations between the velocity measured by TOF or RPC detectors and the momentum reconstructed from the track deflection in the magnetic field \cite{Aga:09}. The respective graphical cuts were adjusted using Monte Carlo simulations in order to include $\sim 99\%$ of the signal for the given particles.
Furthermore, the  misidentified particles were eliminated by rigorous conditions on track correlations following from reaction kinematics, as explained below. 

\begin{figure}
\centering
\includegraphics[width=0.45\textwidth]{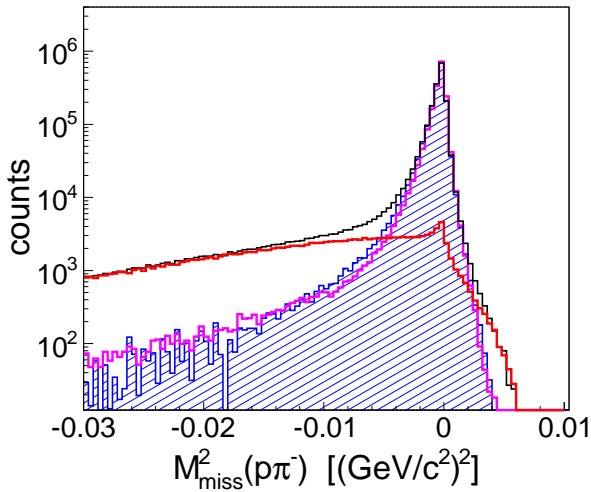}
\caption{$p\pi^{-}$ missing mass squared from the pion-proton system at a beam momentum of $0.685$~GeV/c after elastic scattering selection. Black and red histograms present events from the polyethylene target and contribution of pion reactions on carbon nuclei in the polyethylene target, respectively. Their difference (blue dashed area) corresponds to $\pi^{-}p$ reactions (for details see text). Simulations of $\pi^{-}p$  elastic scattering are shown by the magenta histogram. 
}
\label{fig2}
\end{figure}

The $\pi^{-}p$ elastic scattering was selected demanding co-planarity of the pion and the proton tracks (within $\pm 5^{\circ}$), a condition on the polar emission angles of both tracks: $\tan\theta_{\pi^{-}}$ $\cdot$ $\tan\theta_{p} > 1$ and a cut on the distribution of the pion momentum $p_{CM}$ in the Center of Mass (CM) system  obtained from realistic simulations, as described below. Finally, elastic scattering events were clearly visible in the $\pi^-p$ missing mass (squared) distribution, as shown in Fig. \ref{fig2} by the black histogram.

\begin{figure*}
\centering
 \includegraphics[width=0.48\textwidth]{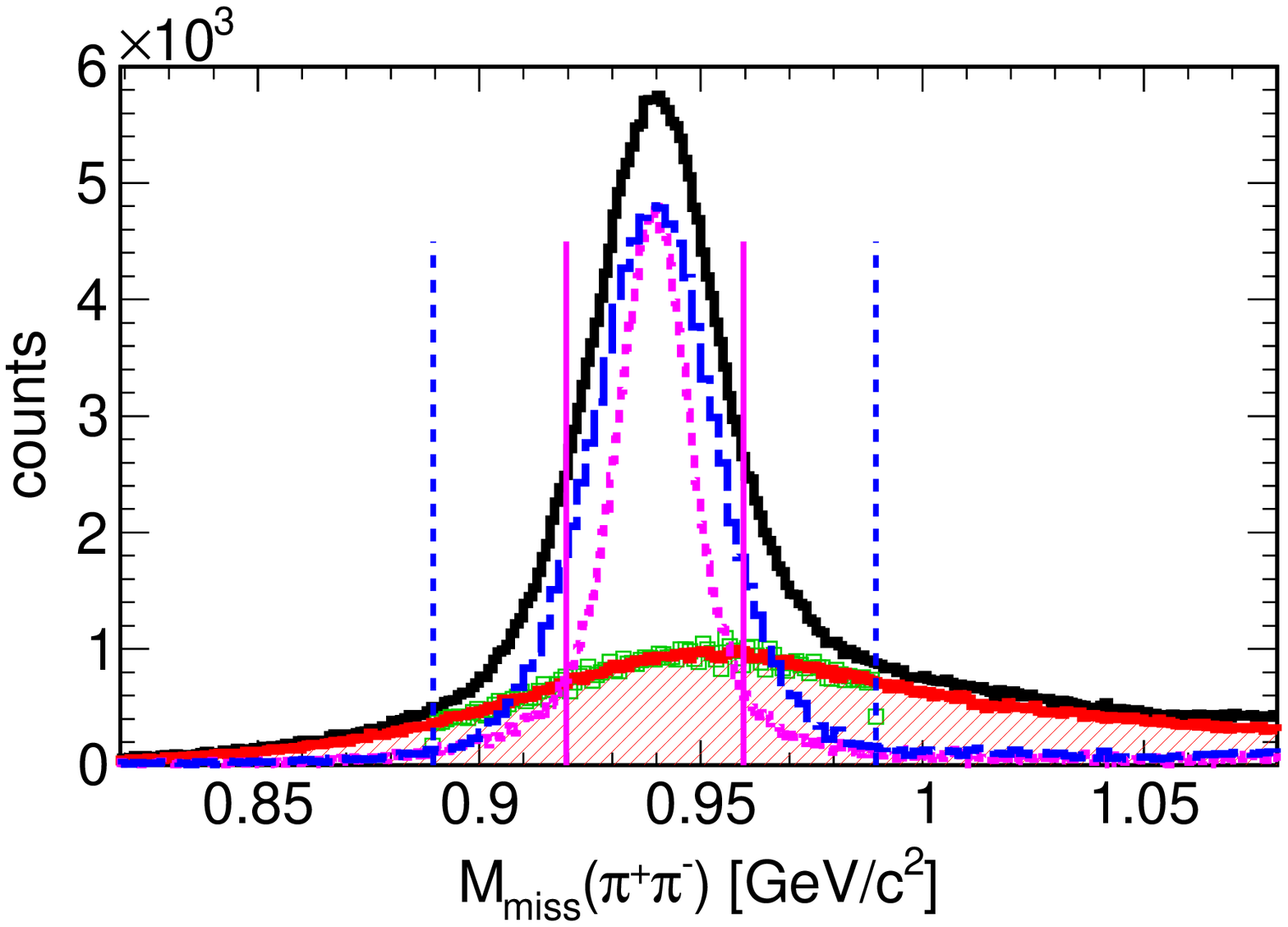} 
 \includegraphics[width=0.48\textwidth]{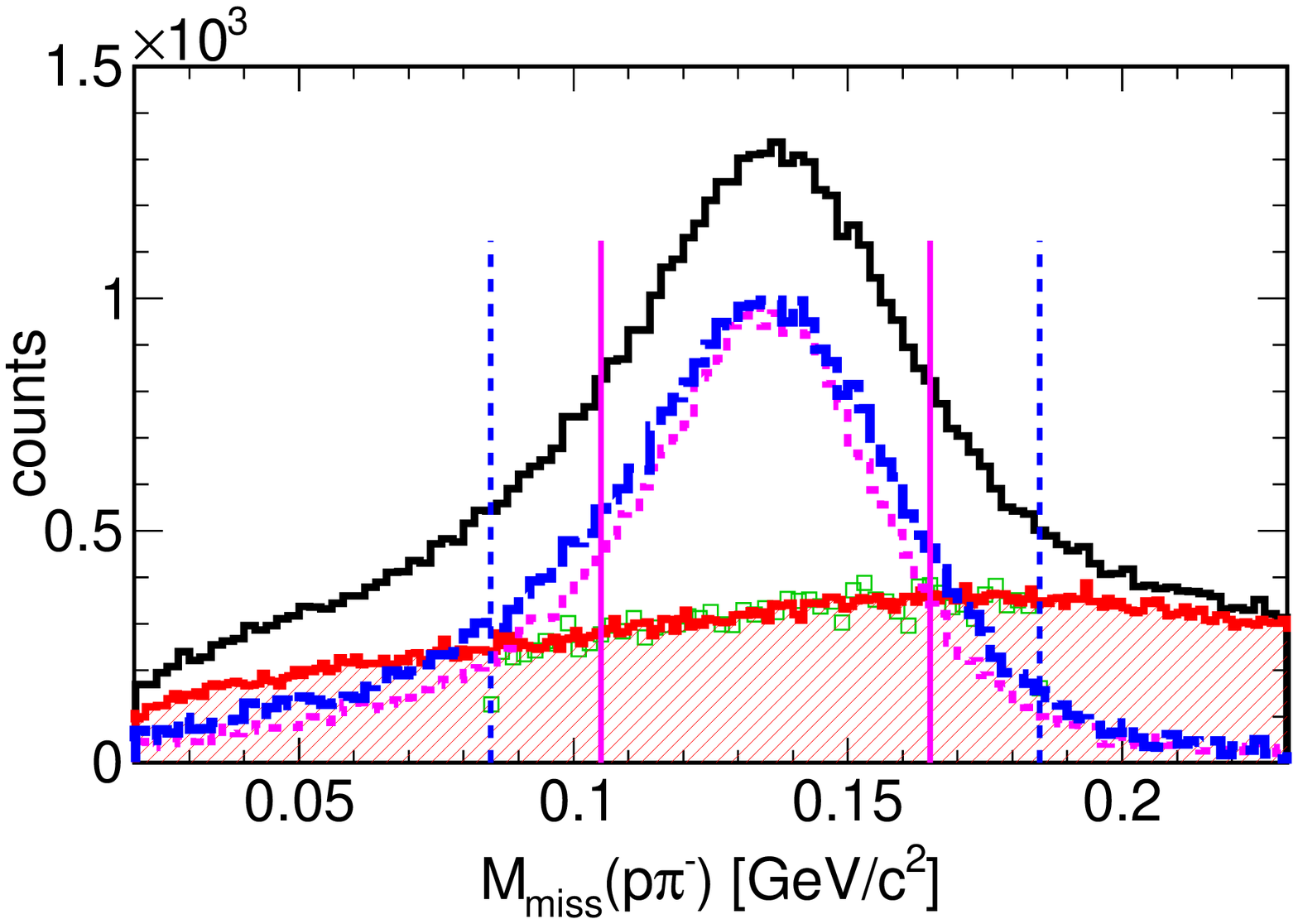}
\caption{Missing masses for two charged pion (left panel) and proton-pion (right panel) systems at a beam momentum of 0.685~GeV/c. Black histogram -  uncorrected data from the polyethylene target, red histogram and red hatched area - events from the carbon target, green open squares (behind red histogram) - events from the polyethylene target matching the events from the carbon target (see text for details), blue dashed histogram - signal from scattering off protons from the polyethylene target, magenta dotted histogram - signal calculated based on information on momentum measured by the pion tracker. The vertical blue dashed lines display a window ($\pm$0.05 GeV/$c^2$) centered around the mass of a neutron and a pion, for the selection of $n\pi^+\pi^-$ and $p\pi^-\pi^0$ events, respectively. The vertical magenta solid lines display a selection window for the signal calculated with the pion tracker information ($\pm$0.02 GeV/$c^2$ around a neutron, and $\pm$0.03 GeV/$c^2$ around a neutral pion, respectively).}
\label{fig3b_0}
\end{figure*}

The identification of two charged pions ($\pi^{+}\pi^{-}$) or a proton and a pion (p$\pi^{-}$) in the final state allows for the complete reconstruction of the kinematics of the exclusive $n\pi^+\pi^-$ and $p\pi^-\pi^0$ channels, respectively, via additional cuts in the respective missing mass distributions around the position of the not detected neutron (Fig. \ref{fig3b_0}, left panel) or  neutral pion (Fig. \ref{fig3b_0}, right panel). In the latter case, the background from elastic scattering was effectively suppressed by veto on the elastic condition, as defined above.  


 In order to obtain a pure sample of events for the above-mentioned exclusive channel analysis, contributions from the $\pi^-C$ reactions must be separated in the sample of events collected with the polyethylene target. Firstly, the relative contribution of the $\pi^-C$ reaction in the polyethylene target was determined with a high precision by an iterative minimization procedure described in details in this section. The input to the procedure consists of: (i) the missing mass distributions of the $\pi^-p$ and the $\pi^+\pi^-$ systems, obtained from the measurements with the polyethylene and the carbon targets, respectively, taken with the same reference beam momentum; (ii) the Monte Carlo simulations of the $\pi^{-}p$ elastic scattering, and the two-pion ($\pi^{+}\pi^{-}$, $\pi^{-}\pi^{0}$) production on the proton target, within the HADES detector acceptance. The simulated channels were reconstructed with the same analysis flow as in the case of the experimental data. In the Monte Carlo simulation, the beam momentum distributions were taken according to the event-by-event pion beam measurements provided by the CERBEROS tracking system (see Fig \ref{fig1}). However, the central values for each pion beam momentum bin  were treated as free parameters, fixed by a fit to the data, as described below. 
 
  In the first step, the distribution of the $\pi^{-}p$ missing mass squared of events selected with the elastic conditions and 
  obtained with the polyethylene target was fitted with the sum of  the $\pi^{-}p$ missing mass squared measured with the carbon target 
  and the Monte Carlo simulations of the elastic reaction ($\pi^-p \to \pi^-p$). Their relative contributions were treated as free parameters of the fit. The minimization was performed by varying also the tracking resolution and the central value of the beam momentum. As a result, a very good description of the elastic peak was achieved (see blue dashed and magenta histograms in Fig. \ref{fig2}), for $p_{beam}=0.685$ GeV/c). The contribution of reactions on carbon was also determined for every pion beam momentum (red histogram in Fig. \ref{fig2}), with an uncertainty of about $1\%$.

In the next step, the minimization procedure was repeated for the events corresponding to the two -pion production channels with the $\pi^{-}p$ and $\pi^{+}\pi^{-}$ missing mass distributions (see Fig.~\ref{fig3b_0}), starting with the parameters from the first step. 
The relative normalization of yields measured on the carbon target was modified only slightly, but the central values of the beam were found to be more sensitive to the expected positions of the $\pi^{0}$ and the neutron peaks ($0.1349$ and $0.9395$ GeV/c$^2$), respectively, as compared to the missing mass peak for the elastic scattering. The determined central values of the beam momentum are: $0.6501$ $\pm 0.002$ GeV/c, $0.6853$ $\pm 0.0025$ GeV/c, $0.7332$ $\pm 0.003$ GeV/c, and $0.786$ $\pm 0.0035$ GeV/c. The errors are related to the uncertainties of the fitting procedure to the aforementioned peak positions, and the variation of fit ranges. The obtained values are lower by $0.005-0.015$ GeV/c than the reference values expected from the magnet settings.  Similar conclusions were derived by the detailed studies of various kinematic constraints derived from elastic scattering only, as discussed in detail in Ref. \cite{Ada:17}. The reason of these discrepancies are not uniquely identified but might be attributed to remanence effects or systematic shifts in the primary beam position on the production target (see discussion in Sec. 4.4 in Ref. \cite{Ada:17}).


Figure~\ref{fig3b_0} shows the missing mass distribution of the $\pi^{-}p$ (right panel) and the $\pi^+\pi^-$ (left panel) systems, as obtained from the analysis of the data collected with the polyethylene target, for the central pion beam momentum $p=0.685$ GeV/c. The total yield (black curves) is separated into the contributions from pion reactions on carbon (red curves) and on protons, within the polyethylene target. The simulated distributions of the subsequent production channels on the proton target are shown by dotted (blue) curves. The magenta curves show, for the sake of comparison, the respective distributions calculated for the pion beam momentum obtained event-by-event from the CERBEROS pion tracker.    

 For the further analysis of kinematic correlations on an event-by-event basis, the separation of $\pi^-p$ from $\pi^-C$ events is necessary. This has been achieved by an event-by-event matching of the carbon signal derived from the polyethylene target ($PE$) with the events from the carbon target ($C$) by comparing  the kinematical characteristics by means of a $\chi^2$ test. 
 The ensemble of events corresponding to reactions on protons is then given as the remaining  events measured on the polyethylene target after subtraction of the carbon-like events found in this procedure.
 The fraction of the carbon contribution in all events collected with the polyethylene target was fixed in this minimization procedure, as described above. 
 
 The events for the matching were grouped into several bins of similar missing mass values. The minimization function $F_{min}$  between tracks measured in reactions with the $PE$ target and the $C$ target is defined as
\begin{equation}
F_{min} = \displaystyle\sum_{i,j,X} w_{i} \left( \frac{X^{PE}_i - X^{C}_j}{X^{C}_j} \right)^{2},
\label{eq_PE_C_MINFUN}
\end{equation}
\noindent
where $i,j$ stand for all combinations of $i$-th event from the polyethylene target with $j$-th event from the carbon target, $X_i$ are the values of the observables : momentum, polar and azimuthal angles of particles, invariant masses, as well as angular observables in the helicity and Gottfried-Jackson reference frames (for definition see Sec. \ref{sec-three-part}), weighted with the empirical weights $w_{i}$. The consistency of the distributions of the observables built from the tagged carbon-like events from the reactions with the $PE$ target and events from the reaction with the $C$ target was carefully investigated for all the observables taken into account. 
In Fig.~\ref{fig3b_0}, the missing mass distributions measured with the carbon target and normalized following the minimization procedure described above (red histograms, $C$ target) are compared with the distributions for events from the polyethylene target tagged in the matching procedure as corresponding to interactions with carbon nuclei (green symbols, carbon-like events from $PE$ target). 
The discrepancy between the signal yields obtained in both procedures was used to estimate the  systematic error of the subtraction procedure, for every observable under consideration. It was found that the distribution of relative errors obtained from the investigated observables is similar to a normal distribution, with a $1 \sigma$ width of $1\%$.

\subsection{\bf {$\pi^{-}p$} elastic scattering and  data normalization}
\label{sec3_nor}
The analysis of the $\pi^{-}p$ elastic scattering events was used to provide the  normalization for the measured yields for the four pion beam momentum settings, using existing  information on the elastic scattering differential cross sections.

\begin{figure*}
\centering
\includegraphics[width=0.40\textwidth]{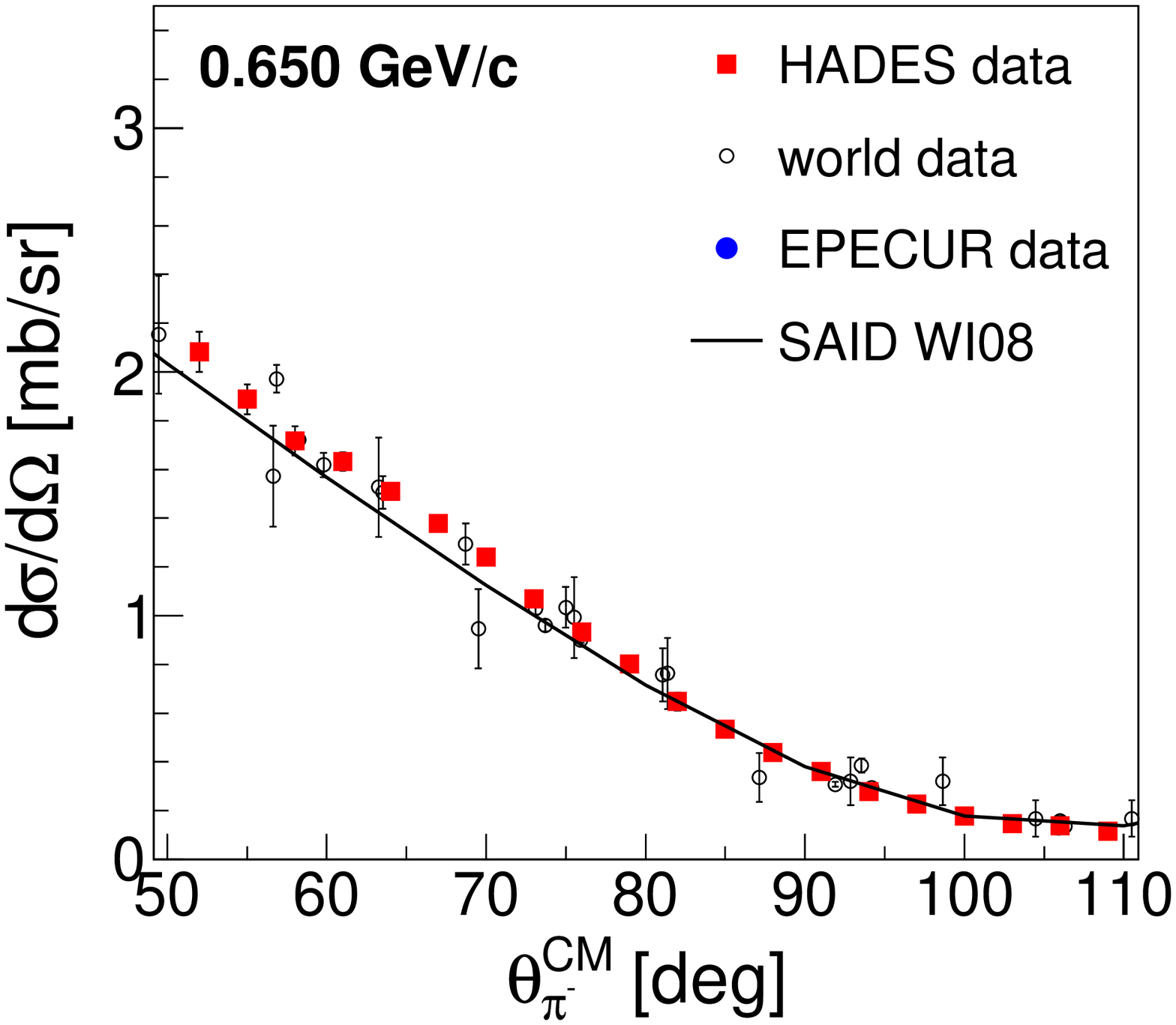}
\includegraphics[width=0.40\textwidth]{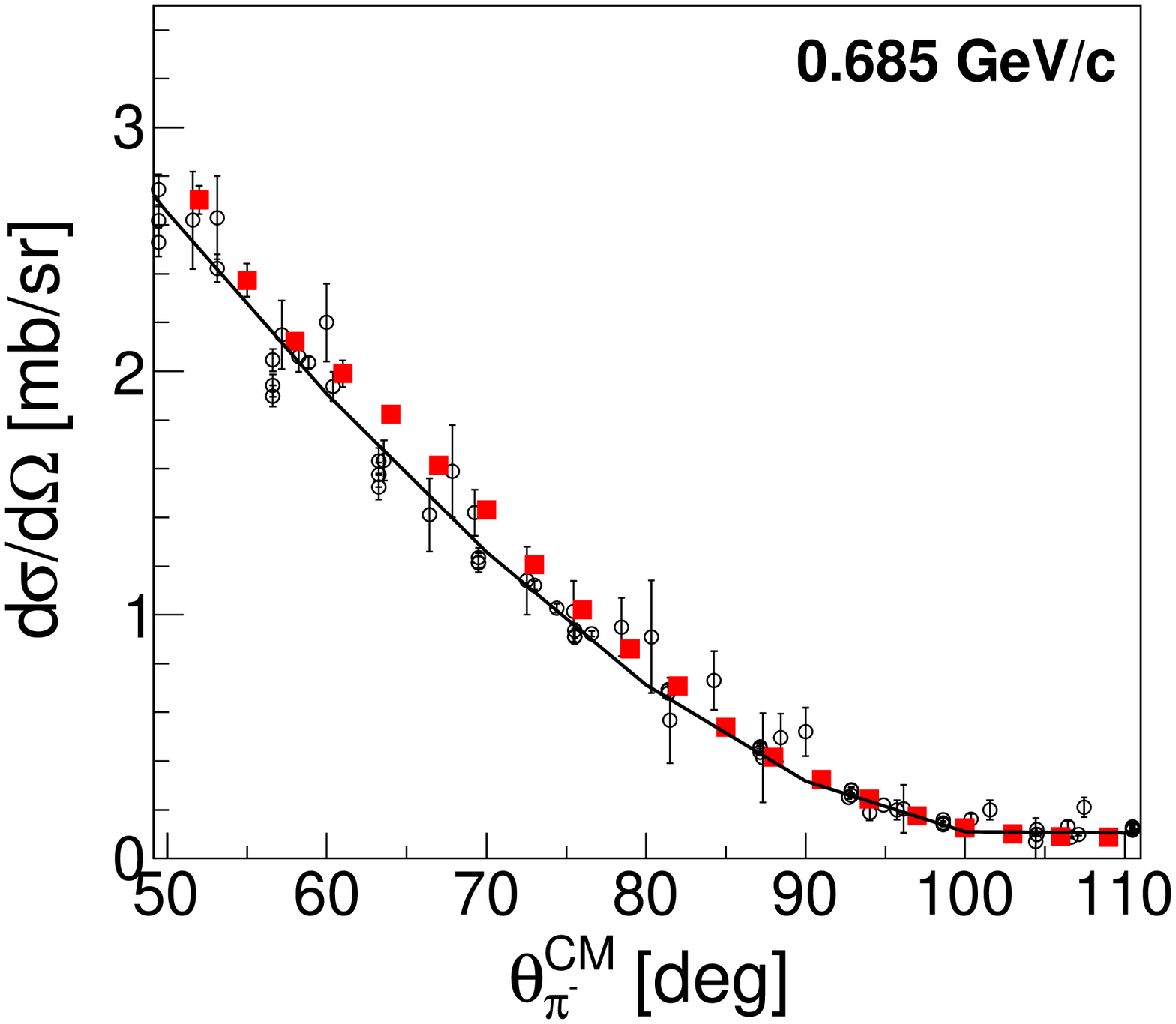}
\includegraphics[width=0.40\textwidth]{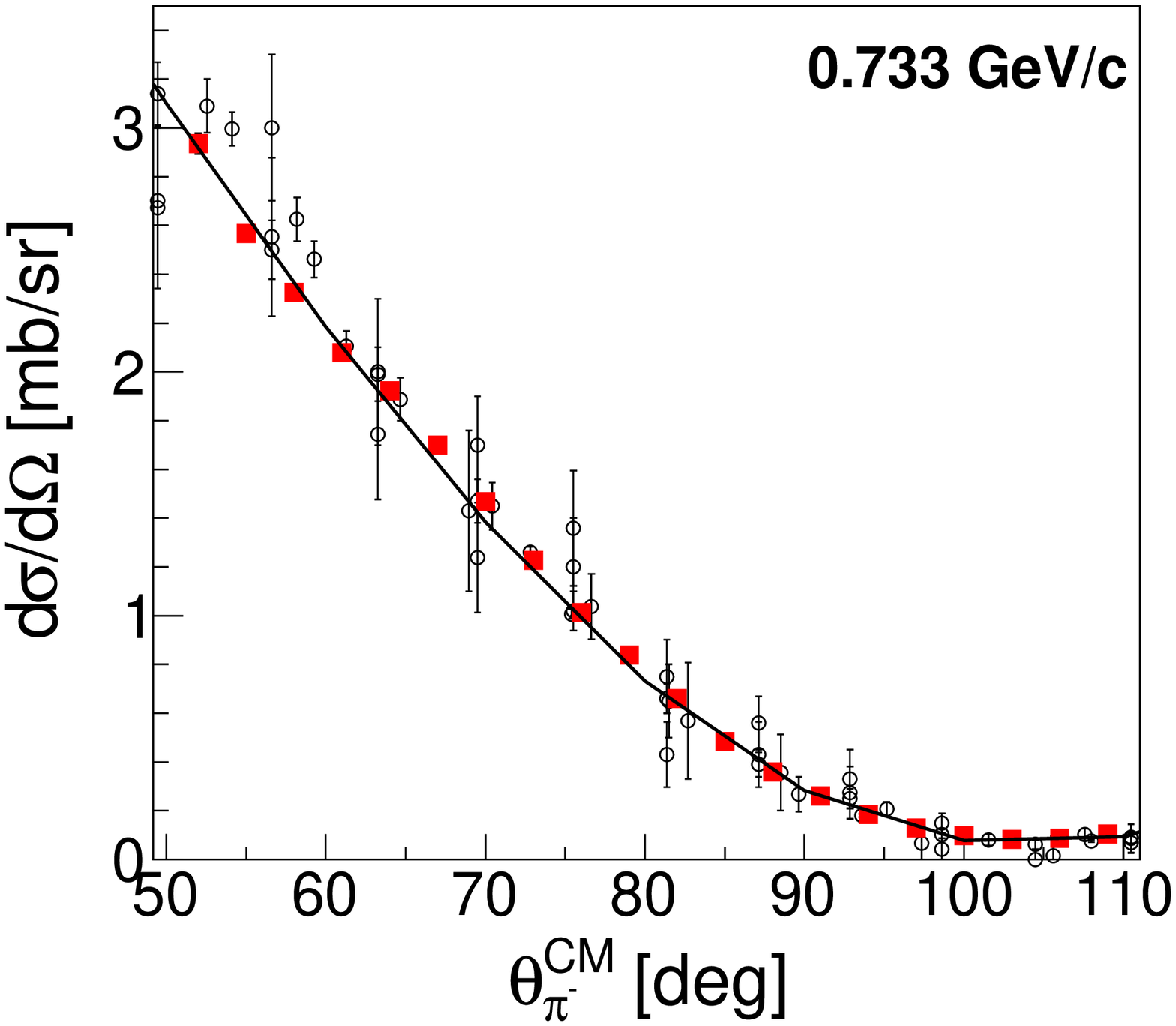}
\includegraphics[width=0.40\textwidth]{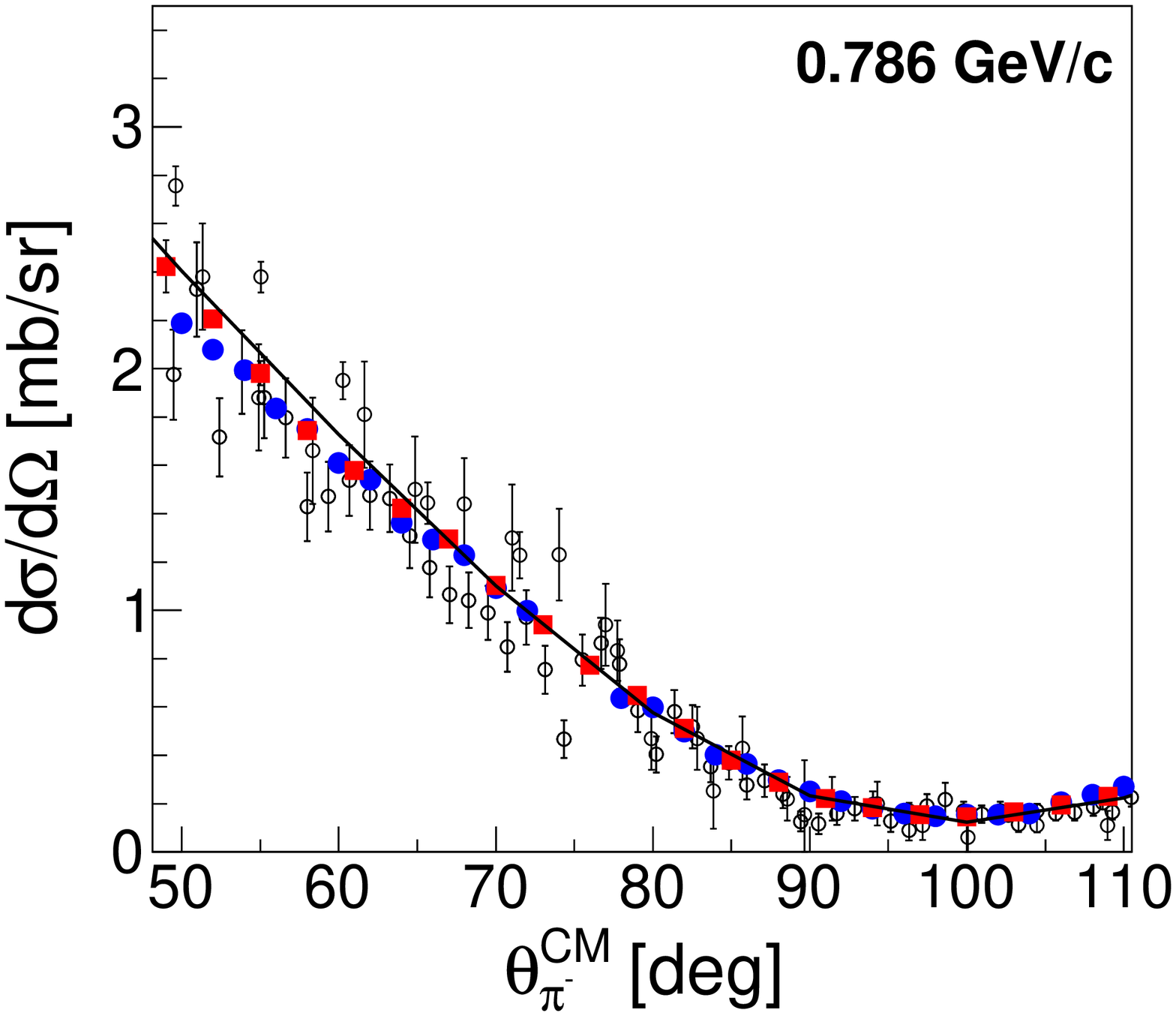}
\caption{$\pi^{-}p$ elastic-scattering cross section at four pion beam momenta $0.650$, $0.685$, $0.733$ and $0.786$ GeV/c. The EPECUR data (blue dots) are available for the highest beam momentum only. The HADES data are presented together with the world data and the current SAID solutions \cite{said}, as specified in the legend. 
}
\label{fig3b_1}
\end{figure*}
 
Acceptance corrections for the measured elastic yields were calculated using a simulation. 
Events were generated in the framework of the PLUTO event generator \cite{Frohlich:2007bi} with an angular parametrization taken from Ref. \cite{Brody1971} and processed through the same analysis and reconstruction procedure as experimental hits. The distribution of elastic scattering events from the Monte Carlo simulation agrees with the measured angular distributions within the HADES acceptance reasonably well. This allows to calculate  a one-dimensional correction defined as the ratio between simulated and reconstructed yields as a function of the scattering angle in the CM frame.  
After applying this correction, the elastic scattering angular distribution was normalized to the average of world data in the $\theta^{CM}_{\pi^{-}}$ range of $59.5^{\circ}-110.5^{\circ}$. The data were selected in the pion beam momentum window $\delta p =\pm 10$ MeV/c centered around the central values given above. The distributions obtained at the four pion beam momenta are presented in Fig.~\ref{fig3b_1}, together with the WI08 SAID solutions \cite{said} and the world data. The 
SAID solution was averaged over $\theta^{CM}_{\pi^{-}}$ bins of  three degrees.

\begin{figure}[!h]
\centering
\includegraphics[width=0.5\textwidth]{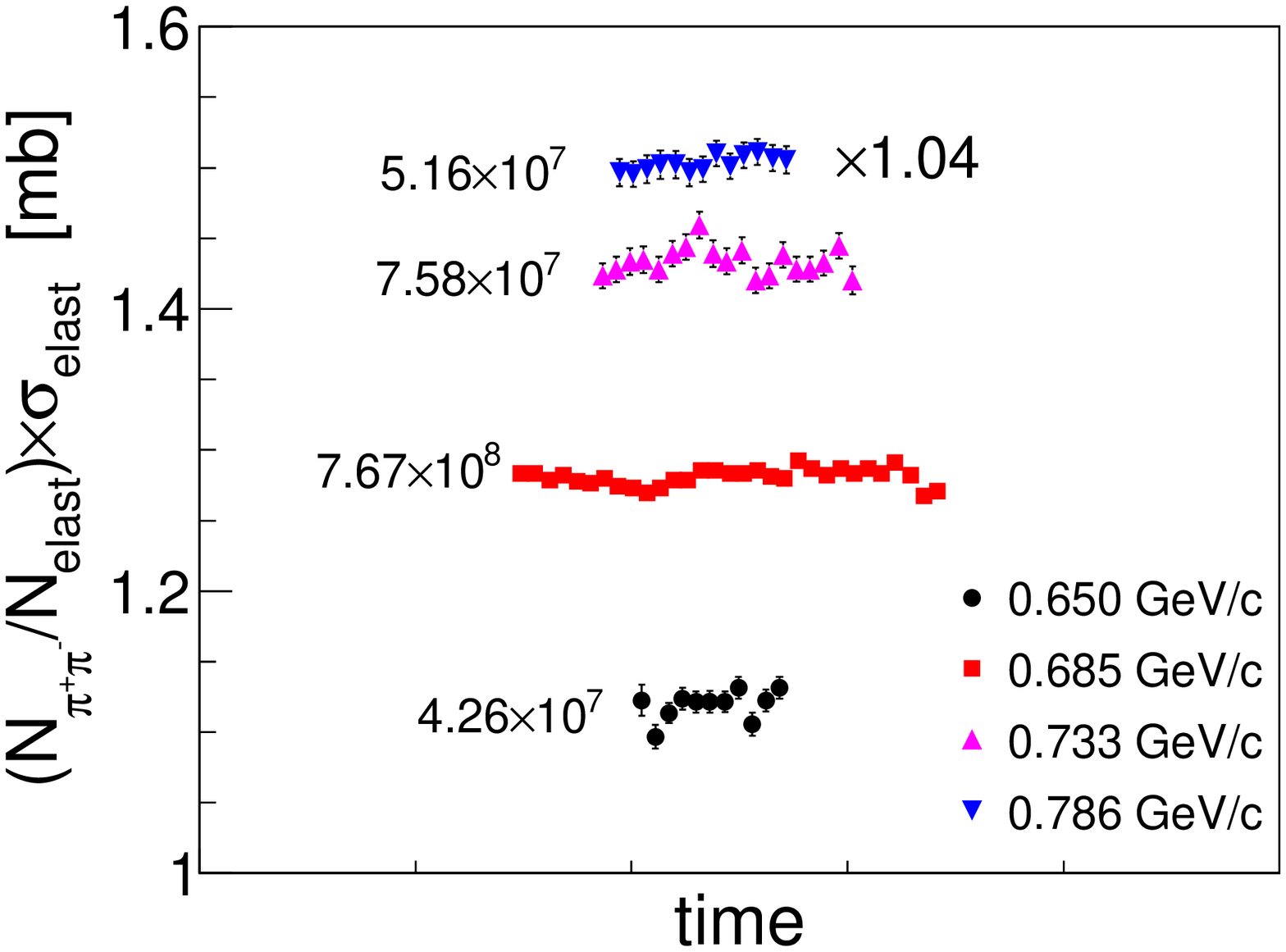}
\caption{Ratio of the number of $\pi^{+}\pi^{-}$ pairs originating from the $n\pi^+\pi^-$ final state to the number of $\pi^{-}p$ pairs from the elastic scattering presented for the whole measurement time. The ratios were calculated for all four momenta (see legend), and corrected for variation of the corresponding elastic scattering cross sections. The rise of the $N_{\pi^{+}\pi^{-}}/N_{elastic}$ as a function of the pion beam momentum is due to the  cross section increase for the two-pion production (see text for details). 
The numbers indicated in the panel refer to the total number of collected events for each momentum under a given trigger condition.  
}
\label{fig3b_2}
\end{figure}

The shape of the distribution measured by HADES at the highest pion beam momentum is in better agreement with the new EPECUR data than with the former world data.
Finally, the following cross sections were chosen for the normalization 
$\sigma_{elast}^{59.5^{\circ}-110.5^{\circ}}:3.60\pm0.07,3.94\pm0.08, 3.87\pm0.13, 3.16\pm0.02$ mb
for the four beam momenta, respectively. The above cross sections agree within the errors with the ones obtained from the SAID solutions, except the cross section for $p=0.685$ GeV/c which is $10\%$ higher.

The systematic uncertainty of the normalization has two main components. 
The first component of the systematic uncertainty is related to the errors of the world data, as given above.
The second one, accounting for the point-to-point variations of the HADES data, was estimated with respect to the averaged world data cross sections, using the following procedure. The differences ($\Delta_{i}$) between the averaged cross section and the HADES data  were calculated for each scattering angle bin $i$ and weighted with the world data errors $w_{i}$ at this point: \\
$$\Delta N_{1}=\sqrt{\sum_{i=59^{\circ}}^{i=110^{\circ}}\Delta_{i}^{2}w_{i}^{2}}.$$
The resulting systematic error $\Delta N_{1}$ was established to be about $1\%$, except $p=0.685$ GeV/c where it was found to be $4\%$. This uncertainty was compared to the one obtained with an alternative method based on differences between elastic scattering yields reconstructed in three independent HADES sector pairs. The respective error agrees with the former one, except for the lowest values of $\theta^{CM}_{\pi^{-}}$, outside the normalization region, where it  was found to be larger, i.e. 9\% at $40^{\circ}$ and decreasing to $2\%$ at $60^{\circ}$. The increase of the systematic error at  smaller scattering pion angles coincides with the strong decrease in the acceptance for pion-proton pairs, with a cut-off around $\theta_{\pi}^{CM}=40^{\circ}$. The region between $40^{\circ}$ and $60^{\circ}$ was therefore excluded from the normalization region. 
  
The relative normalization by means of measured pion-proton elastic scattering has also the  advantage to reduce systematic errors of the total cross sections related to the estimate of the efficiencies of pion and proton track reconstructions needed for corrections.
To check the long term stability of the measured yields, the ratio of the  number of $\pi^{+}\pi^{-}$ pairs originating from the $n\pi^+\pi^-$ final state to the number of $\pi^{-}p$ pairs from elastic scattering ($N_{\pi^{+}\pi^{-}}/N_{elastic}$) has been analysed as a function of time over the whole measurement period. In order to correct for the known variation of the cross section for the elastic scattering as a function of the beam momentum these ratios were multiplied by the respective value of $\sigma_{elast}^{59.5^{\circ}-110.5^{\circ}}$. The result is shown in Fig. \ref{fig3b_2} for $p_{beam}=0.650,0.685,0.733,0.786$ GeV/c. The obtained ratio is very stable in time, for a given beam momentum. Plotted errors are statistical only, and are smallest for $p_{beam}=0.690$ GeV/c where the largest number of events (given in the legend) was collected. The increase of the ratio as a function of beam momentum is due to the changes in the two-pion production cross sections, discussed in more detail in Sec. \ref{cross-sec}.  
The maximum deviations ($2\%$)  were observed for the pion beam momentum of $0.685$ GeV/c, and were used as systematic error related to the two-pion reconstruction efficiency.  On the other hand, the ratio $N_{elastic}/N_{START}$ which was analysed in a similar way  shows variations up to $15\%$ which do not allow for a precise estimate of the luminosity. It is attributed to the 
variations of the intensity of secondary particles, which is very sensitive to the position of the primary beam on the production target. 
Therefore, the relative normalization to the elastic scattering has been chosen as a more accurate approach.  

\subsection{Partial Wave Analysis}
\label{sec3_pwa}

The partial wave analysis of the present data is based on the
Bonn-Gatchina approach developed for the combined analysis of the
pion nucleon scattering and photoproduction. This is a covariant method which treats the reactions with two particle and multi-particle final states on a common basis. 

The amplitude which
describes the transition of the pion-nucleon system with momenta
$k_2$ and $k_1$ into the final channel with a meson and a nucleon with momentum $q_1$ given in the CM frame can be written as:
\begin{equation}
A\!=\!\!\sum\limits_{IJ\xi,\alpha}\!\!\bar u(q_1)
A^{IJ\xi\,,\alpha}_{\gamma_1\ldots\gamma_n}
F^{\gamma_1\ldots\gamma_n}_{\mu_1\ldots\mu_n}(p)
N^{\xi}_{\mu_1\ldots\mu_n}(k^\perp)u(k_1).~~~~
\label{pin_prod}
\end{equation}
Here, $IJ\xi$ are the isospin, the total angular momentum ($J\!=\!n+1/2$) and
naturality of the initial pion-nucleon system, respectively. The tensor
$F^{\gamma_1\ldots\gamma_n}_{\mu_1\ldots\mu_n}(p)$ is the propagator
of the initial $\pi N$ system with the momentum $p=k_1+k_2$. The
tensor $N^{\xi}_{\mu_1\ldots\mu_n}(k^\perp)$ describes the
production vertex which is constructed from the $\gamma$-matrices
and the orbital momentum tensors which depend on the relative
momentum of the initial particles orthogonal to the momentum $p$:
\be
k^\perp=\frac 12(k_1\!-\!k_2)_\nu g_{\mu\nu}^\perp= \frac
12(k_1\!-\!k_2)_\nu \left (g_{\mu\nu}\!-\!\frac{p_\mu
p_\nu}{p^2}\right ).
\ee
The naturality is connected with the orbital momentum $L$ as
$J=L+\frac 12\xi$, and technically is related to the presence of
the $\gamma_5$ matrix in the tensor decomposition.
The explicit form of the tensors $F$ and $N^\xi$ is given
in Ref. \cite{Anisovich:2004zz}. The multi-index $\alpha$
describes the quantum numbers of the final state configurations and
includes isospin, spin and naturality of the intermediate and final
states.

In the case of a two-particle final state with a pseudoscalar meson
and a $J^P=1/2^+$ baryon, the amplitude
$A^{IJ\xi\,,\alpha}_{\gamma_1\ldots\gamma_n}$ depends on the decay
vertex which has the same structure as the production one:
\be
A^{IJ\xi\,,\alpha}_{\gamma_1\ldots\gamma_n}=\tilde
N^{\xi}_{\mu_1\ldots\mu_n}(q^\perp) A_{IJ\xi}(s)C_I.
\ee
Here $C_I$ is the isospin Clebsch-Gordan coefficient, $q^\perp$ is
the relative momentum of the final particles orthogonal to the
momentum $p$ and the tensor $\tilde N$ differs from the tensor $N$ by
the order of the $\gamma$-matrices. In this case, the multi-index
$\alpha$ is a dummy index and the partial wave amplitude depends
only on the total energy squared, $s=p^2$.

In the case of  two pseudoscalar mesons and a baryon in the final
state, the amplitude can be decomposed into the partial waves which
describe the quasi two-particle decay processes. Thus, the initial
system decays into one of the final particles (spectator) and an
intermediate quasi particle which in turn decays into two other
final particles. Therefore, the partial wave amplitudes depend also on the quasi particle energy squared $s_{ij}=(q_i+q_j)^2$.  If the
spectator particle is the pion with momentum $q_2$, the total amplitude has the form
\be
A^{IJ\xi\,,\alpha}_{\gamma_1\ldots\gamma_n}&=&\tilde
G^{\beta_1\ldots\beta_m}_{\gamma_1\ldots\gamma_n}(L_1,\xi_1,q_2^\perp)
F_{\beta_1\ldots\beta_m}^{\mu_1\ldots\mu_m}(q_1\!+\!q_3) \nn
&&\times \tilde N^{\xi_2}_{\mu_1\ldots\mu_m}(q_{13}^\perp)
A^{\alpha}_{IJ\xi}(s,s_{13})C_{I,I_{13}},
\ee
where  $C_{I,I_{13}}$ is the corresponding Clebsch-Gordan
coefficient. The vertex
$G^{\beta_1\ldots\beta_m}_{\gamma_1\ldots\gamma_n}(L,\xi_1,q^\perp_2)$,
which is the only new tensor in this equation, describes the decay
of the initial partial wave into a pion and an intermediate baryon state
with the spin $J_{13}=m+\frac 12$ and naturality $\xi_2$. This
tensor is constructed from the orbital momentum tensors and depends
on the momentum of the spectator particle orthogonal to the momentum
of the initial system. As before, the naturality corresponds to the
number of the $\gamma_5$-matrices in the tensor expression (see Ref. 
\cite{Anisovich:2004zz}). 

The tensor which describes the production
of a scalar meson and the spectator nucleon has the following form:
\be
A^{IJ\xi\,,\alpha}_{\gamma_1\ldots\gamma_n}&=&\tilde
N^{\xi_1}_{\mu_1\ldots\mu_n}(q_{3}^\perp)
A_{IJ\xi}(s,s_{12})C_{I,I_{12}}.
\ee
This expression has a similar form as the single pion production
amplitude. However, due to the positive parity of the scalar mesons,
the naturality changes its sign $\xi_1=(-1)\xi$. For the vector
meson production this equation is modified as:
\be
A^{IJ\xi\,,\alpha}_{\gamma_1\ldots\gamma_n}&=&\tilde
V^{\eta\xi}_{\mu_1\ldots\mu_n}(q_{3}^\perp)\Pi_{\eta\nu}
A^\mu_{IJ\xi}(s,s_{12})q_{12\nu}^\perp C_{I,I_{12}}.
\ee
Here, $\Pi_{\eta\nu}$ is the standard $\rho$-meson propagator; the explicit equation for the tensors
$V^{\eta\xi}_{\mu_1\ldots\mu_n}(q_{3}^\perp)$ which describe the
decay of the intermediate baryon into $\rho$-meson and pion is given
in \cite{Anisovich:2004zz}.

The non-resonance contributions are described by the t-channel
exchange amplitudes taken in the Regezied form:
\be
A=g_1(t)g_2(t)\frac{1+\xi exp(-i\pi\alpha(t))}{\sin(\pi\alpha(t))}
\left (\frac{\nu}{\nu_0} \right )^{\alpha(t)} .
\ee
Here, $\nu=\frac 12 (s-u)$, $\alpha(t)$ is the Reggion trajectory,
and $\xi$ is its signature. The vertices $g_1$ and $g_2$ include
form factors which we parameterize in the exponential form
\be
g_1(t)g_2(t)=\Lambda\,e^{-bt}\,,
\ee
where $\Lambda$ and $b$ are fit parameters.

The partial wave amplitudes are parameterized in the framework of
the $N/D$-based approach described in detail in Ref. 
\cite{Anisovich:2011zz}. This approach can be considered as the
solution of the Bethe-Salpeter equation with the kernel 
\be
K_{ij}=\sum\limits_{\beta=1}^N\frac{g^\beta_ig^\beta_j}{M^2_\beta-s}+f_{ij}.
\ee
The indices $ij$ correspond to the scattering channels, and the
amplitudes are described as the sum of $N$ resonant terms and
non-resonant contributions $f_{ij}$. The quantities $g_{i}^{\beta}$ are decay couplings related to the decay widths of resonances $\Gamma_{\beta,i}(M_{\beta})$ via the relation  
\[
(g_i^{\beta})^2 \rho_{i}^{\beta}(M_{\beta,i})=M_{\beta}\Gamma_{\beta,i}(M_{\beta})
\]
where $\rho^{\beta}(M_{\beta})$ is the phase space for the decay. 
This method satisfies
explicitly the unitarity and analyticity conditions for the two-body
final states. In the case of three-body final states, the unitarity
is satisfied on the level of the quasi two-particle processes. This
approach takes correctly into account such analytical structures of
the amplitudes as poles and cuts. In the majority of cases these
singularities are the dominant singularities and define the energy
behavior of the amplitude. The logarithmic singularities connected
with the triangle diagrams have, as a rule, a rather smooth energy
dependence and can be taken into account by renormalization of the
resonance couplings. However, such a renormalization leads to
the appearance of the coupling phases which were treated as parameters
in the optimization procedure.

The analysis of the HADES data was performed together with photo- and pion-induced data with one and two pseudoscalar mesons in the final state. Amplitudes for the pion-nucleon elastic scattering are taken from SAID solutions. The full list of the fitted reactions with the corresponding references is given on the Bonn-Gatchina web page \cite{bgwp}.
For convenience, we list in
Table~\ref{tab1_list} the fitted data with two pions in the final state. All reactions with the production of two pseudoscalar mesons were fitted in the framework
of the event-by-event likelihood method which allows to take
into account all amplitude correlation in the final phase space. Furthermore, total cross sections for the two pion production channels obtained in \cite{Man:92} were also included in the procedure as an additional constraint (the differential cross sections are not available on the event-by-event basis). 

In the approach, we minimize the function
\be
f=-\sum\limits_j^{N(data)}ln\frac{\sigma_j(PWA)^{data}}{\sum\limits_m^{N(rec\,MC)}\sigma_m(PWA)},
\ee
where $\sigma_j(PWA)^{data}$ is the differential cross section calculated for every fitted data event. The normalization of the minimization function to the sum of Monte Carlo events passed through the detector simulation takes into account the given acceptance of the experimental setup. The quality of the obtained solution can be estimated by comparison of these Monte Carlo events weighted by the final cross section with the measured data.

\begin{table}[h]
\renewcommand{\arraystretch}{1.1}
\bc
\caption{ \label{tab1_list} The reactions, observables and energy ranges
of the two-pion production data used in the PWA.}
\begin{tabular}{|l|c|c|c|c|}
\hline \textbf{Reaction} & \textbf{Observable} & \textbf{W (GeV)} &\textbf{Experiment} \\\hline
$\gamma p\to \pi^0\pi^0 p$ & DCS, Tot & 1.2-1.9 & MAMI \\\hline
$\gamma p\to \pi^0\pi^0 p$ & E   & 1.2-1.9 & MAMI \\\hline
$\gamma p\to \pi^0\pi^0 p$ & DCS,Tot & 1.4-2.38 & CB-ELSA \\\hline
$\gamma p\to \pi^0\pi^0 p$ & $P,H$ & 1.45-1.65 & CB-ELSA \\\hline
$\gamma p\to \pi^0\pi^0 p$ & $T,P_x,P_y$ & 1.45-2.28 & CB-ELSA \\\hline
$\gamma p\to \pi^0\pi^0 p$ & $P_x,P_x^c,P_x^s$ (4D)& 1.45-1.8 & CB-ELSA \\\hline
$\gamma p\to \pi^0\pi^0 p$ & $P_y,P_y^c,P^s_y$ (4D)& 1.45-1.8 & CB-ELSA \\\hline
$\pi^- p\to \pi^0\pi^0 n$ & DCS & 1.29-1.55 & Crystal Ball \\\hline
$\pi^- p\to \pi^+\pi^- n$ & DCS & 1.45-1.55 & HADES \\\hline
$\pi^- p\to \pi^0\pi^- p$ & DCS & 1.45-1.55 & HADES \\\hline
\end{tabular}
\ec
\renewcommand{\arraystretch}{1.0}

\end{table}

To analyze the HADES data we start from the Bonn-Gatchina solution 
described in detail in Ref. \cite{Muller:2019qxg}. In this analysis, the $\rho N$ channel was not directly taken into account. The inelasticities of the fitted states were described by a "black box" with a phase volume taken as $\rho N$ with the lowest possible orbital momentum. In the present study, all $\rho N$ decay channels were introduced explicitly for the resonances with masses below $1.6$ GeV. After including the $\rho N$ channels and HADES data in the fit, all couplings of these states to the "black box" were optimized close to zero and were fixed to this value in the final fit. This means that the widths were fully defined by  the sum of the partial widths of the fitted channels.

The combined analysis allows us to define the initial isospin of all
partial wave amplitudes. For example, the $2\pi^0$ production
reactions do not provide enough information for a unique
decomposition of the $\Delta(1232)\pi$ amplitudes into  initial
states with  fixed isospin. The HADES data provide the needed
information and impose a strong constraints on the $\Delta\pi$ and
$N^*\pi$ amplitudes. In addition,  the HADES data allow us to
extract, with a good precision, the contributions of the $\rho(770)N$
amplitudes which do not contribute to the $2\pi^0$ production
reactions.

In the energy region of the HADES data, the leading contributions to
the reactions are defined by the $\Delta(1232)\pi$, $N(1440)\pi$,
$\rho(770)N$ and $\sigma N$ intermediate states. Here  $\sigma$
describes the energy dependence of the scalar $\pi\pi$ S-wave in the
mass region from the two-pion threshold  up to 0.8 GeV. We also
introduce the contribution from the amplitudes with $N(1535)(\frac{1}{2}^-)\pi$
and $N(1520)(\frac{3}{2}^-)\pi$ intermediate states but did not find any notable
contributions from them.

\subsection{Acceptance and efficiency corrections}
\label{sec3_acc}
\indent To compare the HADES data on $p\pi^0\pi^-$ and  $n\pi^+\pi^-$ final states to the results of PWA fits various differential distributions were studied. They include momentum and angular distributions of the final state particles in the CM frame, invariant masses and angular distributions in the Gottfried-Jackson (GJ) and the Helicity (H) reference frames (for definitions, see Sec. \ref{sec-three-part}). The distributions were compared to the PWA solutions calculated within the HADES acceptance. The agreement between the data and the PWA solutions is generally very good (see figures presented in the next section) and justifies model-driven combined acceptance and efficiency corrections. For this purpose one-dimensional correction functions were computed for all investigated observables. These functions have been obtained as ratios of two PWA solutions given for: (a) the full solid angle ({\em 4$\pi$}), and (b) the HADES acceptance, including all reconstruction cuts like PID and missing mass windows, ({\em ACC}). Finally, experimental distributions were multiplied by the respective correction functions. The averaged correction factors depend slightly on the beam momentum and amount to $6.5-8$ and $9.5-12.5$ for the $n\pi^+\pi^-$ and $p\pi^-\pi^0$ data, respectively. Note that in this procedure no extrapolation outside the HADES acceptance was performed and only the acceptance losses due to incomplete azimuthal coverage where accounted for. Furthermore,  the acceptance regions with a very large correction factor ($>15$), corresponding to low acceptance/efficiency, were excluded. Nevertheless, several angular distributions have a full acceptance coverage in HADES (see below), and can be used to determine the total cross section. The corrected experimental distributions obtained in this way are of more general interest since they can be directly compared to any theoretical model.\\

\indent Systematic uncertainties of the PWA have been estimated by studying several solutions: with truncation at $J^P=3/2^{\pm}$ and at $J^P=5/2^{\pm}$ for the data sets obtained with and without pion tracker with the missing mass selections shown in Fig. \ref{fig3b_0}. Furthermore, two strategies for data normalization were applied in the fits. One based on the cross sections measured in the HADES acceptance only and the other one with the HADES cross sections combined with the cross sections for the two pion production in complete solid angle derived from the PWA analysis of former two-pion experiments, as given in \cite{Man:92} (denoted as PWA Manley).  

These various  solutions were also used to estimate systematic errors of the correction functions introduced above. It appears that correction functions for the given observable calculated with the various PWA solutions are very similar. The correction factors vary from point-to-point and their average spread was estimated to $2\%$, reaching a maximum of $7\%$ for some specific regions of the acceptance. This error was calculated for every bin of investigated differential distributions and was propagated to the total systematic error. The other contributions to systematic errors assigned to each bin arise from the pion-carbon background subtraction ($1-2\%$) and track reconstruction efficiency ($2\%$), discussed in the previous sections.      


\indent For the estimate of the total cross section, angular distributions with the complete HADES coverage were considered for the data sets collected with and without pion tracker. In the case of the $n\pi^+\pi^-$ channel, the following projections were used: $\cos\theta^{n-\pi^{+}}_{n\pi^{-}}$, $\cos\theta^{\pi^{+}-n}_{\pi^{+}\pi^{-}}$, $\cos\theta^{\pi^{+}-n}_{\pi^{+}\pi^{-}}$ in the (H) reference frames, and $\cos\theta^{\pi^{+}}_{\pi^{+}\pi^{-}}$ in the (GJ) reference frames. In the $p\pi^0\pi^-$ channel: $\cos\theta^{\pi^{0}-\pi^{-}}_{p\pi^{0}}$,  $\cos\theta^{\pi^{0}-p}_{\pi^{0}\pi^{-}}$ in the (H) reference frames, and $\cos\theta^{\pi^{0}}_{p\pi^{0}}$, $\cos\theta^{\pi^{0}}_{\pi^{0}\pi^{-}}$ in the (GJ) reference frames (see Figs. \ref{fig:helicity}, \ref{fig:GJ}).
Based on these projections, the total cross sections and the systematic errors related to the extrapolation to the full solid angle were calculated from the acceptance corrected distributions for each reaction channel as the average of the respective integrals and  their dispersion (RMS), respectively.  
Systematic errors related to corrections of losses due to the missing mass cuts were estimated by variation of the window widths and comparing losses in the simulation to the corresponding ones in the data (see table). These errors (about $2-5\%$) were found to be larger (and asymmetric) than the errors related to the extrapolation (about $1-2\%$) and are summarized together with the other sources in Tab. \ref{tab2}. 

\section{Discussion of the results for the $n\pi^+\pi^-$ and $p\pi^-\pi^0$ channels and comparison to the PWA solutions}
\label{sec4}

\begin{table}[!h]
 \centering
\caption{Sources and typical size of systematic effects.}
  \label{tab2}
 \begin{tabular}{|l|c|}
 \hline
 \textbf{Source of uncertainty} & \textbf{Error estimate [\%]} \\\hline
 \multicolumn{2}{|l|}{\textbf{Global errors:}}\\\hline
  PID & 2 \\\hline
  Normalization (world data)& 2-4 \\\hline
 Total cross section extrapolation& 1-2 \\\hline
 Reconstruction procedure of pion & 0.5 \\
 beam momentum & \\ \hline
 Variation of missing mass cuts &  \\
 $n\pi^+\pi^-$ channel  & + (2-5) \\ 
 $p\pi^-\pi^0$ channel &  + (2-3) \\ \hline
\textbf{Total global systematics} & \textbf{4-8} \\\hline\hline
 \multicolumn{2}{|l|}{\textbf{Point-to-point errors:}}\\\hline
 Elastic scattering $\Delta N_{1}$ & 1-4 \\\hline
 Carbon background& 1 \\\hline
 Two-pion reconstruction& 2 \\
  efficiency & \\\hline
 Acceptance corrections & 2-7 \\\hline
  \end{tabular}
  \end{table}

The PWA fits, described in Sec. \ref{sec3_pwa}, have been performed including the HADES data on the $\pi^-p\rightarrow p\pi^0\pi^-$ and $\pi^-p\rightarrow n\pi^+\pi^-$ reaction channels measured at the four pion beam momenta. The differential distributions presented in this section correspond to data obtained with $p_{beam}=0.685$ GeV/c using the pion tracker. The highest statistics was collected for this beam momentum, which had been chosen for the study of the dielectron channel. This data set, therefore, constitutes the most relevant reference and also illustrates well the main conclusions from this analysis. 
The results of the analysis of data obtained for the other beam momenta are included in excitation functions shown at the end of this section. All presented distributions are corrected for the HADES acceptance and reconstruction inefficiency. The yields are converted to cross sections via the relative normalization to the elastic pion-proton scattering, as described in Sec. \ref{sec3_nor}. The distributions are compared to the PWA solution given in the full solid angle. 
The systematic point-to-point errors related to the acceptance corrections and the subtraction of the contribution originating from pion-carbon interactions are added quadratically and displayed as boxes. 
Statistical errors are indicated separately by vertical bars and are usually smaller than the systematic ones. 
In addition to these errors, global errors related to the PID method, the normalization, the cross-section extrapolation, the pion beam momentum reconstruction procedure, are included in the estimate of the total cross sections but not shown in the differential distributions.
 
Table \ref{tab2} summarizes the various types of systematic errors considered in the analysis.

 
\subsection{Two-particle distributions} 
\label{two-part}

\begin{figure*}
\includegraphics[width=0.45\textwidth]{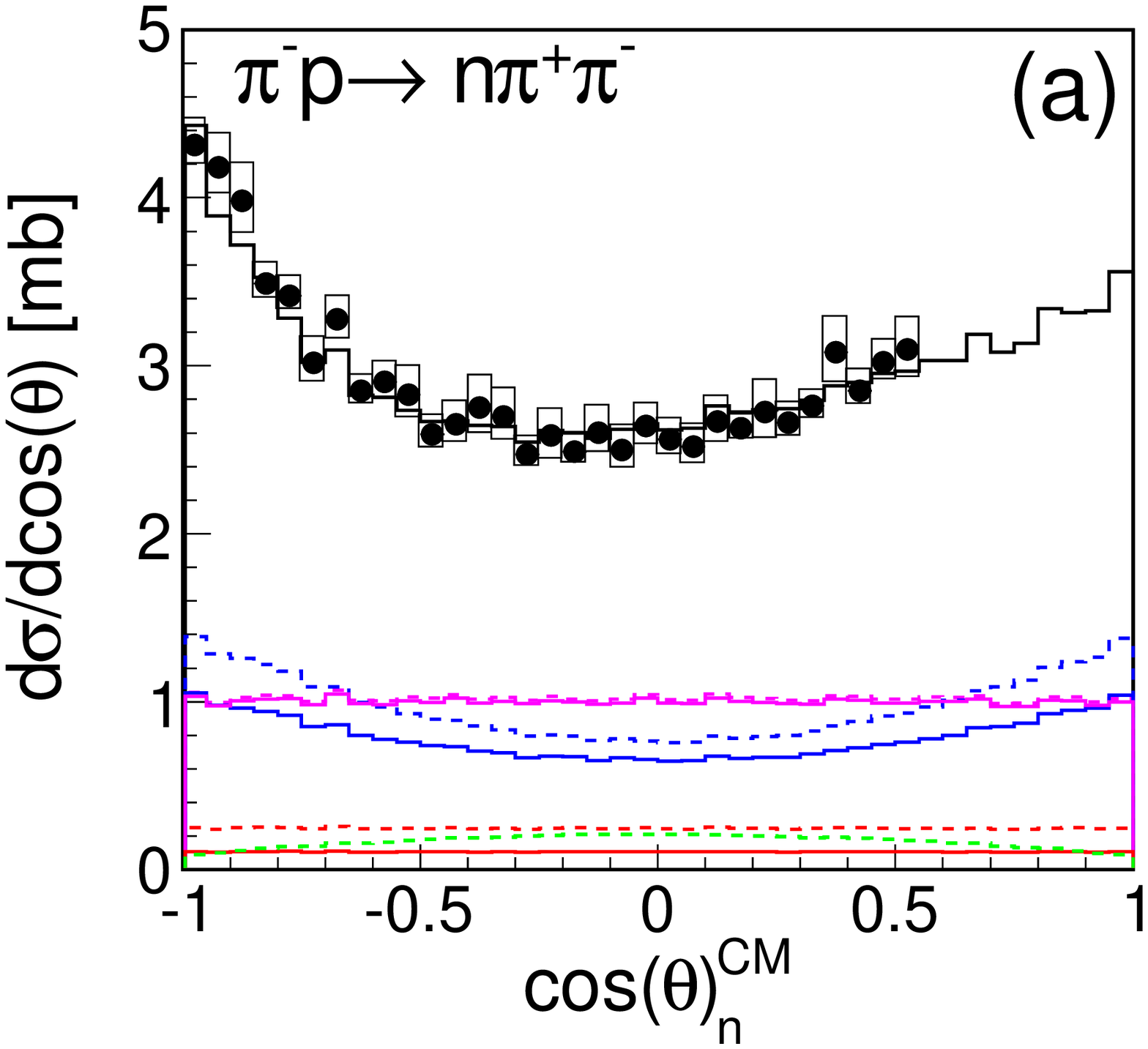}
\includegraphics[width=0.45\textwidth]{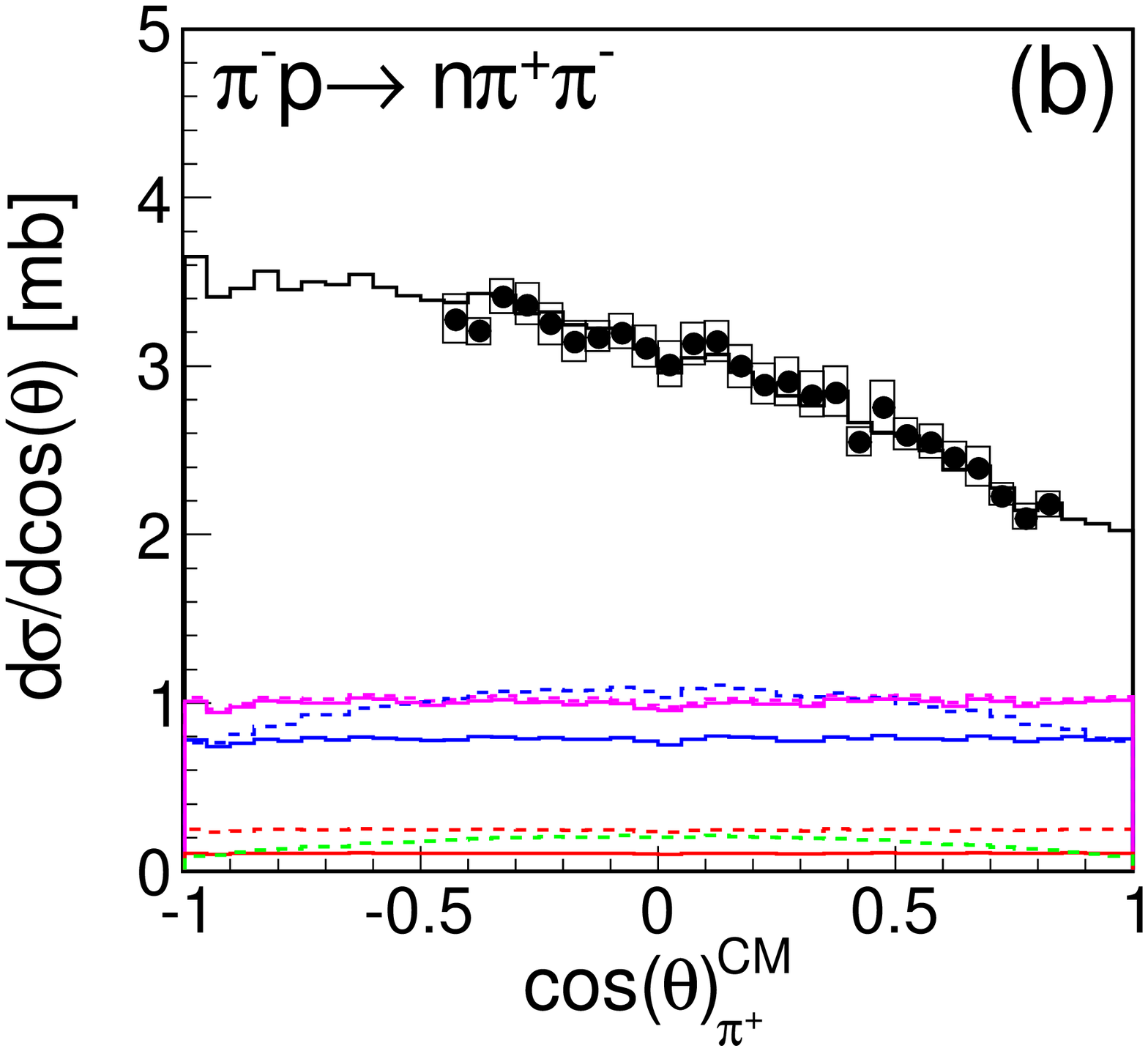}
\includegraphics[width=0.45\textwidth]{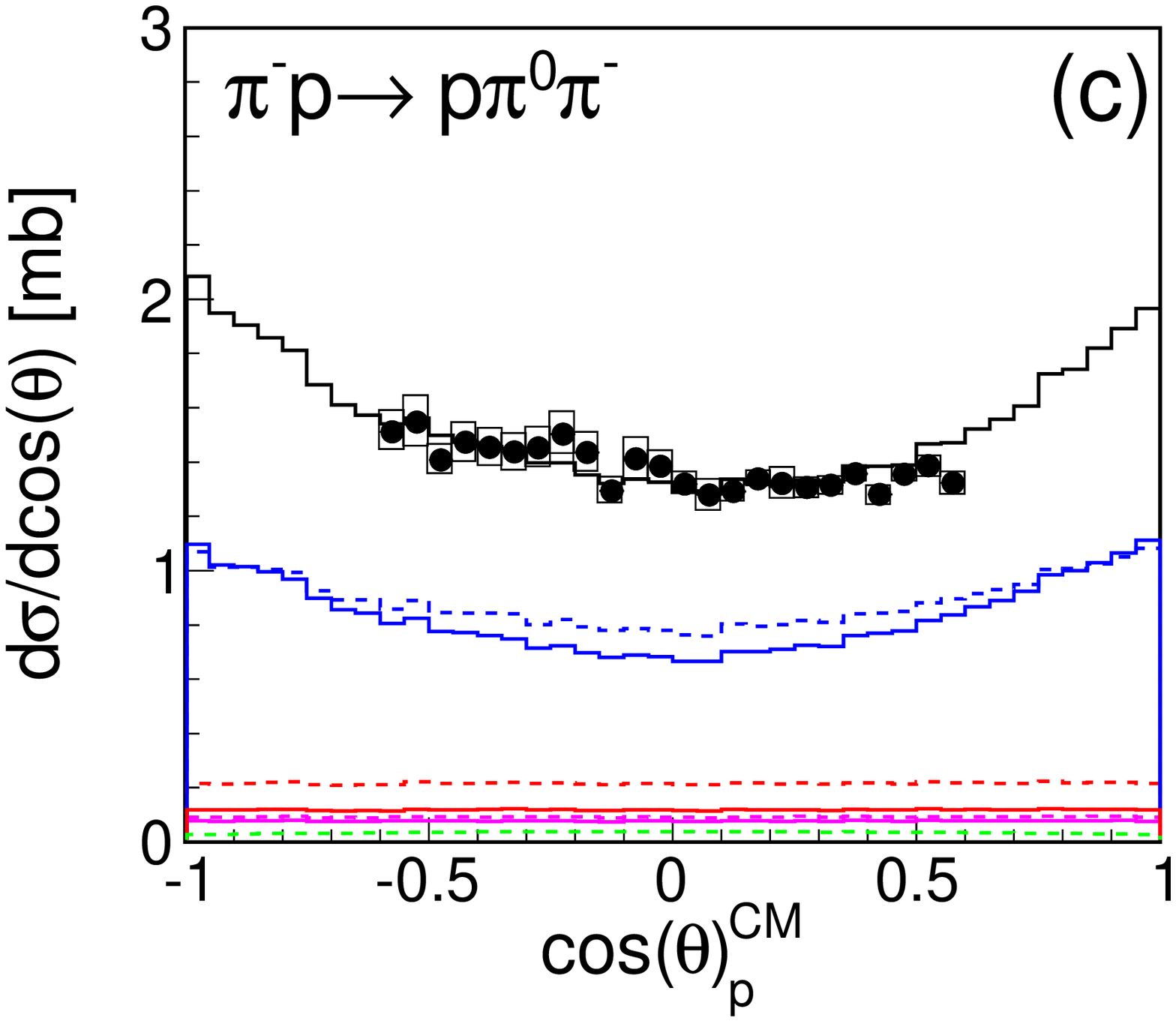}
\includegraphics[width=0.45\textwidth]{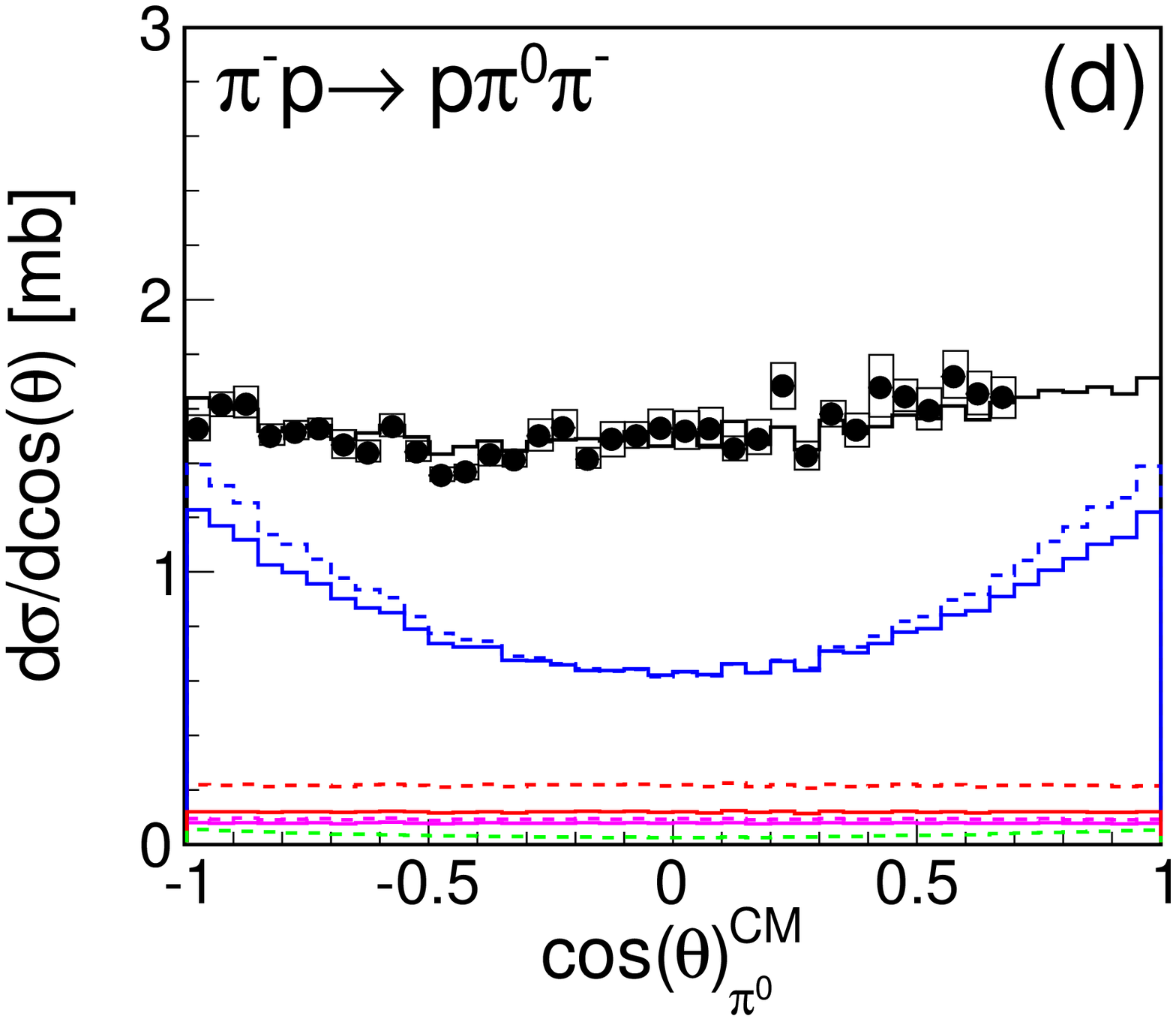}
\includegraphics[width=40pc,height=2.5pc]{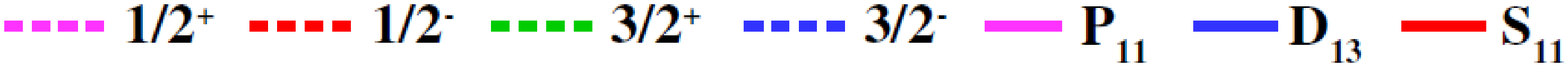}
\caption{(Color online) Angular differential cross sections of the nucleons (left column) and pions (right column) in the CM frame for the $\pi^-p\rightarrow n\pi^+\pi^-$ (upper row) and $\pi^-p\rightarrow p\pi^0\pi^-$ (lower row) reaction channels for $p_{beam}$ = 0.685 GeV/c. Color curves display contributions of partial waves (dashed line histograms) and $I=1/2$ s-channel contributions (solid lines) to the total yield (black solid histogram).}
\label{fig:CM_angles}
\end{figure*}

Within the Bonn-Gatchina framework, the two-pion production is described as a two-step process involving the formation of quasi two-particle final state-isobars. The two-particle states consist of a meson associated with a baryon: $\pi\Delta$ or $N \rho$(I=1), N$\sigma$(I=0). The intermediate states $\rho,~\sigma$ or $\Delta$ decay into the two-body final states $\pi\pi$ or $N\pi$, respectively. In the following discussion, we show differential distributions for such two-particle systems which characterize the first step of the reaction. 

\begin{figure*}
\includegraphics[width=0.45\textwidth]{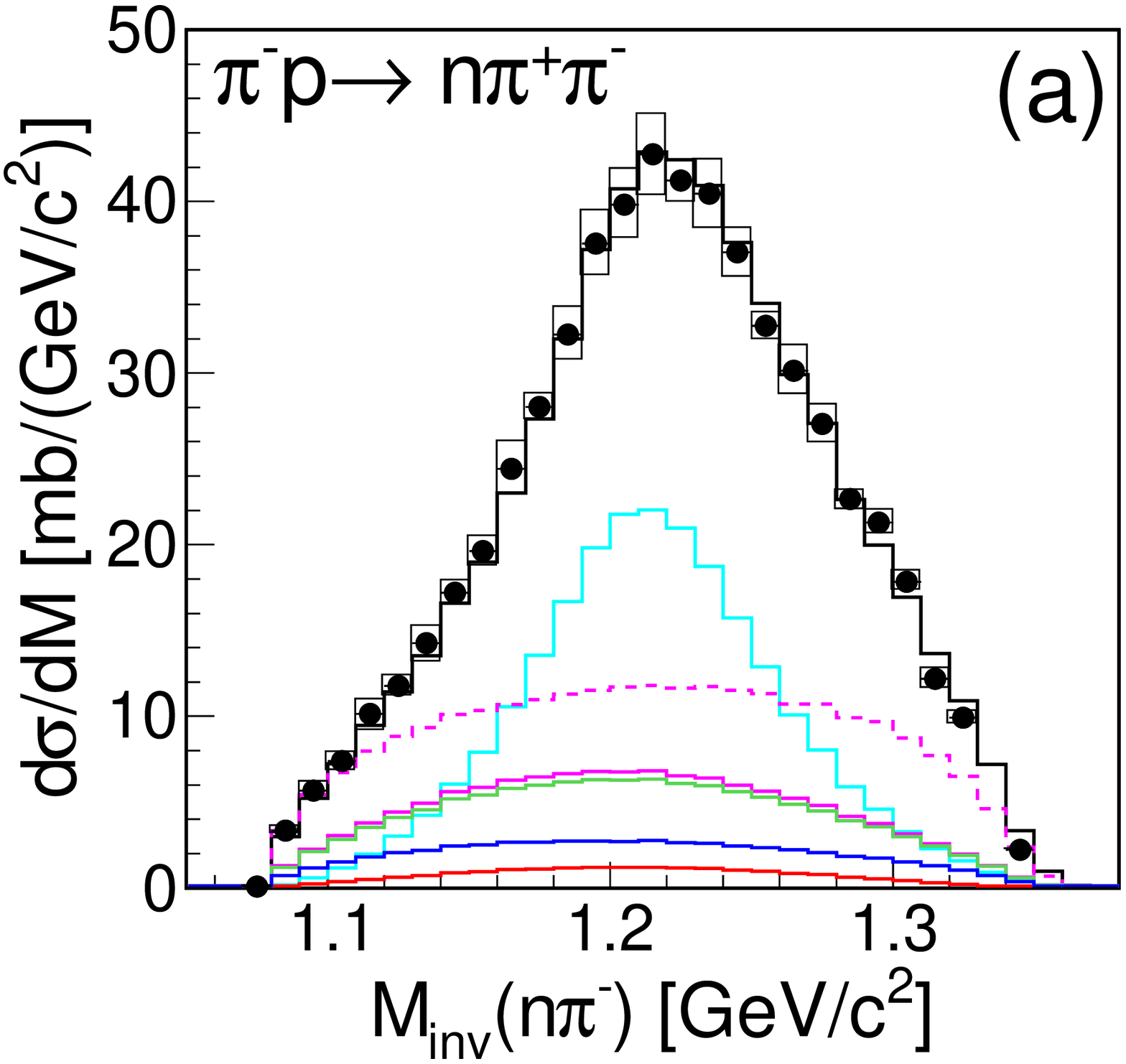}
\includegraphics[width=0.45\textwidth]{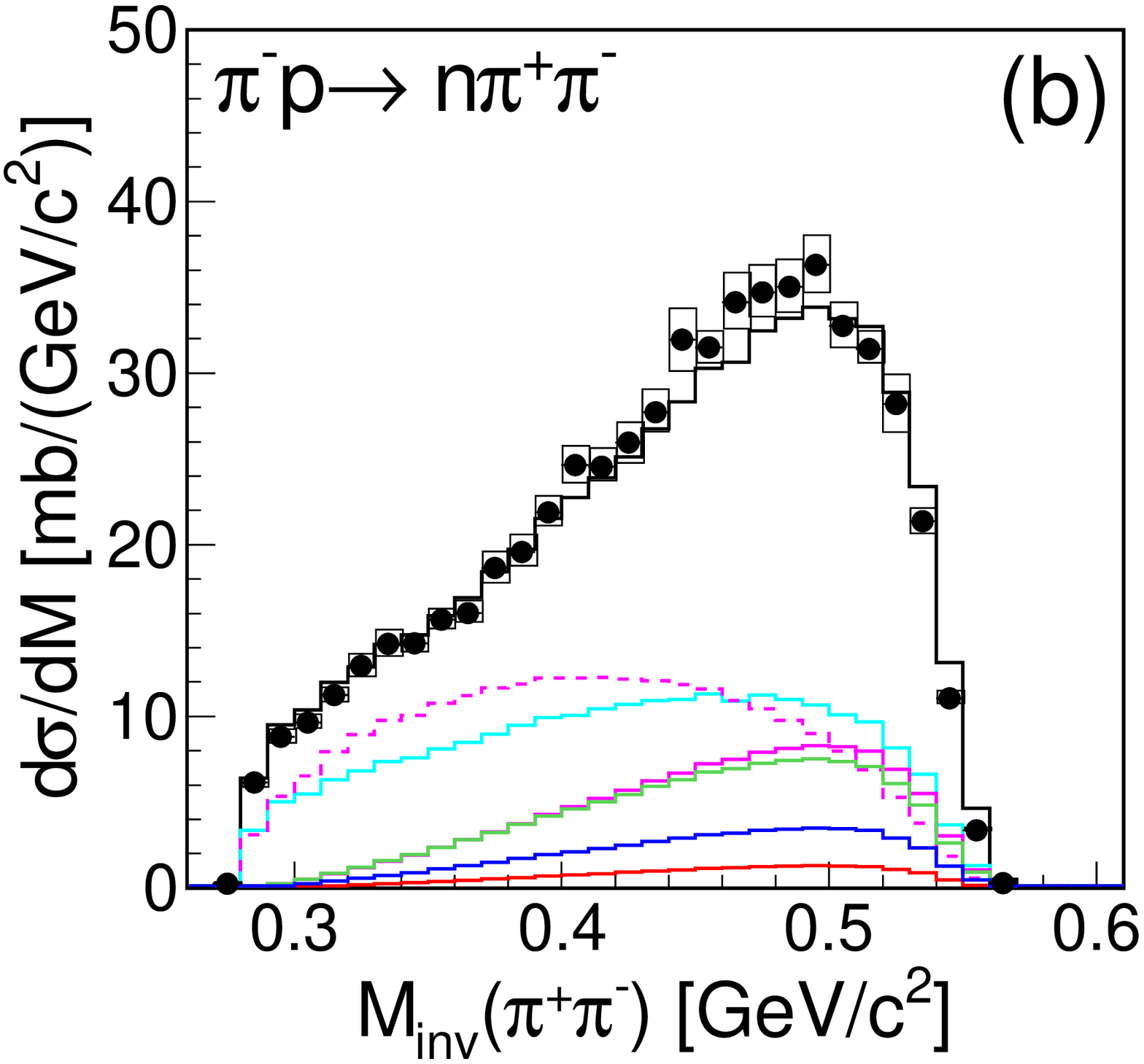}
\includegraphics[width=0.45\textwidth]{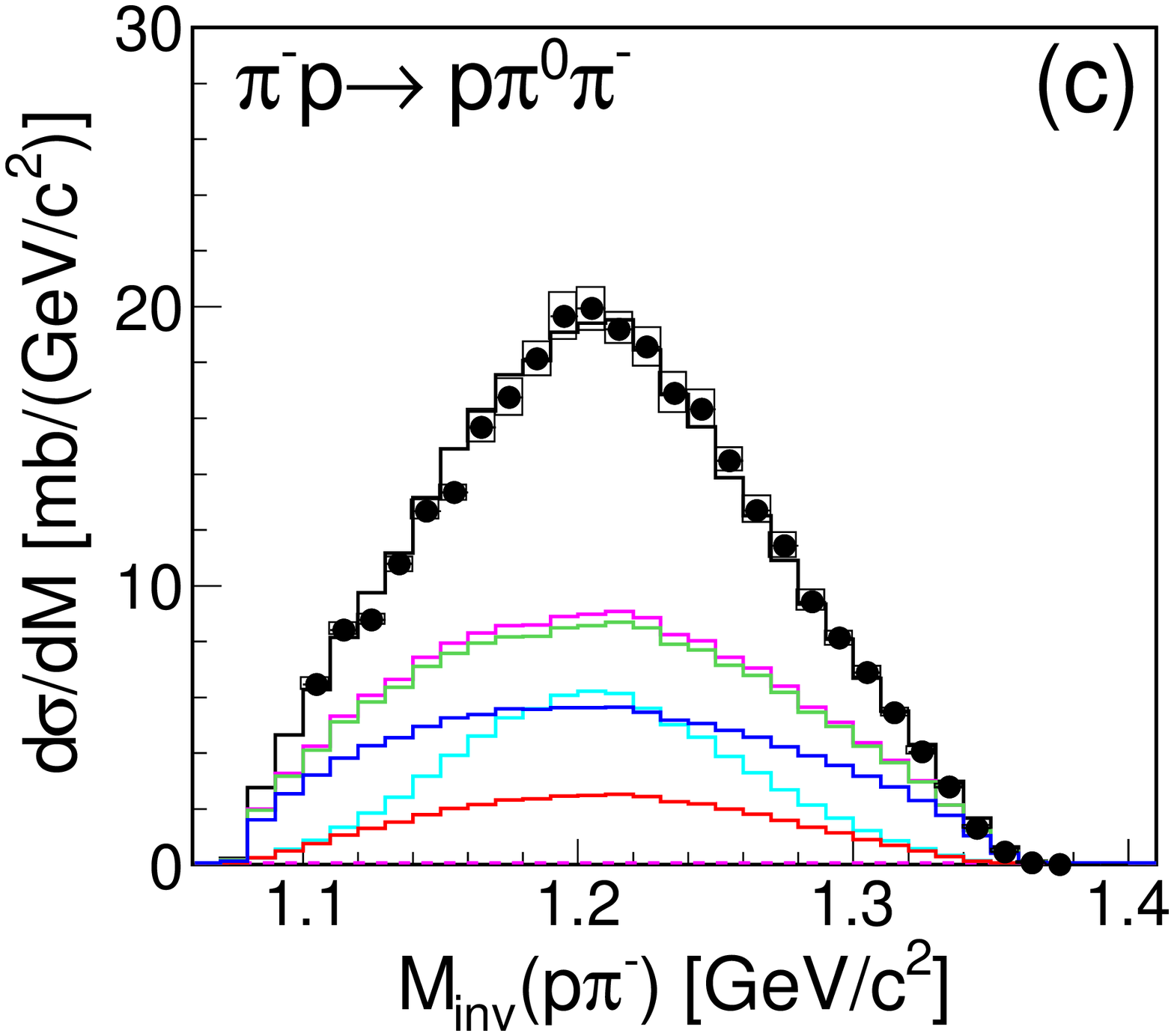}
\includegraphics[width=0.45\textwidth]{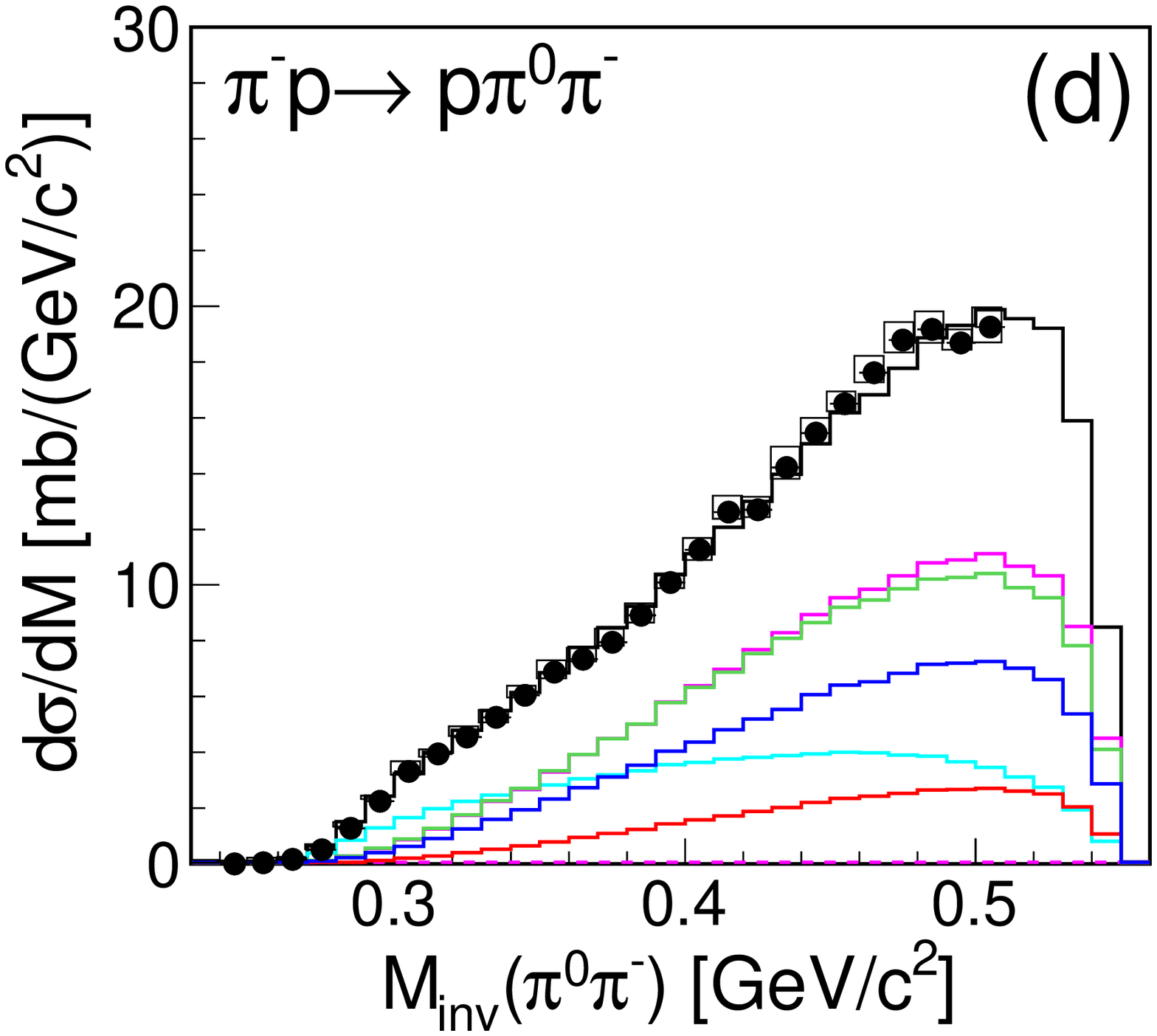}
\includegraphics[width=35pc,height=3.5pc]{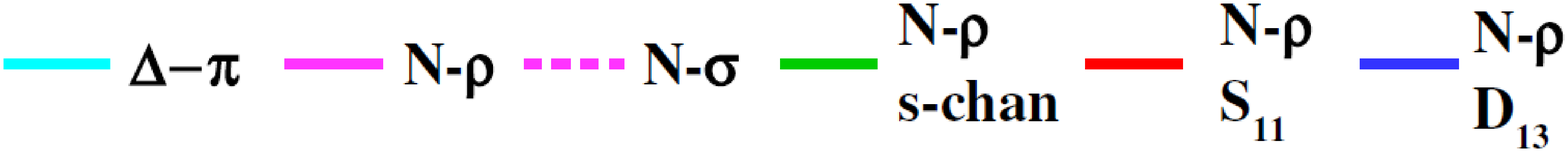}
\caption{\label{ref_frame} Invariant mass distributions of the nucleon-pion (left column) and the two-pion systems (right column) for the $\pi^-p\rightarrow n\pi^+\pi^-$ (upper row) and $\pi^-p\rightarrow p\pi^0\pi^-$ (lower row) reaction channels for $p_{beam}$ = 0.685 GeV/c. Color curves display contributions of various final states to the total yield (black solid line histogram), as indicated in the legend.}
\label{fig:inv_mass}
\end{figure*}

We begin the presentation of the results with the distributions of the polar emission angles in the CM frame of the collision (Fig. \ref{fig:CM_angles}), and the invariant masses (Fig. \ref{fig:inv_mass}). Since the nucleon (pion) and the two-pion (nucleon-pion) systems are emitted back-to-back in the CM frame, the respective angular distributions are closely related (by reflection symmetry). Therefore, the invariant mass distributions for the particle combinations shown in Fig.~\ref{fig:inv_mass} complement the one-particle angular distributions shown in Fig.~\ref{fig:CM_angles}. 

In the figures, the data points (filled circles) are compared to the PWA solution (solid  histograms)  normalized to the experimental cross sections. In this way, we first compare the shapes of the experimental and PWA distributions. The comparison of the total cross sections for the two-pion production, obtained in the PWA, to the cross sections measured by HADES is presented in Sec. \ref{cross-sec}.

The partial wave analysis provides information about all contributing amplitudes which can be organized into various projections. The differential production cross sections are separated in the leading contributions: (a) the total angular momentum and the parity of the initial state ($J^P=1/2^{\pm},3/2^{\pm}$), see Fig.~\ref{fig:CM_angles}, and (b) the type of quasi two-particle states or isobars, see Fig.~\ref{fig:inv_mass}: $\Delta\pi$ (cyan curves), $N\rho$ (violet curves), and $N\sigma $ (dashed violet curves). In case (a), we additionally show contributions from the most important  pion-nucleon waves in the s-channel with $I=1/2$ and fixed total angular momentum $J$ and angular momentum $L$,  $i.e.$ $S_{11}$ ($L=0,J=1/2$)- red curves, $P_{11}$ ($L=1, J=1/2$)-violet curves and $D_{13}$ ($L=2,J N=3/2$)- blue curves.  
The $N^*$ resonances $N(1535)\frac{1}{2}^-$, $N(1440)\frac{1}{2}^+$ and  $N(1520)\frac{3}{2}^-$ contribute respectively to these partial waves.
In case (b), we also present the dominant sources of the  $\rho-$meson production, which is in the focus of this study. There, the following contributions are shown: the total $\rho$ contribution in the s-channel, contributing almost $100\%$ of the total $\rho$ production cross section (green curve), separated into the shares from $S_{11}$ (red curves), and $D_{13}$ (blue curves) partial waves.


Figure \ref{fig:CM_angles} shows distributions for the $n\pi^+\pi^-$ (upper row) and the $p\pi^-\pi^0$ (lower row) final states. They demonstrate the dominance of the $3/2^-$ partial wave (dashed-blue) in the $p\pi^-\pi^0$ final state. For the $n\pi^+\pi^-$ final state we obtain comparable contributions for the  $1/2^+$ (dashed violet curves) and $3/2^-$ waves (dashed blue), respectively. Contributions from the higher partial waves (not shown in the figure) are much smaller.  As one can see,  the different $J^P$ partial waves are dominated by the s-channel $I=1/2$ contributions. The t-channel contributions are found to be much smaller  ($\sim 1-4\%$ depending on the partial wave). 
The  $I=3/2$ components, which are also much smaller than the $I=1/2$ contributions,  play some role via interference effects. This can be seen for example in the case of the $p\pi^-\pi^0$ final state, where the total $J=3/2^-$ partial wave yield is smaller than the $I=1/2$ $D_{13}$ s-channel contribution, due to destructive interference effects with the corresponding $I=3/2$  contribution. 



 \begin{figure*}
\includegraphics[width=0.45\textwidth]{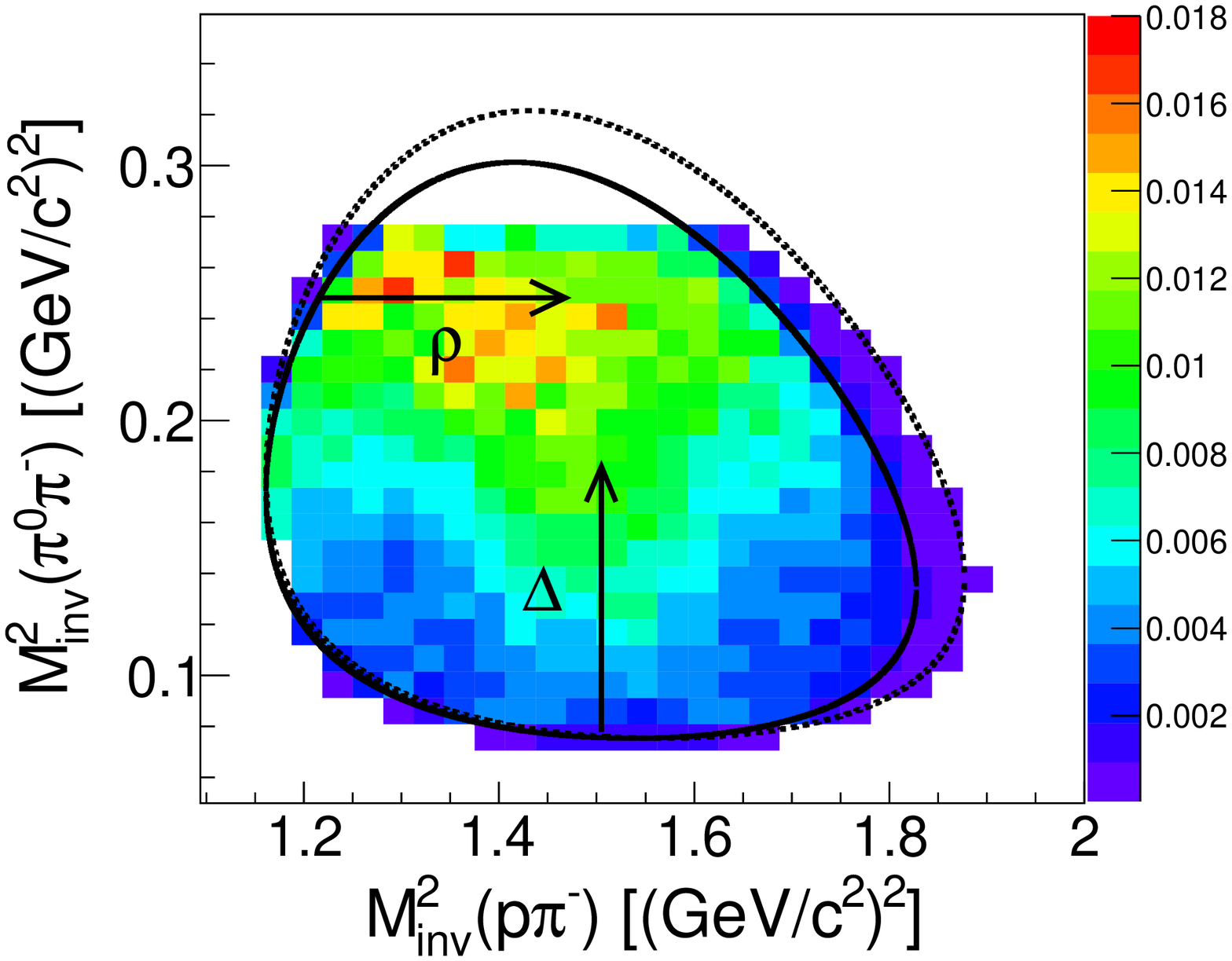}
\includegraphics[width=0.45\textwidth]{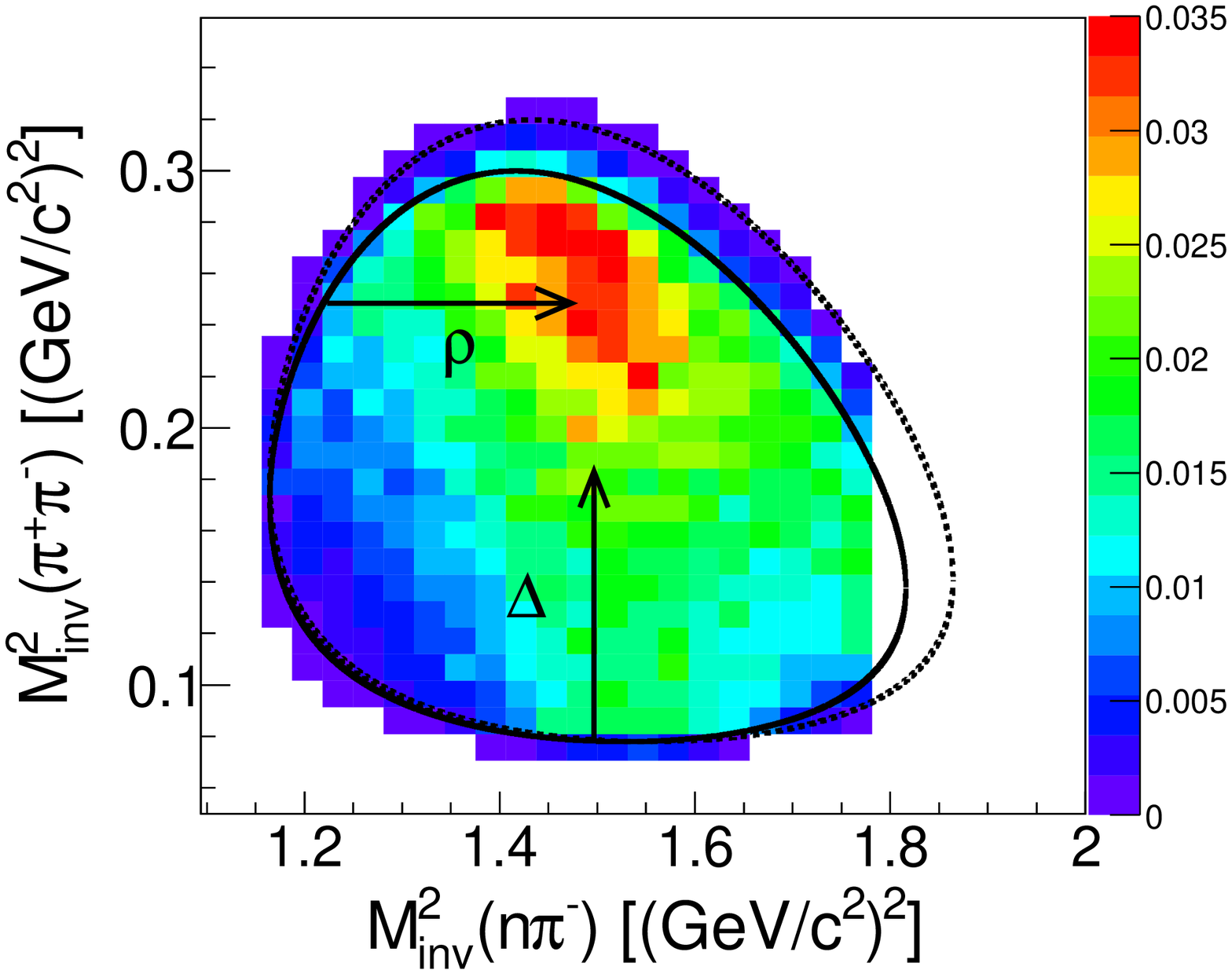}
\caption{Dalitz plots for the $p\pi^-\pi^0$ (left panel) and the $n\pi^+\pi^-$ (right panel) final states for $p_{beam}$ = 0.685 GeV/c with the indicated locations of the $\Delta$ and $\rho-$meson contributions. The distributions are corrected for the HADES acceptance and reconstruction inefficiency. The z-axis represents cross section ([mb]). Theoretical borders of the Dalitz plots are drawn with black solid curves for the fixed (central value) pion beam momentum. The dashed black curve refers to 
$+3\sigma$ of the $\sqrt{s}$ distribution corresponding to the pion beam momentum range as presented in Fig.~\ref{fig1}.}
\label{fig:dalitz_a}
\end{figure*}

\begin{figure*}
\includegraphics[width=0.45\textwidth]{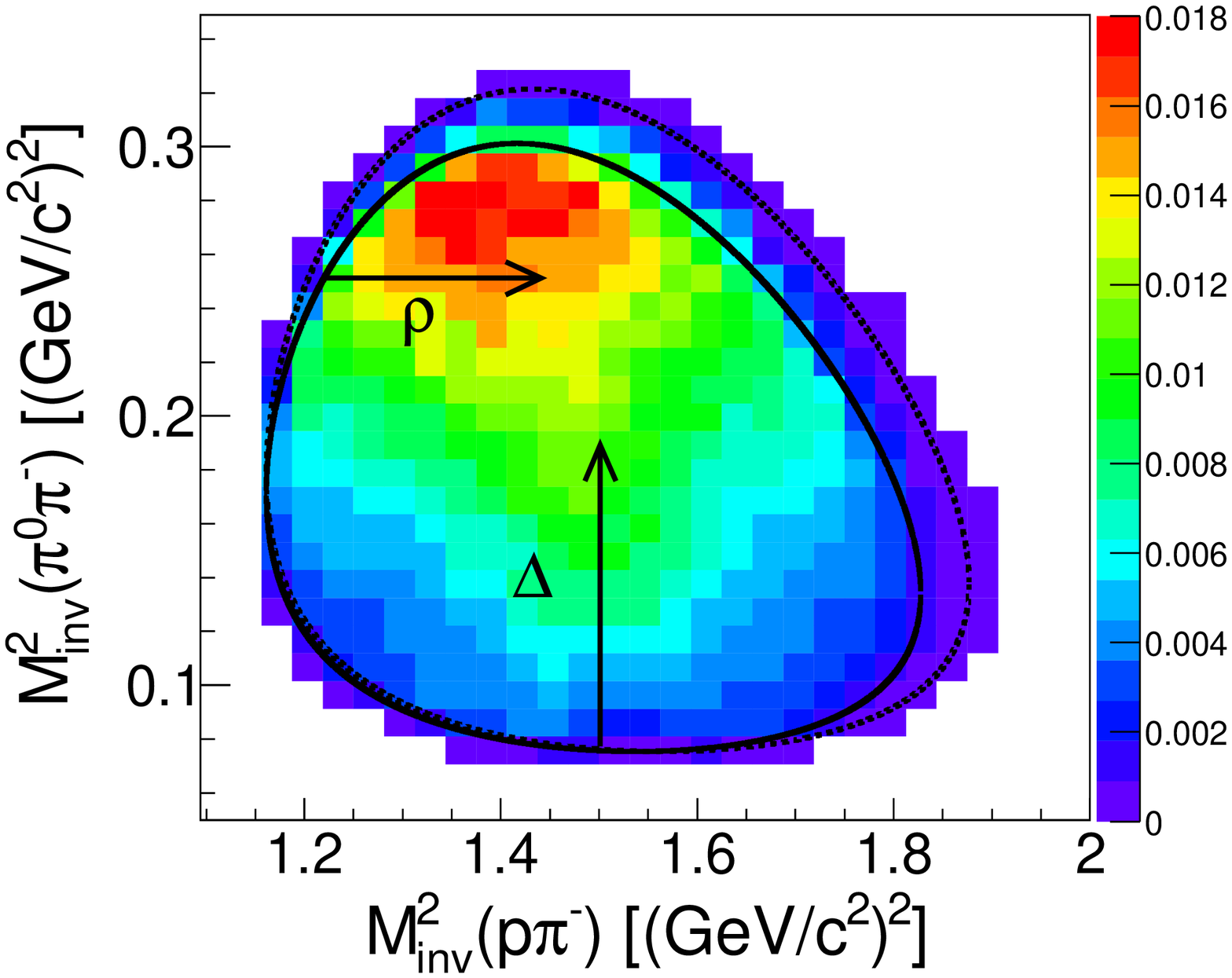}
\includegraphics[width=0.45\textwidth]{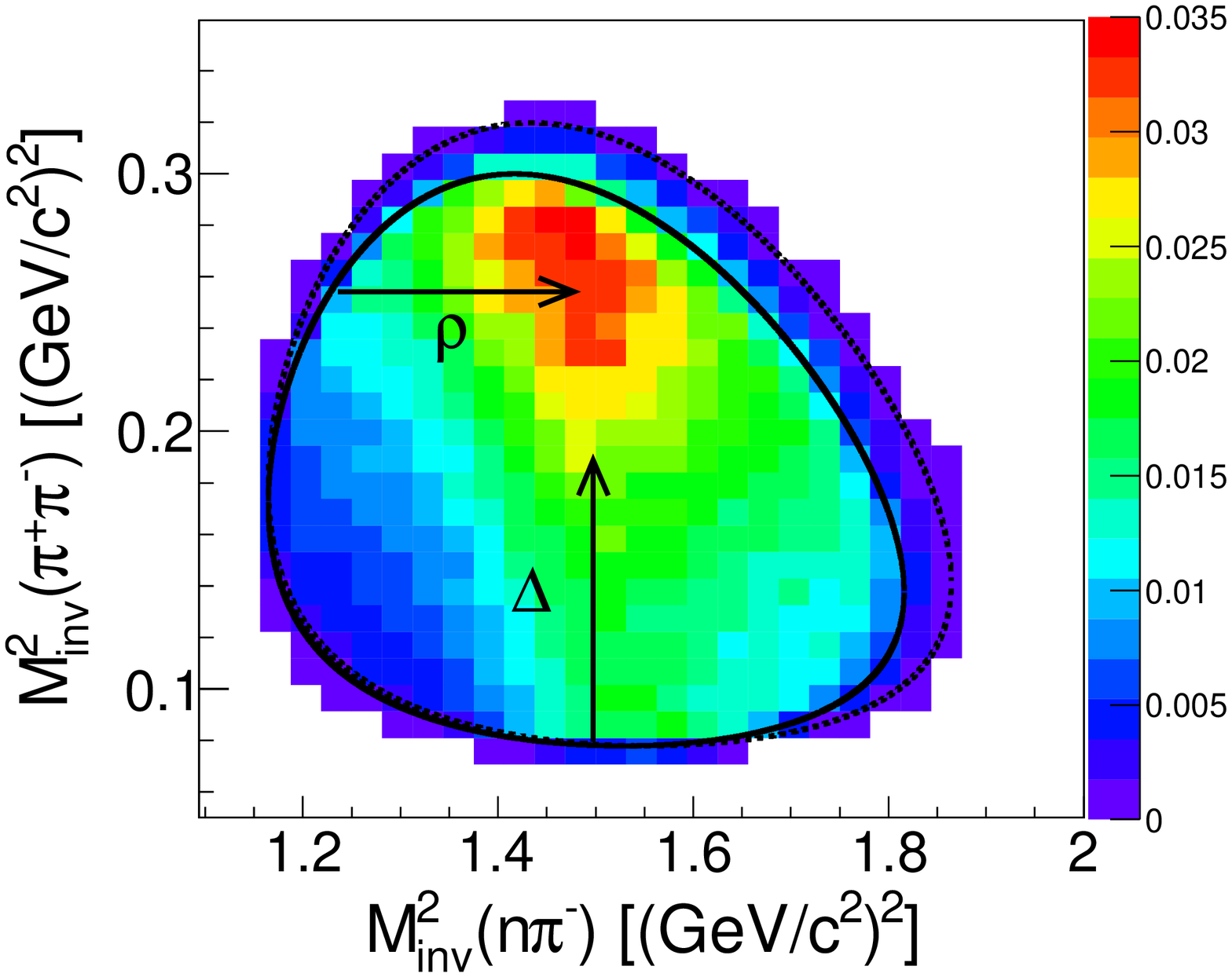}

\caption{Same as Fig.~\ref{fig:dalitz_a} for the distributions from the PWA solutions in the full solid angle.
}
\label{fig:dalitz_b}
\end{figure*}
        
The obtained decomposition into partial waves shows that at HADES energies the most significant contributions originate from $L=0,~1,~2$ ($S,~P,~D$) waves. The respective amplitudes, extracted from our analysis, are associated with $J^P: 1/2^-~(L=0), ~1/2^{+},~ 3/2^+~(L=1)$, and $3/2^-,~5/2^-~(L=2)$ waves (the total parity of pion-nucleon scattering is given by $(-1)^{L+1}$). The truncation at $L=3$ imposed in our analysis is in agreement with the previous PWA analysis of two-pion production \cite{Man:84,Man:92}, and the analysis of inelasticity concluded from pion-nucleon elastic scattering \cite{Hoh:83,Cut:79} at similar energies. The results show that for $\sqrt{s}<1550$ MeV the contributions from $L>3$ are negligible.


The separation of the total cross section into isobar contributions is shown in Fig.~\ref{fig:inv_mass}. It presents the invariant mass distributions of the nucleon-pion (left column) and the two-pion (right column) pairs for the $\pi^-p\rightarrow n\pi^+\pi^-$ (upper row) and the $\pi^-p\rightarrow p\pi^0\pi^-$ (lower row) reaction channels. 
In  the latter case, the dominant contribution is the off-shell s-channel $\rho-$meson production, proceeding predominately via the $D_{13}$, and, to smaller extent, via the $S_{11}$ partial waves. The two-pion mass distributions are cut below the meson pole by the limited phase space ($\sqrt{s}-M_n= 0.55$ GeV/c$^2$). On the other hand, the proton-pion invariant mass distributions are peaked around the $\Delta(1232)$  resonance pole, indicating also the formation of this isobar in the intermediate two-particle state.  
In the $\pi^-p\rightarrow n\pi^+\pi^-$ reaction, the $\Delta\pi$ final state is much  stronger (by about a factor $3$) with respect to the $p\pi^-\pi^0$ final state and dominates over the $\rho N$ channel. Such increase of the $\Delta \pi$ contribution can be understood from the isospin considerations for the $I=1/2$ states. An enhancement by a factor of $2.5$ is expected from the Clebsch-Gordan coefficients for $N^{*0}\rightarrow (\Delta^-\pi^+$, $\Delta^+ \pi^-)$ and for  $N^{*0}\rightarrow (\Delta^0 \pi^0$, $\Delta^+ \pi^-)$ contributing to the $n \pi^+\pi^-$ and $p \pi^-\pi^0$ final states, respectively. Furthermore, as already mentioned (see Fig.  \ref{fig:CM_angles}), in addition to $D_{13}$, also the $P_{11}$, with comparable branch to the $\Delta\pi$, contributes to the $n \pi^+\pi^-$ final state, and consequently, the cross section increases. This difference in the isobar contributions is also reflected in the pion angular distributions for the respective $D_{13}$ and $P_{11}$ components (see Figs. \ref{fig:CM_angles}b and \ref{fig:CM_angles}d). While for the
$n \pi^+\pi^-$ final state the distribution is dominated by $\Delta^-$ emission pattern, leading to an almost flat $\pi^+$ angular distribution for all s-channel contributions, in the case of the  $p \pi^-\pi^0$ final state, a clear anisotropy is observed for $\pi^0$s emitted from the $D_{13}$, due to  the strong contribution from the decay of $N(1520)$ into $\rho N$.   

\begin{figure*}
\includegraphics[width=0.45\textwidth]{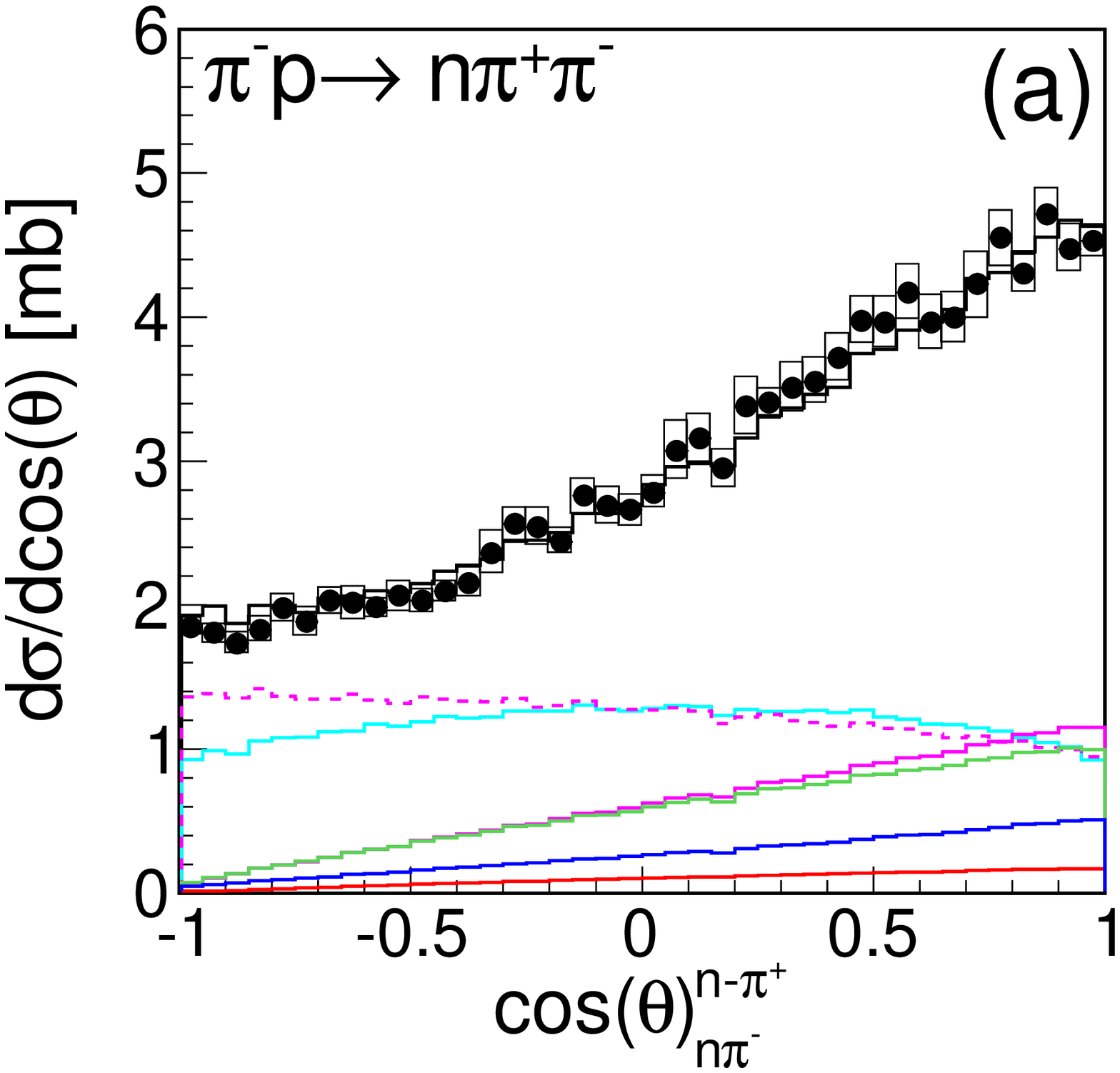}
\includegraphics[width=0.45\textwidth]{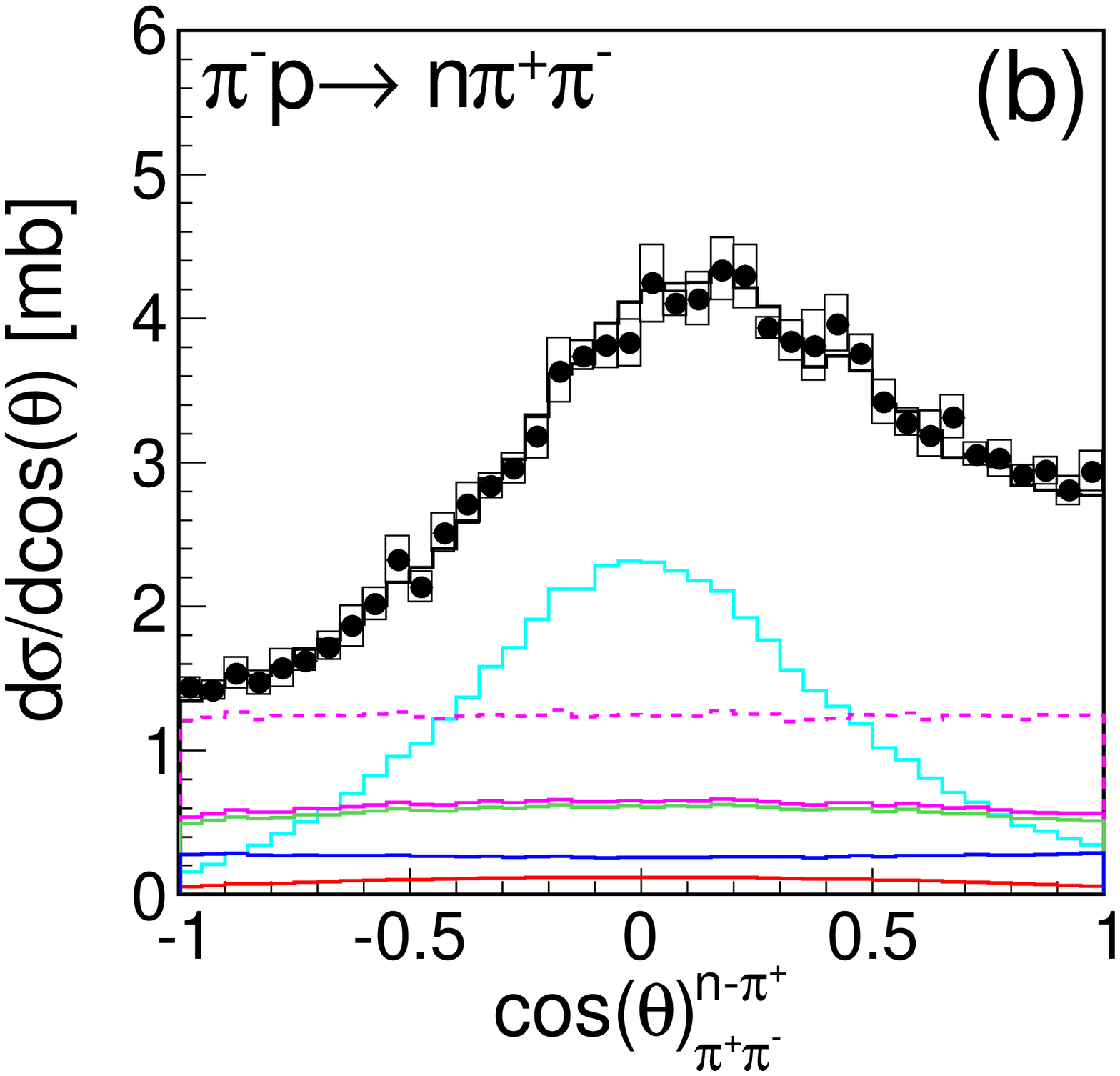}
\includegraphics[width=0.45\textwidth]{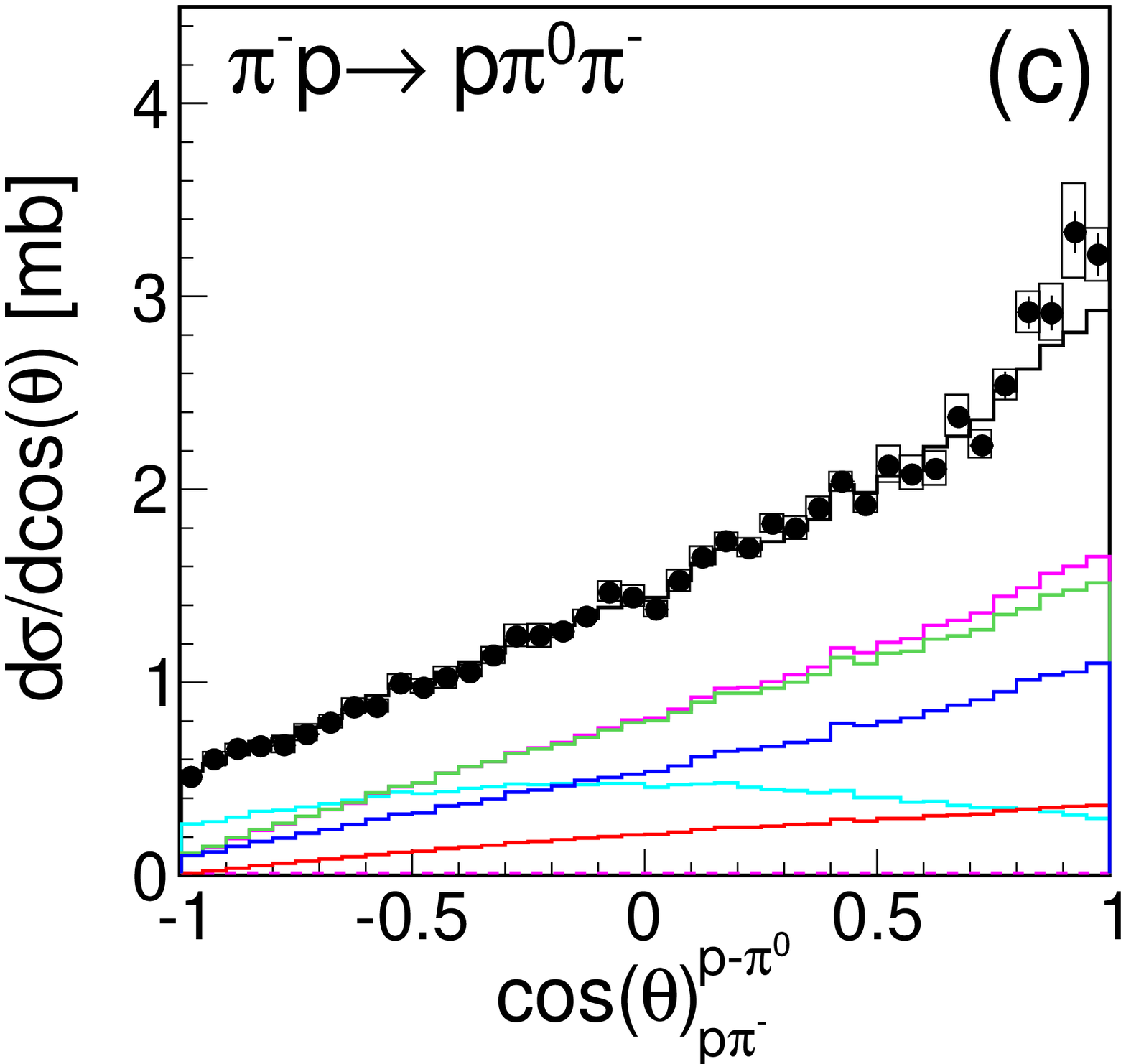}
\includegraphics[width=0.45\textwidth]{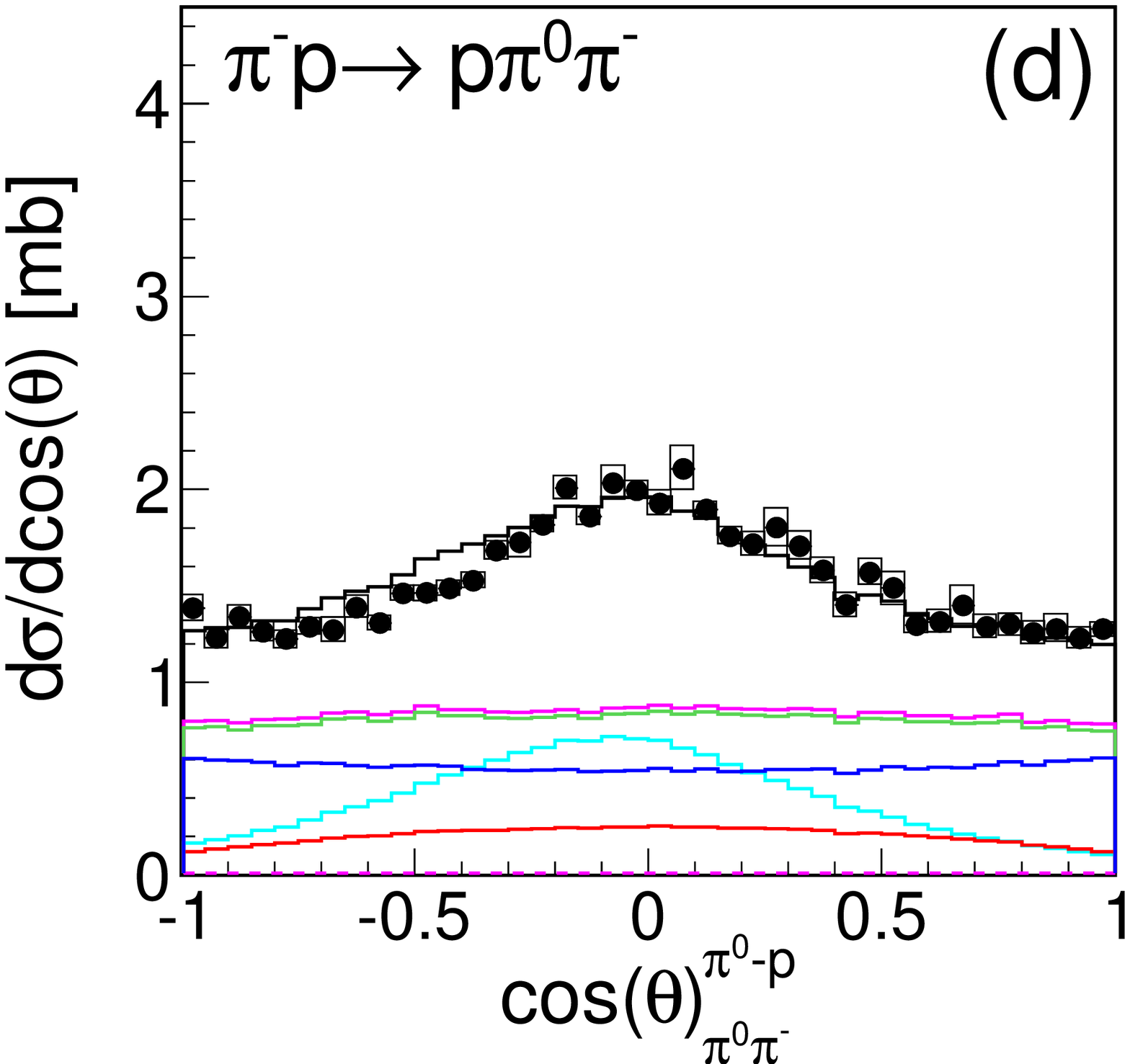}
\includegraphics[width=35pc,height=3.5pc]{leg_set2.eps}
\caption{(Color online) Angular distributions of pions in the nucleon-pion (left) and nucleons in the pion-pion (right) helicity frames for the $\pi^-p\rightarrow n\pi^+\pi^-$ (upper) and $\pi^-p\rightarrow p\pi^0\pi^-$ (lower) reaction channels. The subscript labels the helicity frame, the superscript the angle between the given particles in that frame. The z-axis of the helicity frame is chosen opposite to the neutron (upper panel) and the proton (lower panel) directions. Color curves display various final state contributions (indicated in the legend).}
\label{fig:helicity}
\end{figure*}

In the $\pi^-p\rightarrow n\pi^+\pi^-$ reaction channel also two-pion production in the $I=0$ state ($N\sigma$) is allowed. This isobar production proceeds mainly through the decay branch of the $N(1440)\frac{1}{2}^+$ resonance. As can be seen in the left column of  Fig.~\ref{fig:inv_mass}, the invariant mass distributions for the $N\sigma$ and $\Delta\pi$ contributions have slightly different shapes as compared to the ones for the $\pi^-p\rightarrow p\pi^0\pi^-$ reaction dominated by the $N\rho$ isobar.  The $\sigma$ contribution is also reflected in the angular distribution of two-pions from $J=1/2^+$($D_{13}$).

Finally, in Fig.~\ref{fig:dalitz_a} we show the Dalitz plots for the $p\pi^-\pi^0$ (left panel) and $n\pi^+\pi^-$ (right panel) final states, respectively, with the indicated positions for the expected $\rho$ and $\Delta$ resonance contributions. The upper plots are shown for the HADES data (corrected for the HADES acceptance and reconstruction inefficiency), within the HADES acceptance, while the lower ones are obtained from the PWA in the full solid angle. The solid contours visible in the figure visualize the respective envelopes for the Dalitz distributions assuming a fixed value of $\sqrt{s}$ corresponding to the central beam momentum values $p_{beam}=0.685$ GeV/c. Events distributed outside the envelopes are due to the beam momentum spread and the HADES spectrometer resolution. The distributions show clear enhancements along the indicated positions and good coverage of the detector. The cut-off for the highest $\pi^-\pi^0$ and $n\pi^-$ invariant masses, visible in the Dalitz plots of Fig.~\ref{fig:dalitz_a}, is due to the lack of the detector acceptance at low polar angles essential for the event reconstruction in the case of production close to the kinematic limits. For example, in case of the $p\pi^-\pi^0$ final state, higher energy protons, which would lead to higher invariant masses $M(\pi^0-\pi^-) > 0.27 $ (GeV/$c^2)^2$ , go to polar angles less than 20 degrees in the lab and can thus not be detected in HADES.

\subsection{Three-particle distributions}
\label{sec-three-part}

\begin{figure*}[t]
\includegraphics[width=0.45\textwidth]{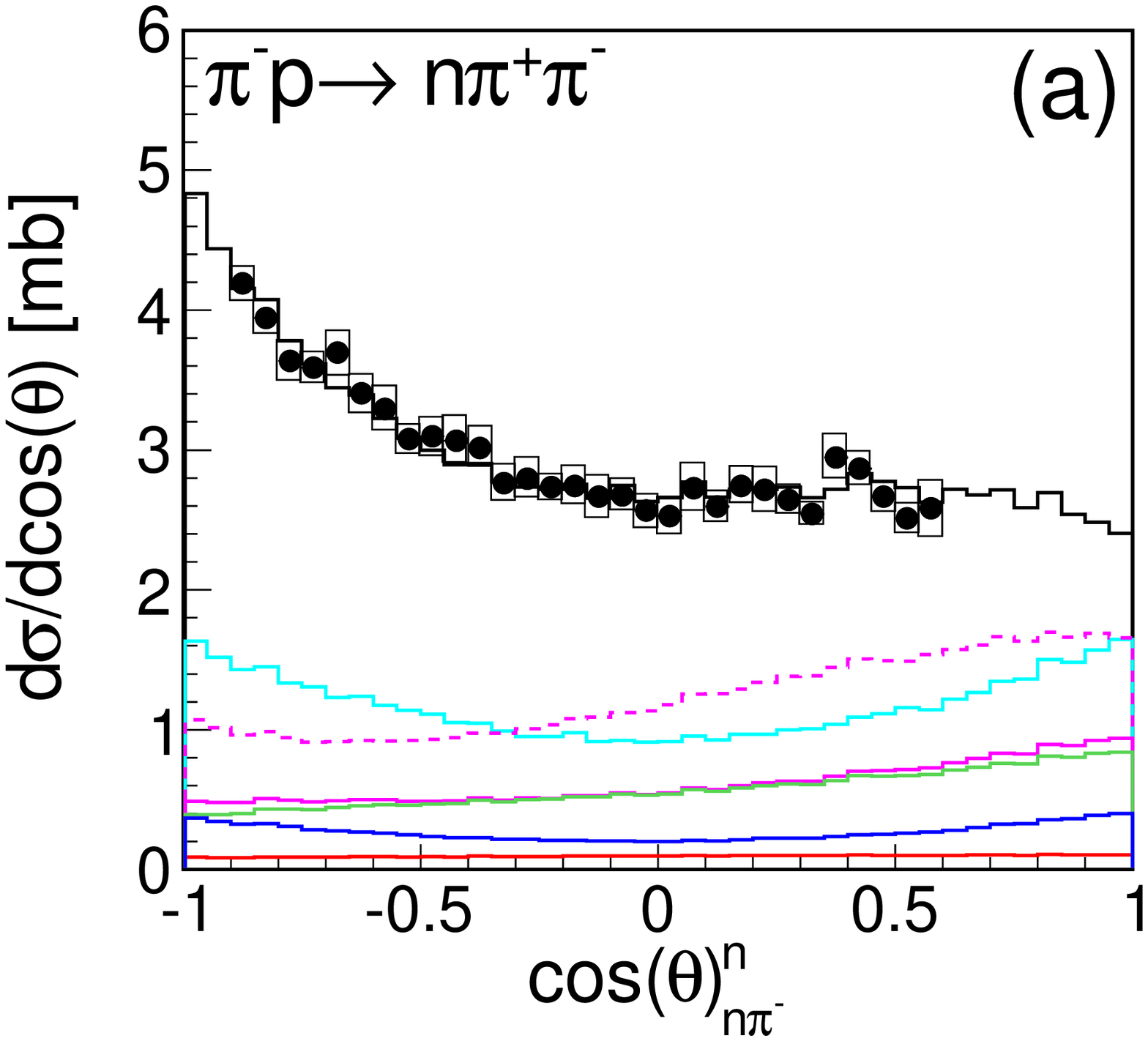}
\includegraphics[width=0.45\textwidth]{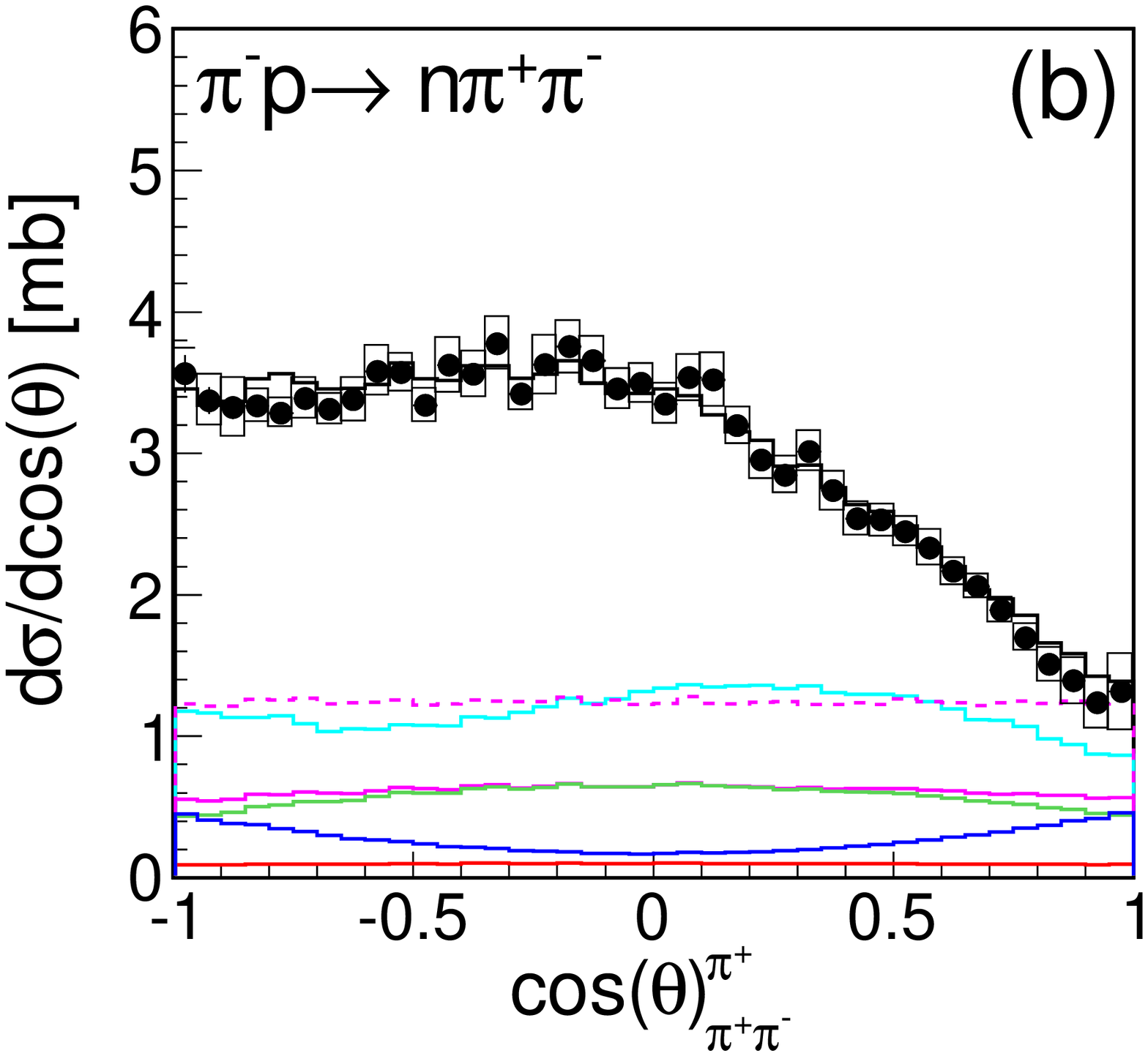}
\includegraphics[width=0.45\textwidth]{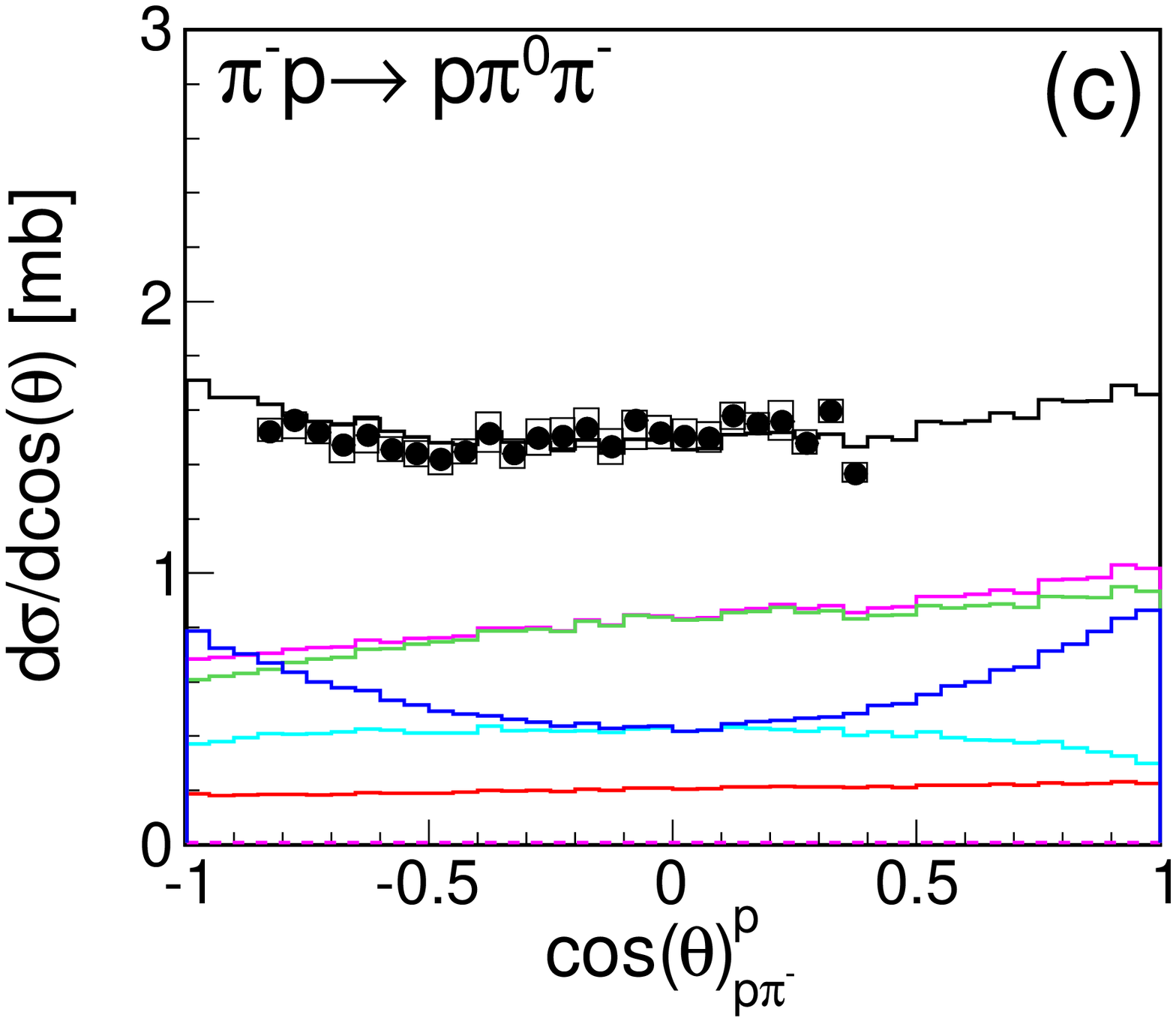}
\includegraphics[width=0.45\textwidth]{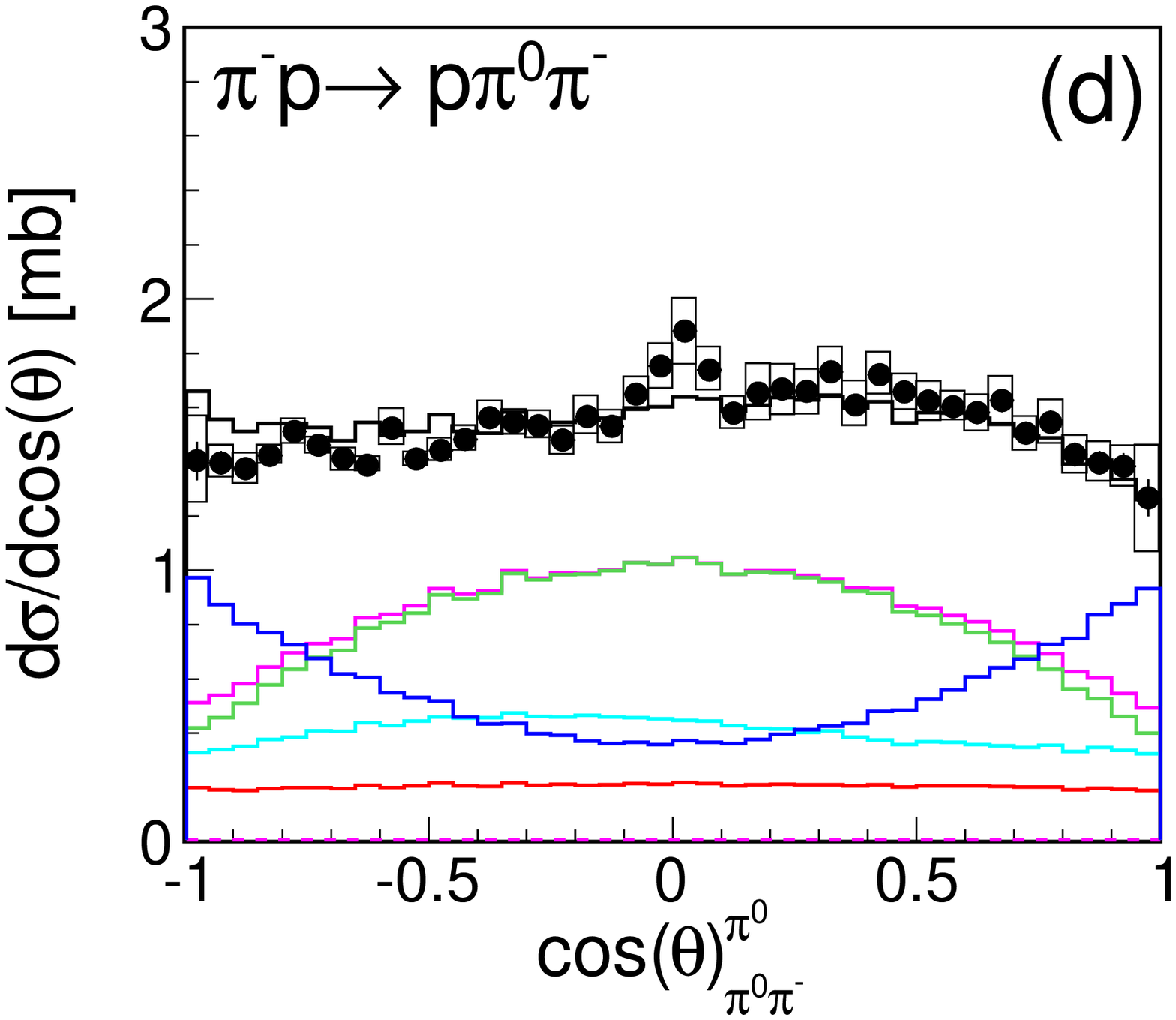}
\includegraphics[width=35pc,height=3.5pc]{leg_set2.eps}
\caption{\label{ref_frame} Angular distributions of incoming pions in the nucleon-pion (left) and in the pion-pion (right) Gottfried-Jackson frames for the $\pi^-p\rightarrow n\pi^+\pi^-$ (upper) and $\pi^-p\rightarrow p\pi^0\pi^-$ (lower) reaction channels. The subscript indicates the two-body reference rest frame and the superscript the angle of the given particle relative to the $\pi$-beam direction. The z-axis of the Gottfried-Jackson rest frame is chosen along the direction of nucleon(left) and pion (right). Color curves display various final state contributions (indicated in the legend).}
\label{fig:GJ}
\end{figure*}

The configurations of three particles in the final state can be studied by angular distributions calculated in the Gottfried-Jackson and the Helicity frames.

The Helicity reference frame (H) is defined as the rest frame of any two final state particles with the z-axis aligned along the direction opposite to the third particle momentum. 
The  polar emission angle of a 
particle, boosted to that frame, is the helicity angle of this particle. Angular distributions in this  particular reference frame (H) are strongly related to 
 the Dalitz plot. They are sensitive to the presence of a resonance  and/or higher partial waves involved in the respective two-particle final state. For example, the appearance of $\rho-$meson production in the $\pi^-p\rightarrow \pi^+\pi^-n$ reaction is expected to produce an enhancement distributed along the high $M^2_{\pi\pi}$ edge of the Dalitz plot spanned by the $M^2_{\pi\pi}$ vs $M^2_{N\pi}$ invariant masses (see Fig. \ref{fig:dalitz_a}). In the pion-nucleon helicity frames the  respective enhancement is expected to show up at large opening angles between the two pions (corresponding to a large $\pi\pi$ invariant mass).  

Indeed, such an enhancement is visible in the distributions plotted in Fig. \ref{fig:helicity} (left column) for small opening angles between the nucleon and the $\pi^+$  in the nucleon-$\pi^-$ helicity frame (indicated by subscripts), which are equivalent to large pion-pion angles. This is visible for the  $\pi^-p\rightarrow n\pi^+\pi^-$ (upper panel) as well as for the $\pi^-p\rightarrow p\pi^0\pi^-$ (lower panel) reaction channels. On the contrary, other channels, like $N\sigma$ or $\Delta\pi$, have more uniform distributions, which demonstrates that the angular distributions in the helicity frame are sensitive to the $N\rho$ production. 

The pion-pion rest system in the H frame, on the other hand, is sensitive to the appearance of the $\Delta$ resonance, which can be clearly seen as a bump in the respective pion-nucleon angular distributions (see right column in Fig.~\ref{fig:helicity}). As already evident from the distributions of the pion-nucleon invariant mass (see Fig. \ref{fig:inv_mass}), the $\Delta(1232)$ is observed in the covered mass range. Consequently, the bump is also observed in the respective angular distribution in the H frame. 

One should emphasize that the presented angular distributions are fully covered by the HADES acceptance. This facilitates acceptance corrections and extraction of the total production cross sections, presented later in this section, as already introduced in Sec. \ref{sec3_acc}. 

In the Gottfried-Jackson reference frame (GJ) the projectile or the target is chosen as the  reference particle boosted to the rest frame of two final state particles. The opening angle between the projectile or the target, and one particle forming the rest frame is calculated. 
The angular distributions plotted in Fig.~\ref{fig:GJ} show the opening angles of the incoming pion projectile with respect to one of the particles (indicated by the superscript) in the two-body reference rest frame (indicated by the subscript).
The angle plotted in the GJ frame is well suited to study the production mechanism. In particular, the two-body scattering of the incoming pion with an exchanged particle to a two-particle final state can be studied by the respective distribution. For example, the t-channel production of a $\rho-$meson with subsequent decay into two pions or of a  $\Delta(1232)$ decaying into a nucleon and a pion 
could be studied in the GJ frame fixed to the two-pion or the pion-nucleon rest frames, respectively.
\begin{figure*}

\includegraphics[width=0.48\textwidth,height=0.44\textwidth]{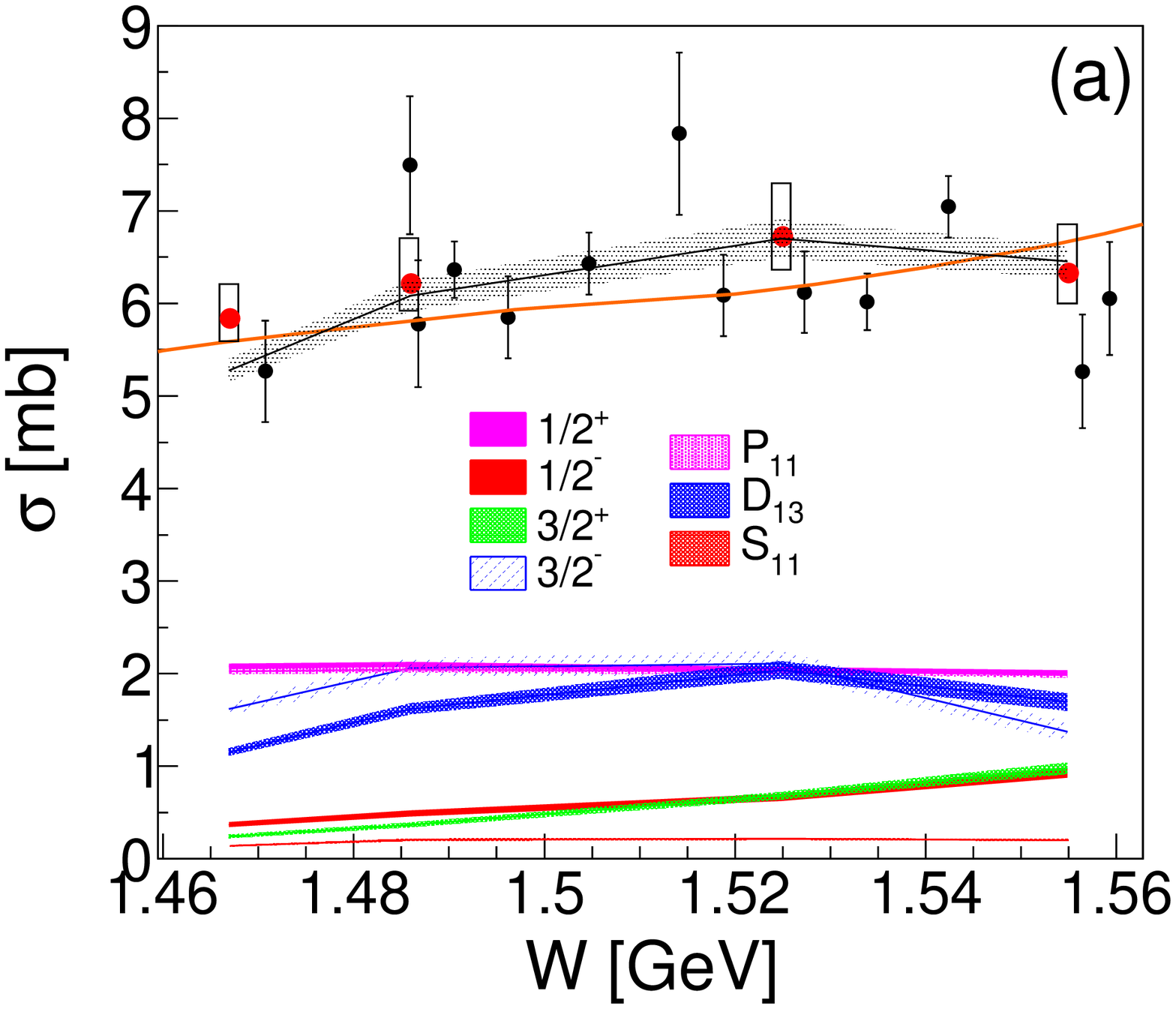}
\includegraphics[width=0.48\textwidth,height=0.44\textwidth]{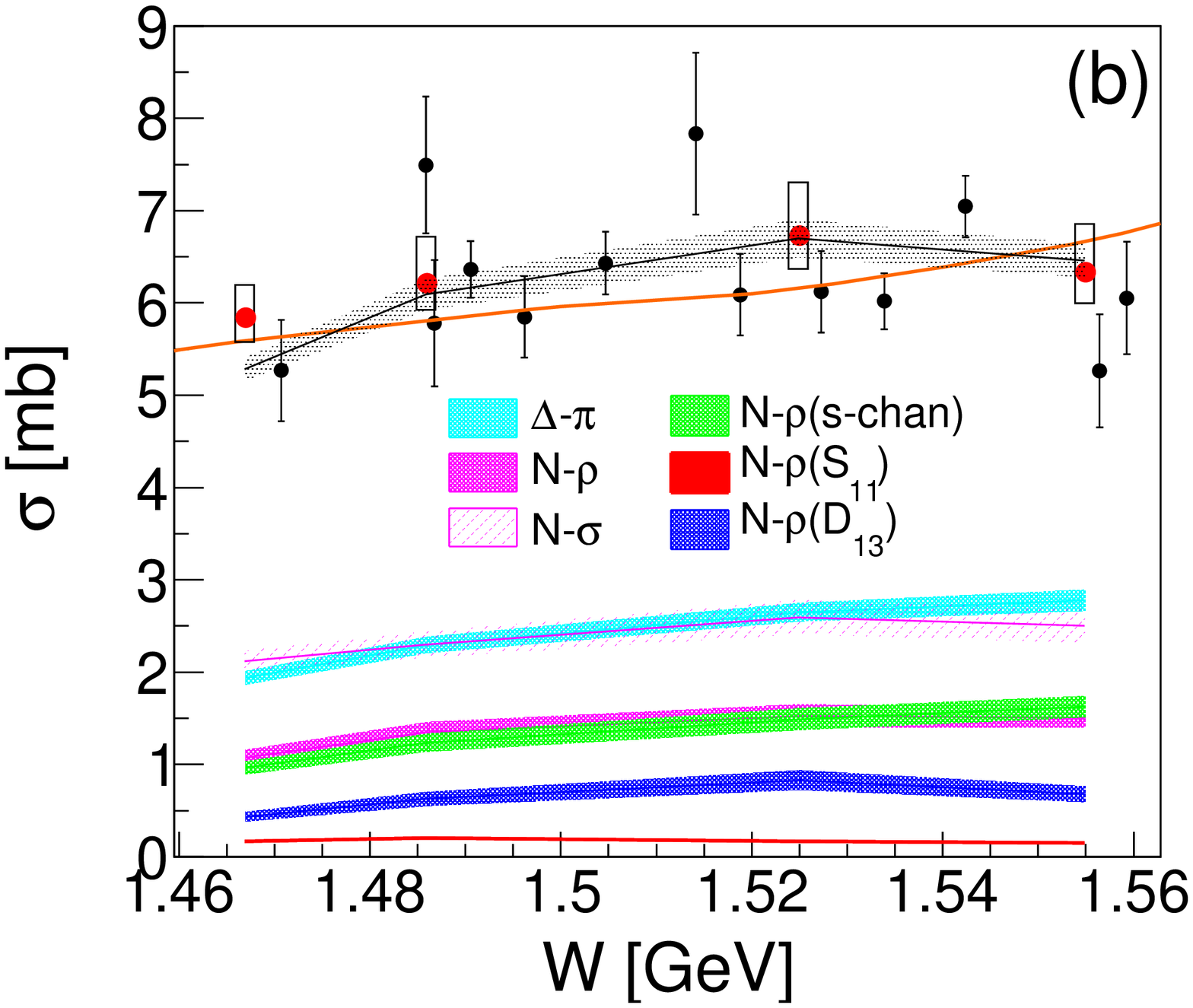}
\includegraphics[width=0.48\textwidth,height=0.44\textwidth]{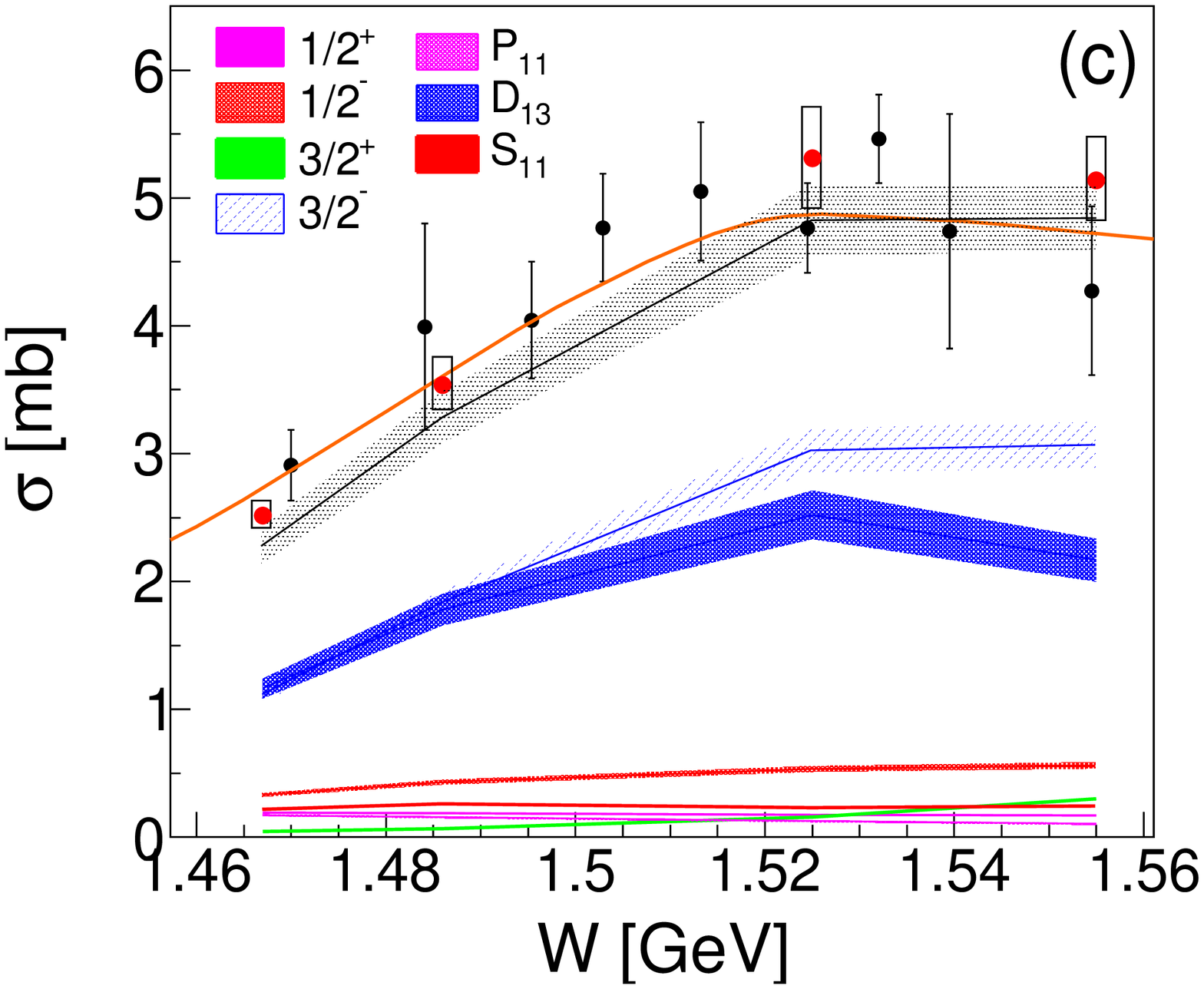}
\includegraphics[width=0.48\textwidth,height=0.44\textwidth]{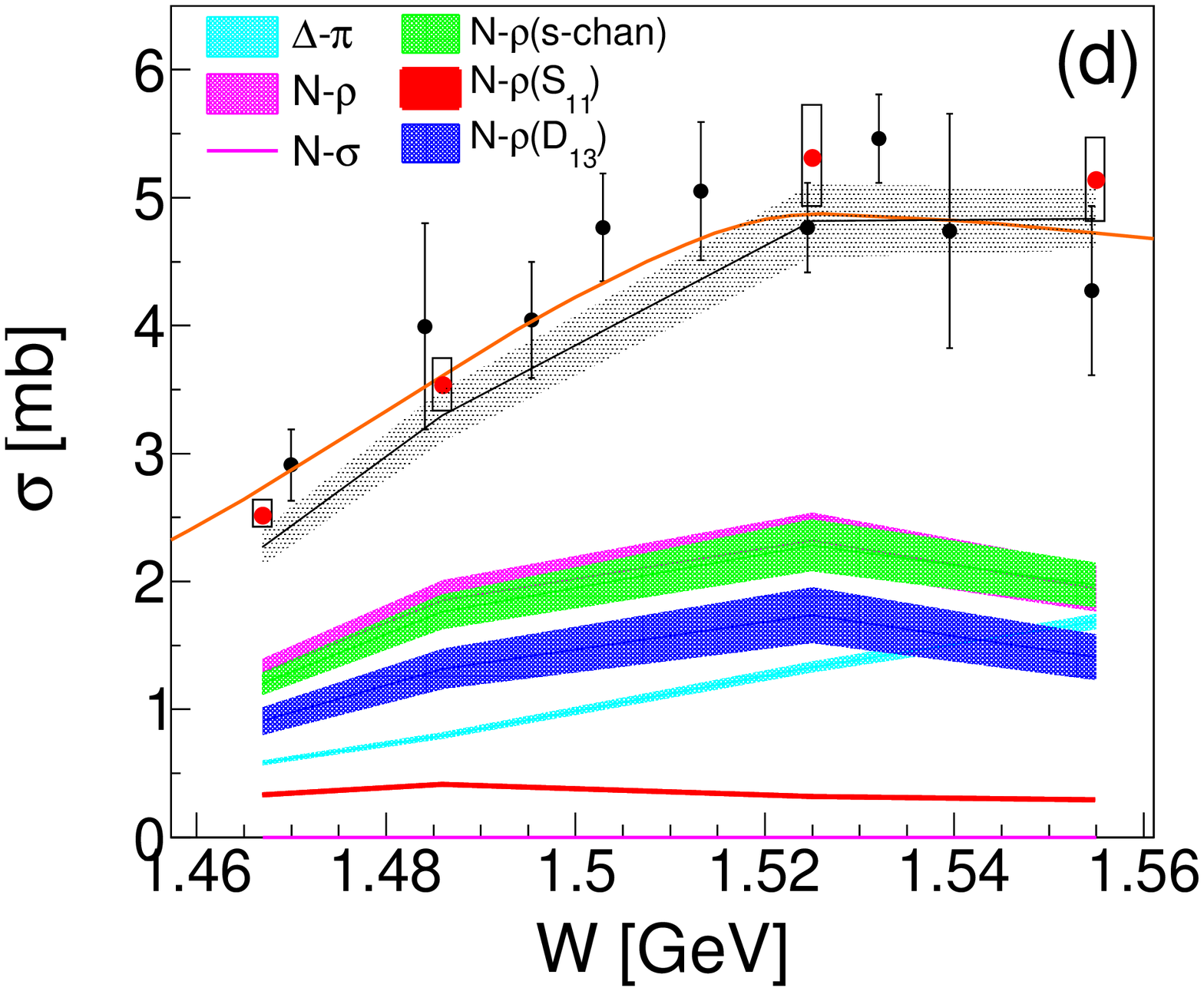}

\includegraphics[width=35pc,height=2.pc]{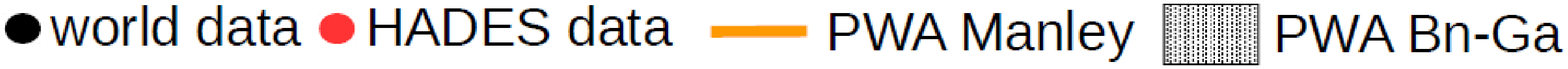}
\caption{\label{ref_frame} Total production cross sections for the two-pion production in the $\pi^-p\rightarrow n\pi^+\pi^-$ (panels a, b) and the $\pi^-p\rightarrow p\pi^0\pi^-$ (panels c, d) reaction channels. Results from this work (red points) and from the other experiments (black points) are shown as a function of the total energy ($W=\sqrt{s}$) in the CM frame. 
The figures in the left column present the subdivision into the $J^P$ partial waves and the $I=1/2$ $N^*$ contributions (see legend). The solid curves are results from the Bonn-Gatchina solution (black curves) and the one obtained in Refs. \cite{Man:84,Man:92} (orange curves), respectively. The curves in the right column display the contributions of the isobar $\Delta\pi$ (cyan curve), $N\sigma$ (dashed violet curve) and $N\rho$ (violet curve) final states. The latter one is subdivided into the coherent sum of s-channels (green curves), $D_{13}$ (blue curve) and $S_{11}$ (red curves) partial waves.}
\label{fig:XS}
\end{figure*}
 The distributions displayed in the left column are plotted for the nucleon-pion systems, while the ones in the right column are for the two-pion system. The upper and lower panels correspond to the $\pi^-p\rightarrow n\pi^+\pi^-$ and $\pi^-p\rightarrow p\pi^0\pi^-$ channels, respectively. They are similar to the CM distributions discussed above (see Fig.~\ref{fig:CM_angles}) but, decomposed into the isobars, reveal interesting features for the $\rho N$ channel. For both final states, the pion distributions emerging from the $\rho N$ final state (Fig.~\ref{fig:GJ}b and \ref{fig:GJ}d),  are anisotropic and convex in shape for the coherent sum of s-channel contributions while they are concave for the dominant $D_{13}$ contribution. This signals strong interferences between the dominant  $I=1/2$ partial waves, $S_{11}$ and $D_{13}$ with also some contribution from $I=3/2$ components.  The latter one will be discussed in more detail in the next section where the partial wave contributions to the total cross section are discussed. One should also point out that the distributions shown in Figs.~\ref{fig:GJ} (b,d) demonstrate a good coverage of HADES for these observables, which allow for a total cross section estimate. 

Summarizing, all the presented angular distributions agree very well with the Bonn-Gatchina PWA solutions. They are consistent with the isobar model assuming formation of the quasi two-particle final states: $\Delta \pi$, $N \rho$ in the $\pi^-p \rightarrow p \pi^-\pi^0$, and additionally $N \sigma$ in the $\pi^-p \rightarrow n\pi^+ \pi^-$ reaction channels. The results show the dominance of the $J^P=1/2^+,3/2^-$ partial waves with $D_{13}$, and $P_{11}$ playing the most important role in the two-pion production, however with visible interferences with $S_{11}$ and smaller $I=3/2$ amplitudes. In particular, a high sensitivity of the angular distributions in the H and GJ frame and of the two-pion invariant masses to the off-shell $\rho-$meson contribution has been found. Furthermore, the broad angular coverage of HADES for measurements of angular distributions in the GJ and the H frames allows for the extraction of the total cross section for two-pion production, presented in the next paragraph.

\subsection{Total cross section}  
\label{cross-sec}

The total cross sections for the two pion production are shown in Fig.~\ref{fig:XS} as a function of the total energy $W=\sqrt{s}$ in the CM system for the $\pi^-p\rightarrow n\pi^+\pi^-$ (upper row) and the $\pi^-p\rightarrow p\pi^0\pi^-$ (bottom row) reaction channels. The red points show the results of this analysis, while the black ones were obtained from the other experiments (see Refs.~\cite{Man:84,Man:92}). The total cross sections from HADES were calculated as explained in Sec.~\ref{sec3_acc} and are also summarized in Tab.~\ref{X_sec}. 
The error bars for the HADES data are dominated by the systematic uncertainties (boxes) while statistical ones are negligible. The errors of the older experiments are statistical only.

 The total cross sections derived from the HADES data agree within the errors with the results of the former experiments. 
\begin{table}[!h]
 \centering
\caption{Total cross sections.}
  \label{X_sec}
 \begin{tabular}{|c|c|c|}
 \hline
 \textbf{p [GeV/c]} & \textbf{$\sigma(n\pi^+\pi^-$) [mb]} & \textbf {$\sigma (p\pi^-\pi^0$) [mb]} \\\hline  
0.650  & 5.84 $(\pm 0.01)_{st}$ $(^{+0.34}_{-0.28})_{sys}$ & 2.52 $(\pm 0.01)_{st}$ $(^{+0.13}_{-0.11})_{sys}$ \\ \hline 
0.685 & 6.21 $(\pm 0.01)_{st}$  $(^{+0.48}_{-0.31})_{sys}$ & 3.54 $(\pm 0.01)_{st}$  $(^{+0.23}_ {-0.21})_{sys}$ \\ \hline 
0.733 & 6.73 $(\pm 0.01)_{st}$  $(^{+0.55}_{-0.39})_{sys}$ & 5.31 $(\pm 0.01)_{st}$  $(^{+0.42}_{-0.40})_{sys}$ \\ \hline
0.786 & 6.33 $(\pm 0.01)_{st}$  $(^{+0.50}_{-0.35})_{sys}$ & 5.14 $(\pm 0.01)_{st}$  $(^{+0.35}_{-0.33})_{sys}$  \\ \hline
 \end{tabular}
 \end{table}
The excitation functions are compared to the Bonn-Gatchina solutions (black curves-denoted as Bn-Ga) and also to the results of the analysis described in Ref.~\cite{Man:84} (orange curve). The dashed band spanned around the Bn-Ga solutions visualizes the errors (RMS) obtained from the various solutions.   

\begin{widetext}

\centering
 \begin{table*}[h]
 \hspace{15mm}
 \caption{Cross sections (in mb) derived for the $p\pi^0\pi^-$ channel at the four incident pion momenta. The contributions of the most important $J^P$ partial waves are given. The three last columns correspond to the s-channel I=1/2 partial waves.}
  \label{tab4}
 \begin{adjustwidth}{-1cm}{}
 \begin{tabular}{|c|c|c|c|c|c|c|c|c|c|c|c|}
 \hline
 \textbf{p [GeV/c]}&\textbf{W [GeV]} & \textbf{Total}&\textbf{1/2}$^{+}$&\textbf{1/2}$^{-}$&\textbf{3/2}$^{+}$&\textbf{3/2}$^{-}$&\textbf{5/2}$^{+}$&\textbf{5/2}$^{-}$&\textbf{S$_{11}$}&\textbf{P$_{11}$}&\textbf{D$_{13}$} \\\hline
 0.650&1.47 & 2.26$\pm$0.14 &0.19$\pm$0.01&0.32$\pm$0.01&0.04$\pm$0.001&1.11$\pm$0.06&0.013$\pm$0.003 &0.02$\pm$0.002&0.21$\pm$0.02 &0.17$\pm$0.01 &1.16$\pm$0.08\\\hline
 0.685&1.49 & 3.28$\pm$0.20 &0.19$\pm$0.01&0.43$\pm$0.02&0.07$\pm$0.003&1.81$\pm$0.10&0.02$\pm$0.004    &0.03$\pm$0.003&0.25$\pm$0.02 &0.16$\pm$0.006&1.78$\pm$0.12\\\hline
 0.733&1.52 & 4.8$\pm$0.29  &0.18$\pm$0.01&0.53$\pm$0.02&0.16$\pm$0.01&3.02$\pm$0.17 &0.04$\pm$0.01   &0.06$\pm$0.01&0.22$\pm$0.02 &0.13$\pm$0.01 &2.52$\pm$0.19\\\hline
 0.786&1.55 &4.83$\pm$0.25  &0.17$\pm$0.01&0.56$\pm$0.02&0.30$\pm$0.04&3.07$\pm$0.18 &0.08$\pm$0.02   &0.11$\pm$0.01&0.23$\pm$0.02  &0.10$\pm$0.01 &2.17$\pm$0.17\\\hline
 \end{tabular}
 \end{adjustwidth}
  \end{table*}
  
\begin{table}[!h]
\caption{Same as Table \ref{tab4} for the $n\pi^+\pi^-$ channel.}
  \label{tab5}
\begin{adjustwidth}{-1cm}{}
 \begin{tabular}{|c|c|c|c|c|c|c|c|c|c|c|c|}
 \hline
 \textbf{p [GeV/c]}&\textbf{W [GeV]} & 
 \textbf{Total}&\textbf{1/2}$^{+}$&\textbf{1/2}$^{-}$&\textbf{3/2}$^{+}$&\textbf{3/2}$^{-}$&\textbf{5/2}$^{+}$&\textbf{5/2}$^{-}$&\textbf{S$_{11}$}&\textbf{P$_{11}$}&\textbf{D$_{13}$} \\\hline
 0.650 & 1.47 &5.27$\pm$0.13 &2.06$\pm$0.03& 0.36$\pm$0.03 & 0.23$\pm$0.02 &1.61$\pm$0.07 & 0.05$\pm$0.01&0.07$\pm$0.01&0.14$\pm$0.01& 2.02$\pm$0.03&1.14$\pm$0.04 \\\hline
 0.685 & 1.49 &6.08$\pm$0.17&2.08$\pm$0.03& 0.48$\pm$0.03 & 0.35$\pm$0.02 &2.05$\pm$0.10 & 0.07$\pm$0.02 &0.10$\pm$0.01&0.20$\pm$0.01 &2.04$\pm$0.03&1.6$\pm$0.06\\\hline
 0.733 & 1.52 &6.70$\pm$0.21 &2.04$\pm$0.03& 0.65$\pm$0.04 & 0.67$\pm$0.04 &2.09$\pm$0.14 & 0.14$\pm$0.03 &0.21$\pm$0.01&0.21$\pm$0.01&2.01$\pm$0.028&2.04$\pm$0.09\\\hline
 0.786 & 1.55 &6.45$\pm$0.18 &1.99$\pm$0.03& 0.93$\pm$0.06 & 0.98$\pm$0.06 &1.37$\pm$0.10 & 0.25$\pm$0.05 &0.38$\pm$0.02&0.2$\pm$ 0.01&1.97$\pm$0.030& 1.68$\pm$-0.10
 \\\hline
  \end{tabular}
  \end{adjustwidth}
 \end{table}
  \end{widetext}

In the left column the total cross section obtained from the PWA are separated into the dominant contributions given by the $1/2^{\pm},3/2^{\pm}$ partial waves contributing to the s-channel (Fig.~\ref{fig:XS} a,c). The total cross section is defined by the incoherent sum of the cross sections from the partial wave amplitudes with a fixed total angular momentum and parity $J^P$. Such amplitudes for fixed $J^P$ are defined as the coherent sum of the respective partial wave amplitudes for $I=1/2$ and $I=3/2$ with the corresponding Clebsch-Gordan coefficients.
The total contribution of the dominant partial waves, listed in Tab.~\ref{tab4},~\ref{tab5}, rises with the incident energy and varies from $83$ to $92\%$ and from $75$ to $89\%$ of the total cross section for the $n\pi^+\pi^-$ and $p\pi^-\pi^0$ final state, respectively. The remaining part of the total cross sections originates from the interferences with t-channel contributions, not included in the partial waves, and from the contributions of the higher partial waves. Tables~\ref{tab4},~\ref{tab5} give a detailed separation of the respective partial waves 
obtained from  the Bonn-Gatchina analysis. The specified errors have been determined from the dispersion of the various PWA solutions, as explained above. The $5/2^{\pm}$ partial waves (not included in Fig.~\ref{fig:XS} c, d) contribute only very little to the total cross sections which justifies the truncation applied in the analysis.

The contributions of the most important $I=1/2$ partial waves: $S_{11},~P_{11},~D_{13}$ are also plotted in Fig.~\ref{fig:XS} for both reaction channels. The $D_{13}$ is dominating the $p\pi^-\pi^0$ final state while in the $n\pi^+\pi^-$ final state the contribution of $P_{11}$ plays a comparable role. The incoherent sum of the $I=1/2$ contributions amounts to $\sim 63\%$ of the total cross section and is roughly constant  for the $n\pi^+\pi^-$ channel, but decreases as a function of the energy from $68$ to $51\%$ for the $p\pi^-\pi^0$ final state. The remaining part of the cross section for the $1/2^{\pm},~3/2^{\pm}$ partial waves is provided by the contribution of the $I=3/2$ amplitudes which increases with the energy for the $p\pi^0\pi^-$ final state, as can be seen in Fig.~\ref{fig:XS} c (compare dashed and full blue lines).        


Tables~\ref{tab6},~\ref{tab7} provide the numerical values of the respective components and their errors deduced from the PWA solutions, as described above. As one can see the $\Delta\pi$ isobar contribution (cyan curve) rises almost linearly with the energy and is the most important contribution in the $n\pi^+\pi^-$ final state. The $N\rho$ channel provides the largest contribution to the $p\pi^-\pi^0$ final state  with the most dominant component originating from the $D_{13}$ partial wave. Furthermore, the $\rho-$meson production in both final states is almost completely determined by s-channels (green curve), as it can be directly concluded from the comparison to the total cross section (violet curve). Similar conclusions on the dominance of the s-channel has also been derived for other final states. The excitation function for the $\rho^-$meson seems to show a resonance-like behaviour with a maximum around the pole of $N(1520)3/2^-$ while the one for the $\rho^0$ is rising more continuously. In the final state with two charged pions also the isoscalar $I=0$ state (dashed violet curve) contributes with a comparable cross section and a rather flat excitation function.   

\begin{widetext}

 \begin{table}[h]
\centering
 \begin{threeparttable}
   \caption{Isobar contributions  (in mb) derived in the PWA for the $p\pi^0\pi^-$ channel at the four incident pion momenta. For the $\rho N$ channel the most important s-channel I=1/2 partial wave contributions ($S_{11},D_{13}$) are given. The last column shows the incoherent sum of the respective $I=3/2$ contributions. }
  \label{tab6}
 \hspace{25mm}
\centering
\begin{tabular}{|c|c|c|c|c|c|c|c|c|}
 \hline
 \textbf{p [GeV/c]}&\textbf{W [GeV]} & \boldmath$\Delta\pi$ &\textbf{N}\boldmath$\rho$&\textbf{N}\boldmath$\sigma$&\textbf{N}\boldmath$\rho$&\textbf{N}\boldmath$\rho$\textbf{($S_{11}$)}&\textbf{N}\boldmath$\rho$\textbf{($D_{13}$)} & \textbf{N}\boldmath$\rho$\textbf{($\Sigma_{\Delta}$)} \\
&&&&&\textbf{(s-channel)}&&& \\\hline
 0.650&1.47 &0.58$\pm$0.02 &1.29$\pm$0.11&0 & 1.2$\pm$0.09 & 0.32$\pm$0.01& 0.91$\pm$ 0.11 &$0.05\pm0.01$
 \\\hline
 0.685&1.49 &0.80$\pm$0.02& 1.85$\pm$0.16 &0& 1.76$\pm$0.14&0.41$\pm$0.01 &1.32$\pm$0.16 &$0.08\pm0.01$
 \\\hline
 0.733&1.52 &1.33$\pm$0.04&2.32$\pm$0.22&0& 2.27$\pm$0.20& 0.32$\pm$0.01& 1.73$\pm$0.22&$0.13\pm0.01$
 \\\hline
 0.786&1.55 &1.69$\pm$0.06& 1.95$\pm$0.18& 0& 1.97$\pm$0.18&0.29$\pm$0.01& 1.41$\pm$0.18&$0.22\pm0.01$
 \\\hline
 \end{tabular}
 \end{threeparttable}
 \end{table}
\hspace{25mm}
\begin{table}[!h]
\centering
\caption{Same as Table \ref{tab6} for the $n\pi^+\pi^-$ channel.}
  \label{tab7}
 \centering
 \begin{tabular}{|c|c|c|c|c|c|c|c|c|}
 \hline
\textbf{p [GeV/c]} &\textbf{W [GeV]} & \boldmath$\Delta\pi$&\textbf{N}\boldmath$\rho$&\textbf{N}\boldmath$\sigma$&\textbf{N}\boldmath$\rho$&\textbf{N}\boldmath$\rho$\textbf{($S_{11}$)}&\textbf{N}\boldmath$\rho$\textbf{($D_{13}$)} & \textbf{N}\boldmath$\rho$\textbf{($\Sigma_{\Delta}$)} \\
 &&&&&\textbf{(s-channel)} &&& \\\hline
 0.650&1.47 & 1.94$\pm$0.07& 1.07$\pm$0.08&2.11$\pm$0.11&0.96$\pm$0.07&0.15$\pm$0.003&0.43$\pm$0.05&$0.1\pm0.01$ \\\hline
 0.685&1.49 & 2.29$\pm$0.09 & 1.35$\pm$0.11&2.29$\pm$0.14&1.24$\pm$0.10&0.19$\pm$0.004&0.63$\pm$0.08 &$0.15\pm 0.01$ \\\hline
 0.733&1.52 & 2.65$\pm$0.10  & 1.53$\pm$0.13&2.59$\pm$0.19&1.49$\pm$0.12&0.15$\pm$0.003&0.83$\pm$0.11&$0.26\pm0.01$\\\hline
 0.786&1.55 & 2.78$\pm$0.12  & 1.51$\pm$0.11& 2.49$\pm$0.18 & 1.62$\pm$0.12& 0.14$\pm$0.003 &0.68$\pm$0.09&$0.42\pm0.01$\\\hline
 \end{tabular}
 \end{table}
\end{widetext}

\subsection{$\boldmath\rho$ meson production}

    One of the main goals of this analysis is to extract the production cross section of the $\rho-$meson and to provide an insight into the reaction mechanism. In particular, the interesting question is the coupling of the $\rho-$meson to $N^*/\Delta$ baryonic resonances characterized by the respective decay branches. Those branches were previously extracted in the analysis of Manley \emph{et al.}  \cite{Man:84,Man:92}, based on the old bubble chamber data. To study this aspect, we have performed a decomposition of the meson production cross section into the dominant $J=3/2^{\pm}$ and $J=1/2^{\pm}$ partial waves and extracted their $I=1/2$, $I=3/2$ components. The results, presented in Tables~\ref{tab6} and \ref{tab7}, show that the most important contributions originate from negative parity $I=1/2$ partial waves and are given by $D_{13}$ and $S_{11}$, for both investigated $\rho-$meson charge states. The $S_{11}$ contribution is approximately constant while the $D_{13}$ contribution is larger and is increasing with the excitation energy. Furthermore, we observe that the incoherent sum of the $I=1/2$ contributions is comparable to the coherent sum of all s-channel contributions for the $p\pi^0\pi^-$ final state. On the other hand, the respective sum is clearly smaller for the $n\pi^+\pi^-$ final state. One should, however, also consider $I=3/2$ contributions to the $1/2^-$ and $3/2^-$ partial waves and the interference effects. The PWA solution show that the main contributions to the $D_{13}$ wave originates from the $N(1520)3/2^-$ resonance (close to $95\%$) in both channels)  while the $I=3/2$ ($\Delta$) is given by the $S_{31}$ and $D_{33}$ partial waves which, although smaller than the respective $I=1/2$ contribution are still important in the coherent sum.  In the case of the full interference even a small partial wave can strongly affect the total contribution. In Tables~\ref{tab6} and \ref{tab7} we present the incoherent sum of the $J^P=1/2^-$ and the $J^P=3/2^-$ partial waves with $I=3/2$ (last column) to quantify their effect with respect to the $I=1/2$ contributions. The contributions are increasing with the energy and amount to around $15-25\%$ and $5-12\%$ with respect to the incoherent sum of $I=1/2$ contributions for the $n\pi^+\pi^-$ and $p\pi^-\pi^0$ final states, respectively.

The importance of the interference effects in the $\rho-$meson production has already been pointed out in the previous section in the interpretation of the pion angular distributions in the (GJ) frame. The opposite interference pattern and the relative contributions of $N^*/\Delta$ to the total coherent sum in both reaction channels can be understood more quantitatively by the isospin decomposition of the total $\rho-$meson production cross section into the respective amplitudes for the $N^*$ and the $\Delta$ components:
\[
\sigma_{\pi^-p \rightarrow N^*/\Delta \rightarrow p \rho^-}^J\sim 1/3(2 A_{N^*}^J+ A_{\Delta}^J)^2,
\]

for the $p\pi^-\pi^0$ final state, and 

  \[
\sigma_{\pi^-p \rightarrow N^*/\Delta \rightarrow n \rho^0}^J\sim 1/3(-\sqrt{2} A_{N*}^J+\sqrt{2} A_{\Delta}^J)^2,
\]
for the $n\pi^+\pi^-$ final state, respectively.

As can be seen from Tables~\ref{tab6} and \ref{tab7}, the $I=1/2$ cross sections are indeed a factor 2 larger for the $p \rho^-$ as compared to the ones for the $n \rho^0$, independent of the energy, in agreement with the above isospin decomposition. On the other hand the contributions of $I=3/2$ are larger for the $n \rho^0$ final state. Furthermore, the sign of $A_{N*}$ has been found in the PWA to be opposite to the one for the $A_{\Delta}$ for the $p\rho^-$ final state, therefore the interference is destructive for this final state while for the other $\rho-$meson charge state it is constructive. From the above expression one can also see that the ratio of the interference term with respect to the incoherent sum of both amplitudes for the two reactions is larger for the $n\rho^0$ final state. It is worth mentioning that the same conclusions on the amplitude signs were drawn from the previous analysis (Ref.~\cite{Man:84}), in agreement with quark models cited in there.

Finally the branching ratios for the $N(1520)3/2^-\rightarrow N
\rho$, and $N(1535)1/2^{-}~\rightarrow~N\rho$  have been extracted from the Bonn-Gatchina
analysis. For the $N(1520)3/2^-$ state the corresponding branching ratio was found to be
$11.8\pm 1.9\%$ for the decay with the orbital moment $L=0$ (S-wave),
and $0.4\pm 0.2\%$ for the decay with the orbital moment $L=2$. The
branching ratio to the S-wave channel appeared to be by almost a factor 2
smaller than the value obtained in the analysis in Ref.~\cite{Man:92}. However, our value is in a good
agreement with the multichannel analysis of Ref.~\cite{Vrana:1999nt} and the analysis of electroproduction data of Ref.~\cite{Mokeev:2012vsa}. The
branching ratio of the $N(1535)1/2^-$ state into the $\rho N$ channel
was found to be notably smaller: $2.7\pm 0.6\%$ for the decay into
S-wave and $0.5\pm0.5\%$ for the decay into D-wave. These numbers
are in a good agreement with the previous results (Refs. 
\cite{Man:92,Vrana:1999nt}). The branching ratios of these two
states into all channels which contribute to the $\pi\pi N$ final
state are listed in Tab.~\ref{tab_br}. Another baryon contributing to the two-pion production in the HADES experiment energy range is the Roper state (N(1440) $1/2^+$), although its coupling to the $\rho N$ channel calculated as residue in the pole position was found to be very small. Due to the small phase-space volume the branching ratio for the decay of the Roper resonance into the $\rho N$ channel appears to be less than 0.2\%. 

\begin{table}[bh!]
 \centering
\caption{The branching ratios (in \%) for the decay of $S_{11}(1535)$ and
$D_{13}(1520)$ into the different $\pi\pi$ channels.}
  \label{tab_br}
 \begin{tabular}{|l|c|c|c|c|c|}
 \hline
 \textbf{State} & \boldmath{$\Delta\pi$} &\boldmath{$\Delta\pi$}& \boldmath{N$\rho$}&\boldmath{N$\rho$}&\boldmath{N$\sigma$}\\

                & \textbf{L=0}       &\textbf{L=2}     &\textbf{L=0} & \textbf{L=2} &  \\
 \hline
$S_{11}(1535)$  &  -            & $3.0\pm 1.0$ & $2.7\pm 0.6$ & $0.5\pm
0.5$&- \\
 \hline
$D_{13}(1520)$  & $12.1\pm 2.1$& $6\pm 2$     & $11.8\pm 1.9$ & $0.4\pm 0.2$ &
$7\pm 3$\\
 \hline
 \end{tabular}
 \end{table}

\section{Conclusion}

The HADES collaboration measured the two-pion production in the exclusive $n\pi^+\pi^-$ and $p\pi^-\pi^0$ final states in pion-proton scattering at incident pion momenta of $p_{beam}=0.650,0.685,0.733$, and $0.786$ GeV/c. These new data have been included in the Bonn-Gatchina PWA accounting for  many other reaction channels measured in various experiments, studying pion- and photo-induced reactions. The solutions allow for the decomposition of the total cross sections into  partial waves with  total angular momentum and parity $J^P$ or into $\Delta\pi$, $N\rho$, $N\sigma$ isobars. The results have been discussed, based on the detailed comparison of the PWA solutions to the measured differential cross sections for $p_{beam}=0.685$ GeV/c and to the measured excitation function. 
We conclude that, in the second resonance region, the two-pion production in the $\pi^- p$ reaction is dominated  by the $J^P=1/2^{\pm}, 3/2^{\pm}$ partial waves with isospin $I=1/2$. The largest contributions are provided by the $D_{13}$ and $D_{13}$, $P_{11}$ partial waves in the $p\pi^-\pi^0$, and the $n\pi^+\pi^-$ final states, respectively. For the $p\pi^-\pi^0$ channel, the $p\rho^-$ contribution is dominating. The situation is different for the $n\pi^+\pi^-$ channel where  the $n\rho^0$ contribution is suppressed, due to smaller isospin coefficients (factor 2), and due to significant contributions of the $\Delta\pi$ and $N\sigma$ final states. Furthermore, we have found that the total cross section for the  $N\rho$ channel is influenced by interference effects between  $I=1/2$ and $I=3/2$ amplitudes, which are constructive for the $n\rho$ and destructive for the $p\rho^-$ case.  

The new data on the $\pi^-p\to \pi^+\pi^-n$ and $\pi^-p\to
\pi^+\pi^-n$ reactions provide an important information about decay properties
of the resonances in the region of center-of-mass energies around $1500$ MeV. In particular, this is a unique source  
to study the decay properties of the resonances into the
$\rho N$ channel. The combined analysis of our data with the Crystal
Ball data measured for the $\pi^-p\to \pi^0\pi^0n$ reaction indeed allows for a precise spin-parity and isotopic decomposition of all partial waves,
which was not possible from the analysis of the Crystal Ball
data alone. As a result, we identify unambiguously  
the contributions of all partial waves to the measured reactions and
determine the branching ratios of the $N(1535) \frac{1}{2}^-$ and
$N(1520) \frac{3}{2}^-$ resonances into the $\rho N$, $\Delta\pi$ and $N\sigma$ channels with good precision. 

This new analysis should be particularly useful for the decay into the $\rho N$ channel $BR=12.2\pm1.9\%$ and $BR=3.2\pm0.7\%$ for the $N(1520) \frac{3}{2}^-$ and $N(1535) \frac{1}{2}^-$ resonances, respectively,  as no information is available in the review of Particle Physics. Our result for the $N(1520)$ is different by a factor 2 from the value found in an earlier analysis of Ref.~\cite{Man:92}. The value of the branching ratio of baryon resonances into the $\rho N$ channel is important for the calculations of the in-medium $\rho$-meson spectral function, which is affected by the coupling to baryon resonances. In particular it confirms the dominant role of S-waves in the decay of these resonances.  It is also important for calculations of dilepton production via baryon resonance decays which rely on the modeling of time-like electromagnetic baryon transition form factors using VDM models.  In particular, the results obtained in this work on the $\rho$-meson production can be directly used for the on-going analysis of the $\pi^- p \rightarrow n e^+e^-$ reaction channel measured in the same energy range.


\newpage 

\vspace{5pc}

\section{Acknowledgments}
\label{sec3}
The collaboration gratefully acknowledges the support by SIP JUC Cracow, Cracow (Poland), National Science Center, 2016/23/P/ST2/040 POLONEZ, 2017/25/N/ST2/00580, 2017/26/M/ST2/00600; TU Darmstadt, Darmstadt (Germany), VH-NG-823, DFG GRK 2128, DFG CRC-TR 211, BMBF:05P18RDFC1; Goethe-University, Frankfurt (Germany) and TU Darmstadt, Darmstadt (Germany), ExtreMe Matter Institute EMMI at GSI Darmstadt; TU München, Garching (Germany), MLL München, DFG EClust 153, GSI TMLRG1316F, BmBF 05P15WOFCA, SFB 1258, DFG FAB898/2-2; NRNU MEPhI Moscow, Moscow (Russia), in framework of Russian Academic Excellence Project 02.a03.21.0005, Ministry of Science and Education of the Russian Federation 3.3380.2017/4.6; JLU Giessen, Giessen (Germany), BMBF:05P12RGGHM; IPN Orsay, Orsay Cedex (France), CNRS/IN2P3; NPI CAS, Rez, Rez (Czech Republic), MSMT LM2015049, OP VVV CZ.02.1.01/0.0/0.0/16 013/0001677, LTT17003. 
The work of A.Sarantsev and V.Nikonov
is supported by the grant from Russian Science Foundation (RSF 16-12-10267).


\newpage

\begin{thebibliography}{99}

\bibitem{Hoh:83}G. H\"{u}hler, Pion-Nucleon Scattering, Landold-B\"{u}rnstein vol I/9b2, (1983).
\bibitem{Cut:79}R.E. Cutkosky et al., Phys. Rev. D {\bf 20}, 2839, (1979).
\bibitem{Arndt:1990bp} 
  R.~A.~Arndt, Z.~j.~Li, L.~D.~Roper, R.~L.~Workman and J.~M.~Ford,
  Phys.\ Rev.\ D {\bf 43}, 2131 (1991).
 
 \bibitem{Arn:06}
  R.~A.~Arndt, W.~J.~Briscoe, I.~I.~Strakovsky and R.~L.~Workman,
  Phys.\ Rev.\ C {\bf 74} (2006) 045205
  doi:10.1103/PhysRevC.74.045205.
 
 \bibitem{Crede:2013sze} 
  V.~Crede and W.~Roberts,
  Rept.\ Prog.\ Phys.\  {\bf 76}, 076301 (2013)
  doi:10.1088/0034-4885/76/7/076301.
 
\bibitem{Anisovich:2011fc} 
  A.~V.~Anisovich, R.~Beck, E.~Klempt, V.~A.~Nikonov, A.~V.~Sarantsev and U.~Thoma,
  Eur.\ Phys.\ J.\ A {\bf 48}, 15 (2012)
  doi:10.1140/epja/i2012-12015-8.

\bibitem{pdg} 
  M.~Tanabashi {\it et al.} [Particle Data Group],
  Phys.\ Rev.\ D {\bf 98}, no. 3, 030001 (2018).
  doi:10.1103/PhysRevD.98.030001.

\bibitem{Man:84}D.M. Manley, R.A. Arndt, Y. Goradia, V.L. Teplitz, Phys. Rev. {\bf D} 30, 904, (1984).
\bibitem{Man:92}D.M. Manley and E.M. Saleski, Phys. Rev. D {\bf 45}, 4002, (1992).

\bibitem{Pra:04}
  S.~Prakhov {\it et al.} [Crystal Ball Collaboration],
  Phys.\ Rev.\ C {\bf 69} (2004) 045202.
  doi:10.1103/PhysRevC.69.045202. 
  
\bibitem{Ker:98}
  M.~Kermani {\it et al.} [CHAOS Collaboration],
  Phys.\ Rev.\ C {\bf 58} (1998) 3431.
  doi:10.1103/PhysRevC.58.3431.

\bibitem{Ale:98}. 
  I.~G.~Alekseev {\it et al.},
  Phys.\ Atom.\ Nucl.\  {\bf 61} (1998) 174
   [Yad.\ Fiz.\  {\bf 61} (1998) 223]
   [Nucl.\ Phys.\ B {\bf 541} (1999) 3]
  doi:10.1016/S0550-3213(98)00796-2.

\bibitem{Sokhoyan:2015fra} 
  V.~Sokhoyan {\it et al.} [CBELSA/TAPS Collaboration],
  Eur.\ Phys.\ J.\ A {\bf 51}, no. 8, 95 (2015)
  Erratum: [Eur.\ Phys.\ J.\ A {\bf 51}, no. 12, 187 (2015)]
  doi:10.1140/epja/i2015-15187-7, 10.1140/epja/i2015-15095-x.
  
\bibitem{Shklyar:2014kra}
  V.~Shklyar, H.~Lenske and U.~Mosel,
  Phys.\ Rev.\ C {\bf 93} (2016) no.4,  045206
  doi:10.1103/PhysRevC.93.045206.
  %
  
  \bibitem{Mokeev:2012vsa}
  V.~I.~Mokeev {\it et al.} [CLAS Collaboration],
  Phys.\ Rev.\ C {\bf 86} (2012) 035203
  doi:10.1103/PhysRevC.86.035203.






\bibitem{Sakurai:1960ju} 
  J.~J.~Sakurai,
  Annals Phys.\  {\bf 11}, 1 (1960).
  doi:10.1016/0003-4916(60)90126-3
  
\bibitem{Kroll:1967it} 
  N.~M.~Kroll, T.~D.~Lee and B.~Zumino,
  Phys.\ Rev.\  {\bf 157}, 1376 (1967).
  doi:10.1103/PhysRev.157.1376


\bibitem{Adamczewski-Musch:2019byl} 
  J.~Adamczewski-Musch {\it et al.},
  Nature Phys.\  {\bf 15}, no. 10, 1040 (2019).
  doi:10.1038/s41567-019-0583-8
  
\bibitem{Rapp:1999ej} 
  R.~Rapp and J.~Wambach,
  Adv.\ Nucl.\ Phys.\  {\bf 25}, 1 (2000)
  doi:10.1007/0-306-47101-9$_1$.
  
\bibitem{Peters:1997va} 
  W.~Peters, M.~Post, H.~Lenske, S.~Leupold and U.~Mosel,
  Nucl.\ Phys.\ A {\bf 632}, 109 (1998)
  doi:10.1016/S0375-9474(98)00803-3.

\bibitem{Post:2000qi} 
  M.~Post, S.~Leupold and U.~Mosel,
  Nucl.\ Phys.\ A {\bf 689}, 753 (2001)
  doi:10.1016/S0375-9474(00)00613-8.
  
\bibitem{Bus:12}
  O.~Buss {\it et al.},
  Phys.\ Rept.\  {\bf 512} (2012) 1
  doi:10.1016/j.physrep.2011.12.001.

\bibitem{Wei:12}
  J.~Weil, H.~van Hees and U.~Mosel,
  Eur.\ Phys.\ J.\ A {\bf 48} (2012) 111
   Erratum: [Eur.\ Phys.\ J.\ A {\bf 48} (2012) 150]
  doi:10.1140/epja/i2012-12111-9, 10.1140/epja/i2012-12150-2.

\bibitem{Bas:98}
  S.~A.~Bass {\it et al.},
  Prog.\ Part.\ Nucl.\ Phys.\  {\bf 41} (1998) 255
   [Prog.\ Part.\ Nucl.\ Phys.\  {\bf 41} (1998) 225]
  doi:10.1016/S0146-6410(98)00058-1
 
  
\bibitem{Bratkovskaya:2013vx}
  E.~L.~Bratkovskaya, J.~Aichelin, M.~Thomere, S.~Vogel and M.~Bleicher,
  Phys.\ Rev.\ C {\bf 87} 064907 (2013)
  doi:10.1103/PhysRevC.87.064907.
  
\bibitem{Staudenmaier:2017vtq} 
  J.~Staudenmaier, J.~Weil, V.~Steinberg, S.~Endres and H.~Petersen,
  Phys.\ Rev.\ C {\bf 98}, no. 5, 054908 (2018)
   doi:10.1103/PhysRevC.98.054908.
  
  
  \bibitem{Adamczewski-Musch:2017hmp}
  J.~Adamczewski-Musch {\it et al.} [HADES Collaboration],
  Phys.\ Rev.\ C {\bf 95}, no.6,  065205 (2017)
   doi:10.1103/PhysRevC.95.065205. 
  
\bibitem{Adamczewski-Musch:2017oij} 
  J.~Adamczewski-Musch {\it et al.} [HADES Collaboration],
  Eur.\ Phys.\ J.\ A {\bf 53}, no. 7, 149 (2017)
   doi:10.1140/epja/i2017-12341-3.
  
    \bibitem{HADES:2011ab} 
  G.~Agakishiev {\it et al.} [HADES Collaboration],
  Eur.\ Phys.\ J.\ A {\bf 48}, 64 (2012) doi:10.1140/epja/i2012-12064-y.
   doi:10.1140/epja/i2012-12064-y.
  
    \bibitem{Agakishiev:2014wqa} 
  G.~Agakishiev {\it et al.},
  Eur.\ Phys.\ J.\ A {\bf 50}, 82 (2014)
  doi:10.1140/epja/i2014-14082-1.

  
\bibitem{Faessler:2000md} 
  A.~Faessler, C.~Fuchs, M.~I.~Krivoruchenko and B.~V.~Martemyanov,
  J.\ Phys.\ G {\bf 29}, 603 (2003)
  doi:10.1088/0954-3899/29/4/302.
  
\bibitem{OConnell:1995nse} 
  H.~B.~O'Connell, B.~C.~Pearce, A.~W.~Thomas and A.~G.~Williams,
  Prog.\ Part.\ Nucl.\ Phys.\  {\bf 39}, 201 (1997)
  doi:10.1016/S0146-6410(97)00044-6 .
  
\bibitem{Friman:1997tc} 
  B.~Friman and H.~J.~Pirner,
  Nucl.\ Phys.\ A {\bf 617}, 496 (1997)
  doi:10.1016/S0375-9474(97)00050-X.
  
\bibitem{Post:2000rf} 
  M.~Post and U.~Mosel,
  Nucl.\ Phys.\ A {\bf 688}, 808 (2001)
  doi:10.1016/S0375-9474(00)00583-2 .
  
\bibitem{Krivoruchenko:2001jk} 
  M.~I.~Krivoruchenko, B.~V.~Martemyanov, A.~Faessler and C.~Fuchs,
  Annals Phys.\  {\bf 296}, 299 (2002)
  doi:10.1006/aphy.2002.6223 .
  
\bibitem{Aga:09}
  G.~Agakishiev {\it et al.} [HADES Collaboration],
  Eur.\ Phys.\ J.\ A {\bf 41} (2009) 243
  doi:10.1140/epja/i2009-10807-5.

\bibitem{Sim:99}
  R.~S.~Simon {\it et al.} [PION Collaboration],
  Prog.\ Part.\ Nucl.\ Phys.\  {\bf 42}, 247 (1999).
  doi:10.1016/S0146-6410(99)00080-0.

\bibitem{Dia:02}
  J.~Dı́az {\it et al.},
  Nucl.\ Instrum.\ Meth.\ A {\bf 478} (2002) 511.
  doi:10.1016/S0168-9002(01)00895-6.

\bibitem{Lal:13}R. Lalik et al., 2013 IEEE Nuclear Science Symposium and Medical Imaging Conference (2013), DOI: 10.1109/NSSMIC.2013.6829440

\bibitem{Ada:17}
  J.~Adamczewski-Musch {\it et al.} [HADES Collaboration],
  Eur.\ Phys.\ J.\ A {\bf 53} (2017) no.9,  188.
  doi:10.1140/epja/i2017-12365-7.

\bibitem{Frohlich:2007bi}
  I.~Fr\"ohlich {\it et al.},
  PoS ACAT {\bf } 076  (2007)
  doi:10.22323/1.050.0076
  [arXiv:0708.2382 [nucl-ex]].
  
  \bibitem{Brody1971}
  A.D. Brody et al.,Phys. Rev  D{\bf 3}, 2619, (1971) 

\bibitem{Przy:16}W. Przygoda, JPS Conf. Proc. {\bf 10}, 010013 (2016).

\bibitem{said}CNS Data Analysis Center, SAID Home Page, http://gwdac.phys.gwu.edu (2018)

\bibitem{epecur}
  I.~G.~Alekseev {\it et al.} [EPECUR Collaboration],
  Phys.\ Rev.\ C {\bf 91} (2015) no.2,  025205
  doi:10.1103/PhysRevC.91.025205.

\bibitem{Anisovich:2004zz}
  A.~Anisovich, E.~Klempt, A.~Sarantsev and U.~Thoma,
  Eur.\ Phys.\ J.\ A {\bf 24}, 111 (2005)
  doi:10.1140/epja/i2004-10125-6
  
\bibitem{Anisovich:2011zz}
  A.~V.~Anisovich, V.~A.~Nikonov, A.~V.~Sarantsev, V.~V.~Anisovich, M.~A.~Matveev, T.~O.~Vulfs, K.~V.~Nikonov and J.~Nyiri,
  Phys.\ Rev.\ D {\bf 84}, 076001 (2011).
  doi:10.1103/PhysRevD.84.076001


\bibitem{bgwp} pwa.hiskp.uni-bonn.de 

\bibitem{Muller:2019qxg} 
  J.~M\"{u}uller {\it et al.} [CBELSA/TAPS Collaboration],
  Phys.\ Lett.\ B {\bf 803} (2020) 135323
  doi:10.1016/j.physletb.2020.135323


\bibitem{Vrana:1999nt}
  T.~P.~Vrana, S.~A.~Dytman and T.~S.~H.~Lee,
  Phys.\ Rept.\  {\bf 328}, 181 (2000)
  doi:10.1016/S0370-1573(99)00108-8
\end{thebibliography}
\end{document}